\documentclass[twocolumn]{aastex62}
\usepackage{amsmath}
\usepackage{graphicx}
\usepackage{natbib}
\usepackage[scaled]{helvet}
\usepackage{epsfig}
\usepackage{url}
\bibpunct{(}{)}{;}{a}{}{,}
\usepackage{textcomp}
\usepackage{mathpazo}
\usepackage{xspace}
\usepackage{hyperref}
\usepackage{apjfonts}
\usepackage{mathrsfs}
\usepackage{verbatim}
\usepackage{tabularx}

\usepackage[utf8]{inputenc}
\usepackage{array}
\usepackage{makecell}

\usepackage{color}
\usepackage[normalem]{ulem}
\newcommand{\bjdtdb}{\ensuremath{\rm {BJD_{TDB}}}}

\newcommand{\teff}{\ensuremath{T_{\rm eff}}\xspace}

\newcommand{\msun}{\ensuremath{\,M_\Sun}}
\newcommand{\rsun}{\ensuremath{\,R_\Sun}}
\newcommand{\lsun}{\ensuremath{\,L_\Sun}}
\newcommand{\mj}{\ensuremath{\,M_{\rm J}}}
\newcommand{\rj}{\ensuremath{\,R_{\rm J}}}

\newcommand{\fave}{\langle F \rangle}
\newcommand{\fluxcgs}{10$^9$ erg s$^{-1}$ cm$^{-2}$}

\newcommand{\loggstar}{\ensuremath{\log{g_\star}}}
\newcommand{\vsini}{\ensuremath{v\sin{I_*}}}
\newcommand{\kms}{\,km\,s$^{-1}$}

\newcommand{\thisstarone}{KELT-25\xspace}
\newcommand{\thisstartwo}{KELT-26\xspace}

\newcommand{\thisplanetone}{KELT-25\,b\xspace}
\newcommand{\thisplanettwo}{KELT-26\,b\xspace}

\newcommand{\thisticone}{TIC 65412605\xspace}
\newcommand{\thistictwo}{TIC 160708862\xspace}

\newcommand{\be}{\begin{equation}}
\newcommand{\ee}{\end{equation}}

\begin{document}

\title{KELT-25b \& KELT-26b: A Hot Jupiter and a Substellar Companion Transiting Young A-Stars Observed by \textit{TESS}\footnote{This paper includes data gathered with the 6.5 meter Magellan Telescopes located at Las Campanas Observatory, Chile.}}

\newcommand{\cfa}{Center for Astrophysics \textbar \ Harvard \& Smithsonian, 60 Garden St, Cambridge, MA 02138, USA}
\newcommand{\andresbello}{Departamento de Ciencias Fisicas, Universidad Andres Bello, Fernandez Concha 700, Las Condes, Santiago, Chile}
\newcommand{\noao}{National Optical Astronomy Observatory, 950 North Cherry Avenue, Tucson, AZ 85719, USA}
\newcommand{\gemini}{Gemini Observatory, Northern Operations Center, 670 N. A’ohoku Place, Hilo, HI 96720, USA}
\newcommand{\sifa}{School of Physics, Sydney Institute for Astronomy (SIfA), The University of
Sydney, NSW 2006, Australia}
\newcommand{\georgemason}{George Mason University, 4400 University Drive MS 3F3, Fairfax, VA 22030, USA}
\newcommand{\recons}{RECONS Institute, Chambersburg, PA 17201, USA}
\newcommand{\gsu}{Department of Physics and Astronomy, Georgia State University, Atlanta, GA 30302, USA}
\newcommand{\umich}{Astronomy Department, University of Michigan, 1085 S University Avenue, Ann Arbor, MI 48109, USA}
\newcommand{\utaustin}{Department of Astronomy, The University of Texas at Austin, Austin, TX 78712, USA}
\newcommand{\MIT}{Department of Physics and Kavli Institute for Astrophysics and Space Research, Massachusetts Institute of Technology, Cambridge, MA 02139, USA}
\newcommand{\MITEPS}{Department of Earth, Atmospheric and Planetary Sciences, Massachusetts Institute of Technology,  Cambridge,  MA 02139, USA}
\newcommand{\louisiana}{Department of Physics and Astronomy, Louisiana State University, Baton Rouge, LA 70803 USA}
\newcommand{\uflorida}{Department of Astronomy, University of Florida, 211 Bryant Space Science Center, Gainesville, FL, 32611, USA}
\newcommand{\riverside}{Department of Earth Sciences, University of California, Riverside, CA 92521, USA}
\newcommand{\nanjing}{School of Astronomy and Space Science, Key Laboratory of Modern Astronomy and
Astrophysics in Ministry of Education, Nanjing University, Nanjing 210046, Jiangsu, China}
\newcommand{\usq}{University of Southern Queensland, Centre for Astrophysics, West Street, Toowoomba, QLD 4350 Australia}
\newcommand{\ames}{NASA Ames Research Center, Moffett Field, CA, 94035, USA}
\newcommand{\geneva}{Observatoire de l’Universit\'e de Gen\`eve, 51 chemin des Maillettes,
1290 Versoix, Switzerland}
\newcommand{\unsw}{
Exoplanetary Science at UNSW, School of Physics, UNSW Sydney, NSW 2052, Australia}
\newcommand{\uw}{Astronomy Department, University of Washington, Seattle, WA 98195 USA}
\newcommand{\warwick}{Department of Physics, University of Warwick, Gibbet Hill Road, Coventry CV4 7AL, UK}
\newcommand{\warwickceh}{Centre for Exoplanets and Habitability, University of Warwick, Gibbet Hill Road, Coventry CV4 7AL, UK}
\newcommand{\princeton}{Department of Astrophysical Sciences, Princeton University, 4 Ivy Lane, Princeton, NJ, 08544, USA}
\newcommand{\liege}{Space Sciences, Technologies and Astrophysics Research (STAR) Institute, Universit\'e de Li\`ege, 19C All\'ee du 6 Ao\^ut, 4000 Li\`ege, Belgium}
\newcommand{\vanderbilt}{Department of Physics and Astronomy, Vanderbilt University, Nashville, TN 37235, USA}
\newcommand{\unm}{Department of Physics and Astronomy, University of New Mexico, 1919 Lomas Blvd NE, Albuquerque, NM 87131, USA}

\newcommand{\fisk}{Department of Physics, Fisk University, 1000 17th Avenue North, Nashville, TN 37208, USA}
\newcommand{\columbia}{Department of Astronomy, Columbia University, 550 West 120th Street, New York, NY 10027, USA}
\newcommand{\toronto}{Dunlap Institute for Astronomy and Astrophysics, University of Toronto, Ontario M5S 3H4, Canada}
\newcommand{\unc}{Department of Physics and Astronomy, University of North Carolina at Chapel Hill, Chapel Hill, NC 27599, USA}
\newcommand{\iac}{Instituto de Astrof\'isica de Canarias (IAC), E-38205 La Laguna, Tenerife, Spain}
\newcommand{\lalaguna}{Departamento de Astrof\'isica, Universidad de La Laguna (ULL), E-38206 La Laguna, Tenerife, Spain}
\newcommand{\louisville}{Department of Physics and Astronomy, University of Louisville, Louisville, KY 40292, USA}
\newcommand{\aavso}{American Association of Variable Star Observers, 49 Bay State Road, Cambridge, MA 02138, USA}
\newcommand{\utokyo}{The University of Tokyo, 7-3-1 Hongo, Bunky\={o}, Tokyo 113-8654, Japan}
\newcommand{\naoj}{National Astronomical Observatory of Japan, 2-21-1 Osawa, Mitaka, Tokyo 181-8588, Japan}
\newcommand{\jstpresto}{JST, PRESTO, 2-21-1 Osawa, Mitaka, Tokyo 181-8588, Japan}
\newcommand{\astrobiojapan}{Astrobiology Center, 2-21-1 Osawa, Mitaka, Tokyo 181-8588, Japan}
\newcommand{\ctio}{Cerro Tololo Inter-American Observatory, Casilla 603, La Serena, Chile}
\newcommand{\nexsci}{Caltech IPAC -- NASA Exoplanet Science Institute 1200 E. California Ave, Pasadena, CA 91125, USA}
\newcommand{\ucsc}{Department of Astronomy and Astrophysics, University of
California, Santa Cruz, CA 95064, USA}
\newcommand{\gsfc}{Exoplanets and Stellar Astrophysics Laboratory, Code 667, NASA Goddard Space Flight Center, Greenbelt, MD 20771, USA}
\newcommand{\sgtinc}{SGT, Inc./NASA AMES Research Center, Mailstop 269-3, Bldg T35C, P.O. Box 1, Moffett Field, CA 94035, USA}
\newcommand{\chile}{Center of Astro-Engineering UC, Pontificia Universidad Cat\'olica de Chile, Av. Vicu\~{n}a Mackenna 4860, 7820436 Macul, Santiago, Chile}
\newcommand{\Pontificia}{Instituto de Astrof\'isica, Pontificia Universidad Cat\'olica de Chile, Av.\ Vicu\~na Mackenna 4860, Macul, Santiago, Chile}
\newcommand{\Millennium}{Millennium Institute for Astrophysics, Chile}
\newcommand{\maxplank}{Max-Planck-Institut f\"ur Astronomie, K\"onigstuhl 17, Heidelberg 69117, Germany}
\newcommand{\utdallas}{Department of Physics, The University of Texas at Dallas, 800 West
Campbell Road, Richardson, TX 75080-3021 USA}
\newcommand{\MauryLewin}{Maury Lewin Astronomical Observatory, Glendora, CA 91741, USA}
\newcommand{\umbc}{University of Maryland, Baltimore County, 1000 Hilltop Circle, Baltimore, MD 21250, USA}
\newcommand{\osu}{Department of Astronomy, The Ohio State University, 140 West 18th Avenue, Columbus, OH 43210, USA}
\newcommand{\MITAA}{Department of Aeronautics and Astronautics, MIT, 77 Massachusetts Avenue, Cambridge, MA 02139, USA}
\newcommand{\openu}{School of Physical Sciences, The Open University, Milton Keynes MK7 6AA, UK}
\newcommand{\swarthmore}{Department of Physics and Astronomy, Swarthmore College, Swarthmore, PA 19081, USA}
\newcommand{\seti}{SETI Institute, Mountain View, CA 94043, USA}
\newcommand{\lehigh}{Department of Physics, Lehigh University, 16 Memorial Drive East, Bethlehem, PA 18015, USA}
\newcommand{\utah}{Department of Physics and Astronomy, University of Utah, 115 South 1400 East, Salt Lake City, UT 84112, USA}
\newcommand{\USNA}{Department of Physics, United States Naval Academy, 572C Holloway Rd., Annapolis, MD 21402, USA}
\newcommand{\DTM}{Department of Terrestrial Magnetism, Carnegie Institution for Science, 5241 Broad Branch Road, NW, Washington, DC 20015, USA}
\newcommand{\UPenn}{The University of Pennsylvania, Department of Physics and Astronomy, Philadelphia, PA, 19104, USA}
\newcommand{\montana}{Department of Physics and Astronomy, University of Montana, 32 Campus Drive, No. 1080, Missoula, MT 59812 USA}
\newcommand{\psu}{Department of Astronomy \& Astrophysics, The Pennsylvania State University, 525 Davey Lab, University Park, PA 16802, USA}
\newcommand{\psust}{Center for Exoplanets and Habitable Worlds, The Pennsylvania State University, 525 Davey Lab, University Park, PA 16802, USA}
\newcommand{\Kutztown}{Department of Physical Sciences, Kutztown University, Kutztown, PA 19530, USA}
\newcommand{\udel}{Department of Physics \& Astronomy, University of Delaware, Newark, DE 19716, USA}
\newcommand{\Westminster}{Department of Physics, Westminster College, New Wilmington, PA 16172}
\newcommand{\steward}{Department of Astronomy and Steward Observatory, University of Arizona, Tucson, AZ 85721, USA}
\newcommand{\saao}{South African Astronomical Observatory, PO Box 9, Observatory, 7935, Cape Town, South Africa}
\newcommand{\salt}{Southern African Large Telescope, PO Box 9, Observatory, 7935, Cape Town, South Africa}
\newcommand{\ssl}{Societ\`{a} Astronomica Lunae, Italy}
\newcommand{\spot}{Spot Observatory, Nashville, TN 37206, USA}
\newcommand{\txamGP}{George P.\ and Cynthia Woods Mitchell Institute for Fundamental Physics and Astronomy, Texas A\&M University, College Station, TX77843 USA}
\newcommand{\txam}{Department of Physics and Astronomy, Texas A\&M university, College Station, TX 77843 USA}
\newcommand{\wellesley}{Department of Astronomy, Wellesley College, Wellesley, MA 02481, USA}
\newcommand{\byu}{Department of Physics and Astronomy, Brigham Young University, Provo, UT 84602, USA}
\newcommand{\Hazelwood}{Hazelwood Observatory, Churchill, Victoria, Australia}
\newcommand{\pest}{Perth Exoplanet Survey Telescope}
\newcommand{\Winer}{Winer Observatory, PO Box 797, Sonoita, AZ 85637, USA}
\newcommand{\icpo}{Ivan Curtis Private Observatory}
\newcommand{\elsauce}{El Sauce Observatory, Chile}
\newcommand{\crow}{Atalaia Group \& CROW Observatory, Portalegre, Portugal}
\newcommand{\dfus}{Dipartimento di Fisica ``E.R.Caianiello'', Universit\`a di Salerno, Via Giovanni Paolo II 132, Fisciano 84084, Italy}
\newcommand{\indfn}{Istituto Nazionale di Fisica Nucleare, Napoli, Italy}
\newcommand{\sotes}{Gabriel Murawski Private Observatory (SOTES)}
\newcommand{\stromlo}{Research School of Astronomy and Astrophysics, Mount Stromlo Observatory, Australian National University, Cotter Road, Weston, ACT, 2611, Australia}
\newcommand{\wyoming}{Department of Physics \& Astronomy, University of Wyoming, 1000 E University Ave, Dept 3905, Laramie, WY 82071, USA}
\newcommand{\thai}{National Astronomical Research Institute of Thailand, 260, Moo 4, T. Donkaew, A. Mae Rim, Chiang Mai, 50180, Thailand}
\newcommand{\asp}{Acton Sky Portal (private observatory), Acton, MA 01720,  USA}
\newcommand{\kyoto}{Department of Physics, Faculty of Science, Kyoto Sangyo University, Kamigamo Motoyama, Kita-ku, Kyoto, 603-8555, Japan}
\newcommand{\chiba}{Planetary Exploration Research Center, Chiba Institute of Technology, 2-17-1 Tsudanuma, Narashino, Chiba 275-0016, Japan}
\newcommand{\berkely}{Department of Astronomy, University of California Berkeley, Berkeley, CA 94720-3411, USA}
\newcommand{\hawaii}{Institute for Astronomy, University of Hawai'i, 2680 Woodlawn Drive, Honolulu, HI 96822, USA}
\newcommand{\carnegie}{The Observatories of the Carnegie Institution for Science, 813 Santa Barbara St., Pasadena, CA 91101, USA}
\newcommand{\brazil}{Instituto de Astronomia, Geof\'isica e Ciencias Atmosf\'ericas, Universidade de S\`ao Paulo, Rua do Mat\~ao 1226, Cidade Universit\~aria, S\'ao Paulo, SP 05508-900, Brazil}
\newcommand{\bhicfa}{Black Hole Initiative at Harvard University, 20 Garden Street, Cambridge, MA 02138, USA}
\newcommand{\lco}{Las Cumbres Observatory, 6740 Cortona Drive, Suite 102, Goleta, CA 93117, USA}
\newcommand{\FFL}{\altaffiliation{Future Faculty Leaders Fellow}}
\newcommand{\torres}{\altaffiliation{Juan Carlos Torres Fellow}}
\newcommand{\sagan}{\altaffiliation{NASA Sagan Fellow}}
\newcommand{\bernoulli}{\altaffiliation{Bernoulli fellow}}
\newcommand{\gruber}{\altaffiliation{Gruber fellow}}
\newcommand{\kavli}{\altaffiliation{Kavli Fellow}}
\newcommand{\peg}{\altaffiliation{51 Pegasi b Fellow}}
\newcommand{\pappalardo}{\altaffiliation{Pappalardo Fellow}}
\newcommand{\hubble}{\altaffiliation{NASA Hubble Fellow}}
\newcommand{\carnegiefellow}{\altaffiliation{Carnegie Fellow}}

\correspondingauthor{Romy Rodr\'iguez Mart\'inez}
\email{rodriguezmartinez.2@buckeyemail.osu.edu}

\author[0000-0003-1445-9923]{Romy Rodr\'iguez Mart\'inez}
\affiliation{\osu}

\author[0000-0003-0395-9869]{B. Scott Gaudi} 
\affiliation{\osu}

\author[0000-0001-8812-0565]{Joseph E. Rodriguez}
\affiliation{\cfa}

\author[0000-0002-4891-3517]{George Zhou}
\affiliation{\cfa}

\author[0000-0002-2919-6786]{Jonathan Labadie-Bartz}
\affiliation{\brazil}

\author[0000-0002-8964-8377]{Samuel N. Quinn}
\affiliation{\cfa}

\author[0000-0003-4464-1371]{Kaloyan Penev} 
\affiliation{\utdallas} 
\author[0000-0001-5603-6895]{Thiam-Guan Tan} 
\affiliation{\pest}

\author[0000-0001-9911-7388]{David W. Latham} 
\affiliation{\cfa} 

\author{Leonardo A. Paredes}  
\affiliation{\gsu}
\author[0000-0003-0497-2651]{John F.\ Kielkopf} 
\affiliation{\louisville}
\author[0000-0003-3216-0626]{Brett Addison}
\affiliation{\usq}
\author{Duncan J. Wright} 
\affiliation{\usq}
\author{Johanna Teske}
\altaffiliation{NASA Hubble Fellow} 
\affiliation{\carnegie}
\author[0000-0002-2532-2853]{Steve B.\ Howell} 
\affiliation{\ames}
\author{David Ciardi} 
\affiliation{\nexsci}
\author[0000-0002-0619-7639]{Carl Ziegler} 
\affiliation{\toronto}
\author[0000-0002-3481-9052]{Keivan G. Stassun} 
\affiliation{\vanderbilt}
\affiliation{\fisk}
\author[0000-0002-5099-8185]{Marshall C.\ Johnson} 
\affiliation{\lco}
\author[0000-0003-3773-5142]{Jason D. Eastman} 
\affiliation{\cfa}
\author[0000-0001-5016-3359]{Robert J.\ Siverd} 
\affiliation{\gemini}
\author[0000-0002-9539-4203]{Thomas G.\ Beatty} 
\affiliation{\steward}
\author[0000-0002-0514-5538]{Luke Bouma} 
\affiliation{\princeton}
\author{Timothy Bedding}
\affiliation{\sifa}
\author[0000-0002-3827-8417]{Joshua Pepper} 
\affiliation{\lehigh}
\author[0000-0002-4265-047X]{Joshua Winn}
\affiliation{\princeton}
\author[0000-0003-2527-1598]{Michael B. Lund} 
\affiliation{\nexsci}
\author[0000-0001-6213-8804]{Steven~Villanueva~Jr.} 
\pappalardo
\affiliation{\MIT}
\author[0000-0002-5951-8328]{Daniel J.\ Stevens}
\altaffiliation{Eberly Fellow}
\affiliation{\psu}
\affiliation{\psust}
\author[0000-0002-4625-7333]{Eric L.\ N.\ Jensen} 
\affiliation{\swarthmore}
\author{Coleman Kilby} 
\affiliation{\swarthmore}
\author[0000-0002-5226-787X]{Jeffrey D. Crane} 
\affiliation{\DTM}
\affiliation{\carnegie}
\author{Andrei Tokovinin}  
\affiliation{\ctio}
\author[0000-0002-0885-7215]{Mark E. Everett} 
\affiliation{\noao}
\author[0000-0002-7595-0970]{C.G. Tinney}
\affiliation{\unsw}
\author[0000-0002-9113-7162]{Michael Fausnaugh} 
\affiliation{\MIT}

\author[0000-0003-2995-4767]{David H.\ Cohen} 
\affiliation{\swarthmore}
\author{Daniel Bayliss}
\affiliation{\warwick}
\affiliation{\warwickceh}
\author[0000-0001-6637-5401]{Allyson Bieryla} 
\affiliation{\cfa}
\author[0000-0002-1617-8917]{Phillip A. Cargile}
\affiliation{\cfa}
\author[0000-0001-6588-9574]{Karen A. Collins}
\affiliation{\cfa}
\author[0000-0003-2239-0567]{Dennis M.\ Conti} 
\affiliation{\aavso}
\author[0000-0001-8020-7121]{Knicole D.\ Col\'on}
\affiliation{\gsfc}
\author{Ivan A.\ Curtis}
\affiliation{\icpo}
\author{D.\ L.\ Depoy}
\affiliation{\txamGP}
\affiliation{\txam}
\author[0000-0002-5674-2404]{Phil Evans}
\affiliation{\elsauce}
\author[0000-0002-2457-7889]{Dax L. Feliz}
\affiliation{\vanderbilt}
\author{Joao Gregorio}
\affiliation{\crow}
\author{Jason Rothenberg}
\affiliation{\wyoming}
\author[0000-0001-5160-4486]{David J. James}
\affiliation{\cfa}
\affiliation{\bhicfa}
\author[0000-0003-0634-8449]{Michael D.\ Joner}
\affiliation{\byu}
\author[0000-0002-4236-9020]{Rudolf B. Kuhn}
\affiliation{\saao}
\affiliation{\salt}
\author{Mark Manner}
\affiliation{\spot}
\author[0000-0002-1910-7065]{Somayeh Khakpash}
\affiliation{\lehigh}
\author{Jennifer L.\ Marshall}  
\affiliation{\txamGP}
\affiliation{\txam}
\author[0000-0001-9504-1486]{Kim K.\ McLeod}
\affiliation{\wellesley}
\author[0000-0001-7506-5640]{Matthew T.\ Penny}
\affiliation{\louisiana}
\author[0000-0002-5005-1215]{Phillip A.\ Reed}
\affiliation{\Kutztown}
\author{Howard M. Relles} 
\affiliation{\cfa}
\author{Denise C.\ Stephens}
\affiliation{\byu}
\author[0000-0003-2163-1437]{Chris Stockdale}
\affiliation{\Hazelwood}
\author{Mark Trueblood} 
\affiliation{\Winer}
\author{Pat Trueblood} 
\affiliation{\Winer}
\author[0000-0003-4554-5592]{Xinyu Yao}
\affiliation{\lehigh}
\author{Roberto Zambelli}
\affiliation{\ssl}

\author{Roland Vanderspek} 
\affiliation{\MIT}
\author[0000-0002-6892-6948]{Sara Seager}
\affiliation{\MIT}
\affiliation{\MITEPS}
\affiliation{\MITAA}
\author{Jon M. Jenkins}
\affiliation{\ames}

\author{Todd J. Henry}  
\affiliation{\gsu}
\author{Hodari-Sadiki James}  
\affiliation{\gsu}
\author{Wei-Chun Jao}  
\affiliation{\gsu}
\author[0000-0002-6937-9034]{Sharon Xuesong Wang}
\affiliation{\carnegie}
\author{Paul Butler} 
\affiliation{\DTM}
\author{Ian Thompson}
\affiliation{\carnegie}

\author{Stephen Shectman} 
\affiliation{\carnegie}

\author{Robert Wittenmyer} 
\affiliation{\usq}
\author{Brendan P. Bowler}
\affiliation{\utaustin}
\author[0000-0002-1160-7970]{Jonathan Horner}
\affiliation{\usq}
\author{Stephen R. Kane}
\affiliation{\riverside}
\author{Matthew W. Mengel}
\affiliation{\usq}
\author{Timothy D. Morton}
\affiliation{\uflorida}
\author{Jack Okumura}
\affiliation{\usq}
\author{Peter Plavchan}
\affiliation{\georgemason}
\author{Hui Zhang}
\affiliation{\nanjing}

\author[0000-0003-1038-9702]{Nicholas J. Scott} 
\affiliation{\ames}
\author[0000-0001-7233-7508]{Rachel A. Matson} 
\affiliation{\ames}

\author[0000-0003-3654-1602]{Andrew W. Mann}
\affiliation{\unc}

\author[0000-0003-2313-467X]{Diana Dragomir} 
\affiliation{\unm}

\author{Max Günther} 
\altaffiliation{Juan Carlos Torres Fellow}
\affiliation{\MIT}

\author{Eric B. Ting} 
\affiliation{\ames}

\author{Ana Glidden} 
\affiliation{\MIT}
\author{Elisa V. Quintana} 
\affiliation{\gsfc}



\shortauthors{Rodriguez Martinez et al.}

\begin{abstract}
We present the discoveries of KELT-25b (TIC 65412605, TOI-626.01) and KELT-26b (TIC 160708862, TOI-1337.01), two transiting companions orbiting relatively bright, early A-stars.  The transit signals were initially detected by the KELT survey, and subsequently confirmed by \textit{TESS} photometry. KELT-25b is on a 4.40-day orbit around the V = 9.66 star CD-24 5016 ($T_{\rm eff} = 8280^{+440}_{-180}$ K, $M_{\star}$ = $2.18^{+0.12}_{-0.11}$ $M_{\odot}$), while KELT-26b is on a 3.34-day orbit around the V = 9.95 star HD 134004 ($T_{\rm eff}$ =$8640^{+500}_{-240}$ K, $M_{\star}$ = $1.93^{+0.14}_{-0.16}$ $M_{\odot}$), which is likely an Am star. We have confirmed the sub-stellar nature of both companions through detailed characterization of each system using ground-based and \textit{TESS} photometry, radial velocity measurements, Doppler Tomography, and high-resolution imaging. For KELT-25, we determine a companion radius of $R_{\rm P}$ = $1.64^{+0.039}_{-0.043}$ $R_{\rm J}$, and a 3-sigma upper limit on the companion's mass of $\sim64~M_{\rm J}$. For KELT-26b, we infer a planetary mass and radius of $M_{\rm P}$ = $1.41^{+0.43}_{-0.51}$ $M_{\rm J}$ and $R_{\rm P}$ = $1.940^{+0.060}_{-0.058}$ $R_{\rm J}$. From Doppler Tomographic observations, we find KELT-26b to reside in a highly misaligned orbit. This conclusion is weakly corroborated by a subtle asymmetry in the transit light curve from the \textit{TESS} data. KELT-25b appears to be in a well-aligned, prograde orbit, and the system is likely a member of a cluster or moving group.  

\end{abstract}

\keywords{planetary systems, planets and satellites: detection,  stars: individual (\thisstarone,\thisstartwo)}

\section{Introduction} 
The field of exoplanets has grown tremendously since the first detection of a transiting exoplanet around a bright star \citep{charbonneau:2000,henry:2000} two decades ago. Thousands of planets\footnote{Almost 4100, as of 2019 (https://exoplanetarchive.ipac.caltech.edu/)} have been validated orbiting stars of almost every spectral type, and span a wide range of masses, orbits and likely compositions. Our knowledge of the demographics of short-period giant exoplanets quickly expanded with the advent of dedicated wide-field transit surveys from the ground such as The Hungarian-made Automated Telescope Network (HATNet; \citealp{Bakos:2007}), the HATSouth survey \citep{bakos:2013}, the Wide Angle Search for Planets (WASP/SuperWASP; \citealp{pollacco:2006,colliercameron:2009}), the Qatar Exoplanet Survey (QES; \citealp{alsubai:2011}), XO \citep{mccullough:2005}, the Trans-Atlantic Exoplanet Survey network (TrES; \citealp{Alonso:2007}), and the Kilodegree Extremely Little Telescope (KELT; \citealp{Pepper:2007,Pepper:2012}) to mention a few.

The {\it Kepler} space telescope \citep{borucki:2010} provided the first statistical survey of a large number of transiting planets over a broad region of radius and orbital period.  {\it Kepler} transformed our understanding of the population of relatively short period ($P \la 100$~days) planets. Later, the next generation of dedicated wide-field surveys came online, such as the Next Generation Transit Survey (NGTS; \citealt{bayliss:2018}) and the Multi-site All-Sky CAmeRA (MASCARA: \citealt{lesage:2014}). In addition, based largely on arguments presented in \citet{Gould:2003b} and \citet{Blake:2007},  targeted ground-based surveys were initiated such as MEarth (\citealp{nutzman:2008, Charbonneau:2009,berta:2012}), TRAPPIST \citep{Gillon:2017}, and SPECULOOS \citep{delrez:2018}, which primarily  concentrate their efforts on the search for exoplanets around low-mass stars. 

Massive and hot stars are typically avoided by transit surveys because planets around them induce weaker photometric signals (as a result of the lower planet-star radius and mass ratios). More importantly, the host stars have fewer spectral lines, and the lines they do have are significantly broadened by their fast rotation. A large fraction of stars above the Kraft break ($T_{\rm eff}\ga 6250\rm{K}$; \citealp{Kraft:1967}) are observed to rotate significantly faster (\vsini $\geq$ 10 km $\rm s^{-1}$) than cooler stars.  This is because stars with $T_{\rm eff}\ga 6250\rm{K}$ have essentially no convective envelopes, and thus do not slow down due to magnetic braking over their evolution. They therefore essentially retain their high primordial spin rates. The relative paucity of spectral features coupled with their faster rotation rates make candidate planetary companions transiting hot stars harder to confirm using the radial velocity method (see e.g., \citealt{Johnson:2018} and \citealt{dholakia:2019}).
To attempt to circumvent the challenges in searching for planets around massive stars on the main sequence, a number of radial velocity surveys have studied ``retired A-stars’’ -- cooler, evolved stars that have moved off the main sequence. Such surveys have yielded a number of discoveries (e.g., \citealt{johnson:2007,johnson:2011}), but have only been able to sample a relatively small number of target stars due to the focussed nature of radial velocity studies.

Despite the observational challenges posed by hot stars, they provide opportunities to study the most massive, highly-irradiated, close-in planets -- in particular the relatively new category of ``Ultra Hot Jupiters'' \citep{CollierCameron:2010}. Ultra Hot Jupiters provide a unique opportunity for the detailed atmospheric characterization of highly irradiated giant planets.  They have high equilibrium temperatures, allowing one to probe extreme conditions that are not present in the Solar System. In particular, many of these Ultra Hot Jupiters have day-side temperatures that are hot enough to disassociate all molecules, leaving atomic metals as the dominant species on the day side \citep{Gaudi:2017}.  Furthermore, the exceptionally high day-side temperatures of these planets imply that they are typically close to thermodynamic equilibrium \citep{kitzmann:2018}, making the interpretation of observations much simpler, and resulting in atmospheres that are likely quite different than typical hot Jupiters \citep{kitzmann:2018,bell:2018,lothringer:2018}.  In addition, A-type hosts present opportunities to test the effects of host star mass and binarity, and short evolutionary timescales, on giant planet formation, evolution, and engulfment by their host stars. For example, a recent study found that most giant planets around A-type stars are eventually engulfed by their host stars \citep{stephan:2018}, and a related study determined the observable effects of engulfment \citep{stephan:2019}.  In general, the high temperatures and high scale heights of Ultra Hot Jupiters make follow-up observations much easier, allowing one to test models of hot Jupiter atmospheres (e.g., \citealt{cauley:2019}; \citealt{hoeijmakers:2019}).

Studying the planetary population of massive (A) stars is also interesting for other reasons.  First, the large amount of high-energy radiation they emit helps to test theories of planet atmospheric evaporation.  Second, the fact that A stars tend to be rapidly rotating means that they are typically oblate, resulting in significant gravity darkening, which can be used to estimate the true (not just projected) spin-orbit angle of the planet \citep{Barnes:2009}. The oblateness of the host star, combined with certain orbital alignments of the planet, can result in relatively short precession times of the planetary orbit \citep{Johnson:2015}. Finally, their short lifetimes imply that the lifetimes of close-in planets orbiting such stars is likely to be much shorter than that of such planets orbiting low-mass stars \citep{stephan:2018}.

\citealt{royer:2007} found that typical rotational values (\vsini) are greater than 100 km s$^{-1}$ for stars in the the B9-F2 spectral range (for reference, the Sun's rotational velocity is only 1.6 $\pm$ 0.3 km s$^{-1}$; \citealp{pavlenko:2012}). Because of the difficulties in confirming planets orbiting these fast rotators, KELT routinely uses a combination of radial velocity (RV) and Doppler Tomographic (DT) observations \citep{CollierCameron2010}. DT measures the distortion in the spectral lines of a star caused by the transiting planet blocking the light from the star with different projected velocities as it crosses the disk of the star. Doppler tomographic observations can help confirm that the planet transits the target star and is not, for example, a signal from a nearby eclipsing binary, although confirming the planetary nature of the occultor also requires an appropriately stringent upper limit on its mass (see, e.g., \citealp{Bieryla:2015} for a discussion of validating planets orbiting rapidly rotating stars using Doppler Tomography). The advantage of this technique is that it is better suited to faster stellar rotations, thereby providing an alternative way to confirm planets around hot stars. Many discovery papers have demonstrated that A-type stars are the most suitable for the measurement of the Rossiter-McLaughlin (RM) effect because they have the optimal combination of $R_p/R_s$ and rotation -- two parameters to which the RM signal is proportional (see also \citealp{Gaudi:2007}, and Table \ref{tab:Astarplanets} of this paper).

\begin{figure}[ht!]
    \vspace{.0in}
    \centering\includegraphics[width=\linewidth, clip]{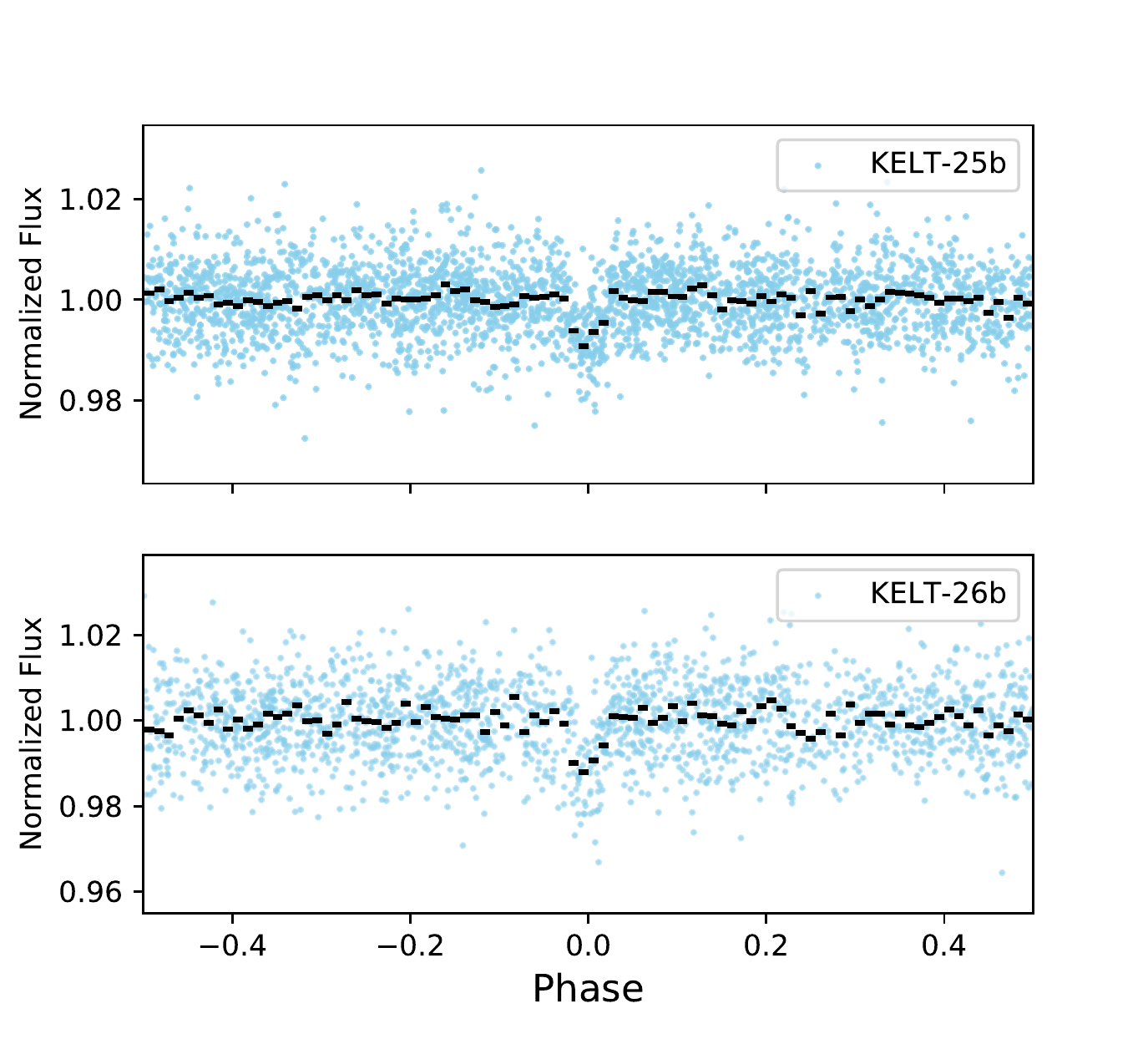}
    \caption{Discovery light curves for \thisplanetone (\textbf{Top}) and \thisplanettwo (\textbf{Bottom}) from the KELT-South telescope. The light curves are phase-folded on the preliminary orbital periods of 4.401093 days (\thisplanetone) and 3.344886 days (\thisplanettwo). The blue points represent the data and the black points are the data binned at intervals of 0.01 in phase.} 
    \label{fig:DiscoveryLC}
\end{figure}

With Doppler Tomography, we can not only validate new planets, but also determine their projected spin-orbit angles, $\lambda$. From the current sample of over 200 systems\footnote{\url{https://www.astro.keele.ac.uk/jkt/tepcat/obliquity.html}} with measured $\lambda$, a dichotomy in stellar temperature has emerged: giant planets around cool stars ($T_{\rm eff}$ < 6200 K) have lower obliquities than those around hot stars \citep{Schlaufman:2010,Winn:2010, Dawson:2015}. In addition, the most massive planets ($M_{p} > 3 M_{J}$) tend to have lower spin-orbit angles \citep{Hebrard:2010}. These results are important because they help constrain planet formation and migration mechanisms for hot Jupiters. One area of ongoing research is the question of whether hot Jupiters form ``in situ'' or at larger separations from their host and then migrate into their present observed locations via planet-disk or planet-planet dynamical or secular interactions, or secular interactions with a distant planetary, brown dwarf, or stellar companion (e.g., \citealp{Dawson:2018}). Measurements of $\lambda$ provide insights into these formation channels since different theories predict different values of the distribution of spin-orbit angles. 

Giant planets at close separations are highly irradiated by their host stars, and this intense radiation can significantly impact their properties. One possible consequence is radius inflation: it has been observed that hot Jupiters' radii appear to increase with increasing incident radiation from the host star \citep{Demory:2011,howell:2019}, although it is not clear whether this is caused by the radiation re-inflating the planet, or because it simply slows the cooling and contraction process of the planets, which may be hot and thus inflated upon formation. 

In this paper, we present the discoveries of a sub-stellar companion and likely planet (\thisplanetone)  and a hot Jupiter (\thisplanettwo\footnote{Whilst we were writing this manuscript, we noted a paper on arXiv in which WASP announced the discovery of WASP-178b \citep{Hellier:2019}.  We had already collected and analyzed the data needed to confirm \thisplanettwo, and other than to confirm that \thisstartwo and WASP-178 had the same coordinates, we did not read or use the results of \citet{Hellier:2019} in any way. We therefore claim an independent discovery (regardless of whether or not they are the same planet).  If they are the same planet, we do not, of course, claim to be the first to have made the detection.}) both orbiting bright, early A-stars, first identified as candidates in KELT data and subsequently observed by the \textit{TESS} mission. Although the \textit{TESS} mission's main science driver is to measure small planets, simulations have estimated yields of thousands of giant planets, which include Jupiter-sized planets around bright stars (e.g. \citealt{Barclay:2018}). \textit{TESS}'s expected yield complements the discoveries made by KELT and other ground-based surveys and advances our understanding of giant planets around hot stars. 

\begin{table}
\scriptsize
\setlength{\tabcolsep}{1pt}
\centering
\caption{Literature Properties for \thisstarone and \thisstartwo \label{tab:LitProps}}
\begin{tabular}{llccc}
\hline 
\hline 
Parameter & Description &KELT-25&KELT-26 & Source\\
\hline 
Other identifiers& & CD-24 5016 & HD 134004 &  \\
& & TIC 65412605 & TIC 160708862 &  \\
& & TYC 6528-1639-1 & TYC 7829-2324-1 &  \\
$\alpha_{J2000}$\dotfill	&Right Ascension (RA)\dotfill & 07:12:29.55004 & 15:09:04.89304 & 1	\\
$\delta_{J2000}$\dotfill	&Declination (Dec)\dotfill & -24:57:12.82193& -42:42:17.78910 & 1	\\

\\
$l$\dotfill     & Galactic Longitude\dotfill & $237.5346109\degr$ & $328.1938368\degr$ & 1\\
$b$\dotfill     & Galactic Latitude\dotfill & $-6.79674034\degr$ & $+13.3150904\degr$ & 1\\
\\
B$_T$\dotfill	&Tycho B$_T$ mag.\dotfill & 9.841$\pm$0.019 &  10.083 $\pm$ 0.028	& 2	\\
V$_T$\dotfill	&Tycho V$_T$ mag.\dotfill & 9.655$\pm$0.018	&  9.961 $\pm$ 0.033	& 2	\\
${\rm G}$\dotfill   & Gaia $G$ mag.\dotfill &9.5960$\pm$0.0003& 9.912 $\pm$ 0.020& 1\\
\\
J\dotfill			& 2MASS J mag.\dotfill & 9.362 $\pm$ 0.03 &9.775 $\pm$ 0.030	& 3	\\
H\dotfill			& 2MASS H mag.\dotfill & 9.273 $\pm$ 0.02 &9.735 $\pm$ 0.020	    &  3	\\
K$_S$\dotfill			& 2MASS ${\rm K_S}$ mag.\dotfill & 9.248 $\pm$ 0.02& 9.703 $\pm$ 0.020 & 3	\\
\\
\textit{WISE1}\dotfill		& \textit{WISE1} mag.\dotfill & 9.213 $\pm$ 0.022 & 9.670  $\pm$ 0.030 & 4	\\
\textit{WISE2}\dotfill		& \textit{WISE2} mag.\dotfill & 9.245 $\pm$ 0.02 &9.683 $\pm$ 0.030  &  4 \\
\textit{WISE3}\dotfill		& \textit{WISE3} mag.\dotfill &  9.302 $\pm$ 0.033& 9.645 $\pm$ 0.043 &4	\\
\\
$\mu_{\alpha}$\dotfill		& Gaia DR2 proper motion\dotfill & $-2.276 \pm 0.06$ & $-10.011 \pm 0.122$ & 1 \\
                    & \hspace{3pt} in RA (mas yr$^{-1}$)	&  \\
$\mu_{\delta}$\dotfill		& Gaia DR2 proper motion\dotfill 	&  $0.338 \pm 0.075$ & $-5.652
 \pm 0.097$&  1 \\
                    & \hspace{3pt} in DEC (mas yr$^{-1}$) &  \\
$\pi^\dagger$\dotfill & Gaia Parallax (mas) \dotfill & 2.342$\pm$  0.043$^{\dagger}$ &  2.394 $\pm$  0.060$^{\dagger}$&  1 \\
$RV$\dotfill & Absolute radial \hspace{9pt}\dotfill  &  35.472$\pm 1.011$ & $-24.140\pm 0.045$ & This work \\
     & \hspace{3pt} velocity (\kms)  & \\
$d$\dotfill & Distance (pc)\dotfill & $427.0\pm 7.8$ & $417.7\pm10.5$ &  1 \\
$U^{*}$\dotfill & Space Velocity (\kms)\dotfill & $-13.40\pm0.58$  &$-24.63\pm0.37$ & \S\ref{sec:uvw} \\
$V$\dotfill       & Space Velocity (\kms)\dotfill &$-13.83\pm0.84$ &$6.75\pm0.58$ & \S\ref{sec:uvw} \\
$W$\dotfill       & Space Velocity (\kms)\dotfill &$-1.24\pm0.19$ &$
0.78\pm0.20$ & \S\ref{sec:uvw} \\
\hline
\end{tabular}
\begin{flushleft}
 \footnotesize{ \textbf{\textsc{NOTES:}}
 $\dagger$ Parallaxes here are corrected for the 82 $\mu$as offset reported in \citet{Stassun:2018}.\\
 $^*U$ is in the direction of the Galactic center. \\
 References: $^1$\citet{Gaia:2018},$^2$\citet{Hog:2000},$^3$\citet{Cutri:2003}, $^4$\citet{Zacharias:2017}
}
\end{flushleft}
\end{table}

\setlength{\tabcolsep}{2pt}

\begin{table*}
 \centering
 \caption{Follow-up photometric observations of KELT-25b and KELT-26b \label{tab:followup}}
 \begin{tabular}{ccrcccccc}
    \hline
    \hline
Target & Observatory & Date (UT) & Diameter (m) & Filter & FOV & Pixel Scale  & Exposure (s) \\
    \hline
    
KELT-25b & PEST & 2019 January 18 & 0.3 & $Rc$ & 31$\arcmin$ $\times$ 21$\arcmin$ &  1.2$\arcsec$ & 30 \\
KELT-25b & \textit{TESS} & \makecell{2019 January 7 -\\2019 February 7 }& 0.105 & \makecell{\textit{TESS} \\(600-1000nm)} & 24$^{\circ}$ $\times$ 24$^{\circ}$ &  21$\arcsec$ & 1800 \\

\hline 
 
KELT-26b & PEST & 2016 August 26 & 0.3 & $I$ & 31$\arcmin$ $\times$ 21$\arcmin$ &  1.2$\arcsec$ & 30& \\

KELT-26b & \makecell{Mt. Kent\\ CDK700} & 2018 March 20 & 0.6858 &$r^{\prime}$ & 27.3$\arcmin$ $\times$ 27.3$\arcmin$ &  0.4$\arcsec$ & 65 &\\

KELT-26b & \textit{TESS} & \makecell{2019 April 22 -\\ 2019 May 20} & 0.105 & \makecell{\textit{TESS}\\ (600-1000nm)} & 24$^{\circ}$ $\times$ 24$^{\circ}$ &  21$\arcsec$ & 1800 \\

        \hline
        \hline
 \end{tabular}

\begin{flushleft}
  \end{flushleft}
\end{table*}

\section{Discovery and Follow-up Observations}
\label{Obs}

\subsection{KELT Discovery}
\label{sec:KELT}

One survey that has contributed significantly to the discovery and study of Ultra Hot Jupiters is the Kilodegree Extremely Little Telescope survey (KELT\footnote{\url{https://keltsurvey.org}}; \citealp{Pepper:2007}). KELT observes $\sim$85\% of the sky targeting bright stars in the magnitude range 7.5 < V < 12, filling in the gap between radial velocity surveys and other transit surveys, which generally focus on brighter and fainter stars, respectively. KELT consists of two observatories: KELT-South (KS, \citealp{Pepper:2012}), located in Sutherland, South Africa, which surveys most of the southern hemisphere, and KELT-North (KN), which observes the northern hemisphere from Sonoita, Arizona. KN and KS have been successful at finding giant planets around hot stars, discovering 24 planets, of which 18 -- including the ones confirmed in this paper -- orbit A and F stars.

Using two separate 42mm-aperture telescopes, the KELT survey observes over 85\% of the entire sky with a 20$-$30 minute cadence. Each observing site has a Mamiya 645 80mm f/1.9 42mm lens with a 4k$\times$4k Apogee CCD on a Paramount ME mount. The KELT telescopes have a $23\arcsec$ pixel scale and a $26\arcdeg\times26\arcdeg$ field of view. The original goal of the KELT survey was to discover hot Jupiters orbiting bright ($7<V [mag]<12$) host stars, which are amenable to detailed characterization through transmission or eclipse spectroscopy. More recently, KELT has become a significant contributor to the understanding of planets around early-type stars, specifically with the discovery of 6 transiting hot Jupiter around A-stars, including the planets presented here \citep{Zhou:2016, Gaudi:2017, Lund:2017, Johnson:2018, Siverd:2018}. 

The likely planetary companion orbiting \thisticone\ (hereafter \thisplanetone) was first identified as a planet candidate following the reduction of KELT-South field KS35. \thisstarone\ is located at $\alpha_{\rm J2000} =$ 07$^{h}$ 12$^{m}$ 29$\fs$55, $\delta_{\rm J2000} =$ $-$24$\degr$57$\arcmin$12$\arcsec$82 \citep{Gaia:2018} and the KELT-South survey field 35 is centered at $\alpha_{\rm J2000}=07^h40^m12\fs0$, $\delta_{\rm J2000}=-20\arcdeg00\arcmin00\farcs0$. KS35 was observed 2,860 times from UT 2013 May 10 until UT 2017 October 1. The image reduction and candidate selection pipeline are described in detail in \citet{Siverd:2012} and \citet{Kuhn:2016}, and are briefly summarized here. To reduce the raw survey images, KELT uses a modified, image subtraction pipeline based on the ISIS software \citep{Alard:1998,Alard:2000}. The list of sources identified in the KELT fields are then cross-matched to the Tycho-2 \citep{Hog:2000} and UCAC4 \citep{Zacharias:2013} catalogs to obtain their proper motions. We then implement reduced proper motion (RPM) cuts to identify and exclude giants before searching for transit signals \citep{gould:2003a,collier-cameron:2007}. Finally, we search for transit-like features in all stars that passed the RPM cuts using the Box-fitting Least Squares (BLS) algorithm \citep{kovacs:2002}.

The planet orbiting \thistictwo (hereafter \thisplanettwo) is located at $\alpha_{J2000}=$ 15$^{h}$ 09$^{m}$ 04$\fs$89304, $\delta_{J2000} =$ $-$42$\degr$ 42$\arcmin$17$\arcsec$78910 \citep{Gaia:2018} in the KELT-South KS37 field, which is centered at $\alpha_{J2000}=$  15$^{h}$ 07$^{m}$ 12$\fs$00, $\delta_{J2000}$ -53$\degr$00$\arcmin$00$\arcsec$00. This field
was observed 2,085 times from UT 2013 September 5 until UT 2015 September 11. 

Both companions were identified as candidates from a periodicity search using the BLS algorithm. \thisplanetone was initially identified with a BLS orbital period of 4.40 days and a transit depth of 0.66\%, while \thisplanettwo was detected with a period of 3.34 days and a 1.1\% transit depth. The phase-folded KELT discovery light curves are shown in Figure \ref{fig:DiscoveryLC}. See Table~\ref{tab:LitProps} for literature information about the stellar hosts \thisstarone and \thisstartwo. 

\subsection{TESS Photometry}
\label{sec:tess}

The NASA \emph{Transiting Exoplanet Survey Satellite} (\textit{TESS}, \citealp{Ricker:2015}) was launched on April 18, 2018 with the primary goal of discovering and characterizing small ($R_{p} < 4R_{\oplus}$) exoplanets around bright, nearby stars. Presently, dozens of exoplanets have already been validated, while almost a thousand candidates await confirmation. The confirmed systems include a few giant planets on short-period orbits \citep{Brahm:2018, Nielsen:2019,Rodriguez:2019A}. 

\thisstarone was observed in Sector 7 by Camera 2 of the \textit{TESS} spacecraft between UT 2019 January 7 and February 7. We made use of the 30-minute cadence Full Frame Images (FFI) made available by the Mikulski Archive for Space Telescopes (MAST) and calibrated by the Science Processing Operations Center (SPOC) pipeline \citep{Jenkins:2016}. Cutouts of $10\times10$\,pixels were extracted from the FFIs around each target star via the MAST \emph{TESScut} tool, and aperture photometry was performed using the \emph{lightkurve} package \citep{lightkurve,barentsen19}. The target star aperture encompassing pixels around each target star with fluxes higher than 90\% of pixels in the cutout, and pixels nearby that do not encompass adjacent stars were used in the evaluation of the background flux. We accounted for dilution within the \textit{TESS} aperture by computing for and removing the light contribution of known nearby stars, as per their \textit{TESS} band magnitudes in TIC.  

\begin{figure*}[!ht]
\centering
\includegraphics[width=0.8\textwidth]{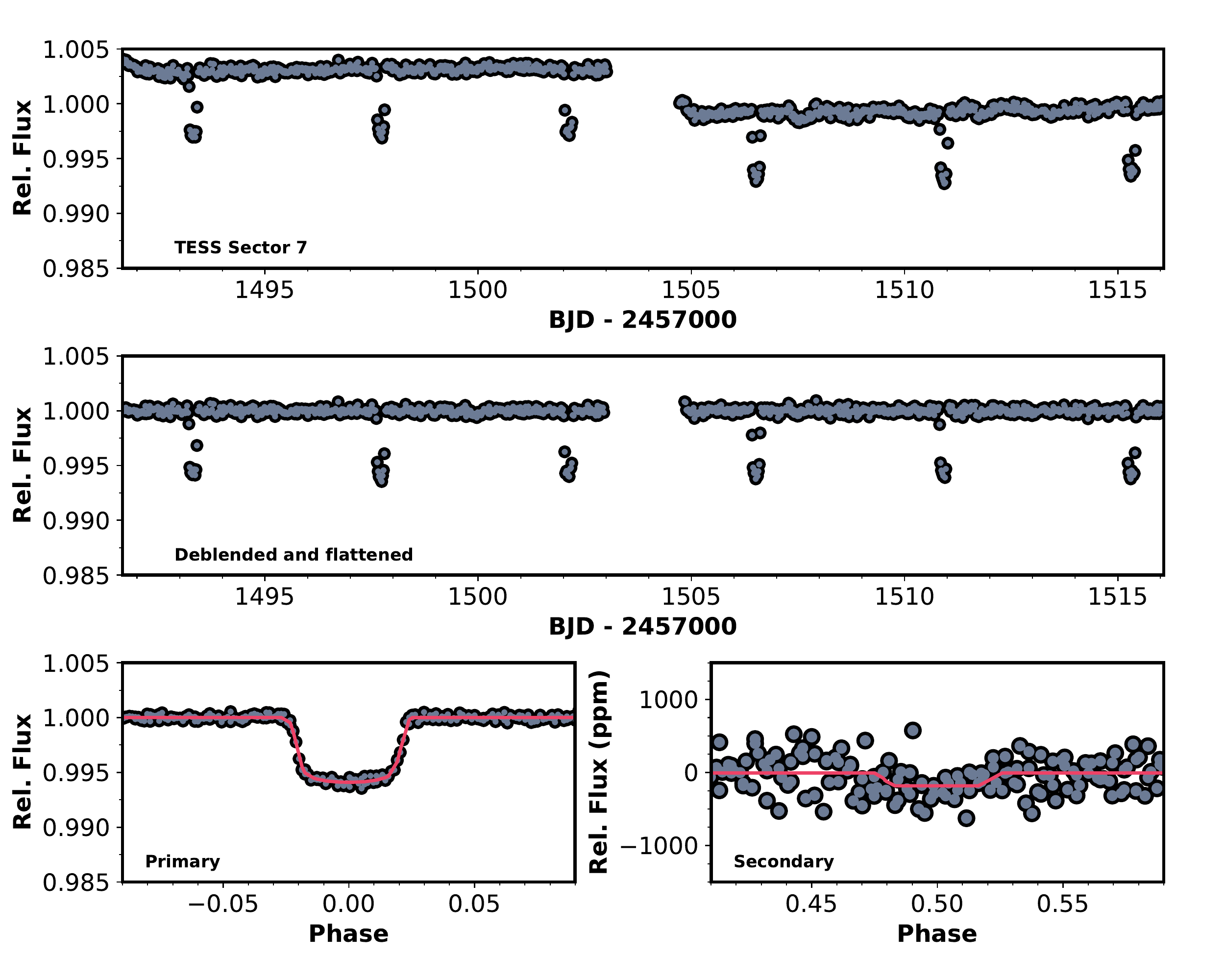}
\caption{\textit{TESS} light curve of \thisstarone. The \textbf{Top} panel shows the raw light curve; the \textbf{middle} panel shows the detrended light curve. The Bottom \textbf{Left} panel shows the transit and best-fit EXOFASTv2 model to the detrended light curve, phase-folded on the orbital period. The Bottom \textbf{Right} plot shows the region of the secondary eclipse, which is clearly detected in \textit{TESS}.}
\label{fig:tess25}
\end{figure*}

\begin{figure*}[!ht]
\centering
\includegraphics[width=0.8\textwidth]{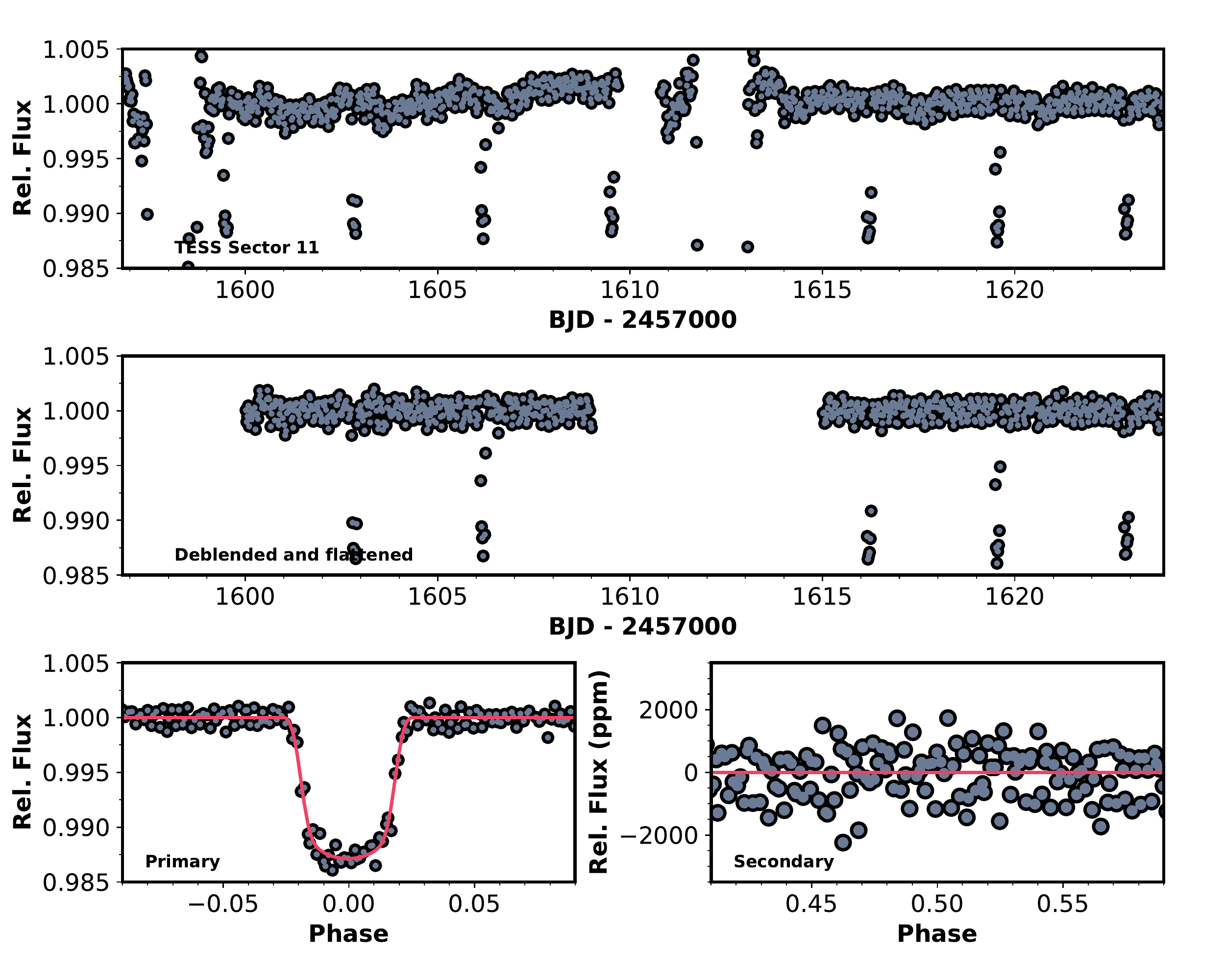}
\caption{\textit{TESS} light curve of \thisstartwo. The \textbf{Top} panel shows the raw light curve; the \textbf{middle} panel shows the detrended light curve. The Bottom \textbf{Left} panel shows the transit and best-fit EXOFASTv2 model to the detrended light curve, phase-folded on the orbital period. The Bottom \textbf{Right} plot shows the region of the secondary eclipse. While the secondary eclipse is not detected in the \textit{TESS} data, the primary phase-folded transit shows evidence for a slight asymmetry, which is likely real and may be caused by gravity darkening of the star. However, the primary also shows periodic photometric oscillations at a period that is nearly commensurate (1:18) with the period of the planet.  This variability may also be causing the slight asymmetry.
}
\label{fig:tess26}
\end{figure*}

Similarly, \thisstartwo was observed by \textit{TESS} between UT 2019 April 22 and May 20 during Sector 11 of the mission. The light curve extraction for \thisstartwo was the same as that described above. Figures~\ref{fig:tess25} and~\ref{fig:tess26} show the raw and reduced \textit{TESS} light curves for \thisplanetone and \thisplanettwo.

\begin{figure}[!ht]
\vspace{.1in}
\centering
\includegraphics[width=1\linewidth]{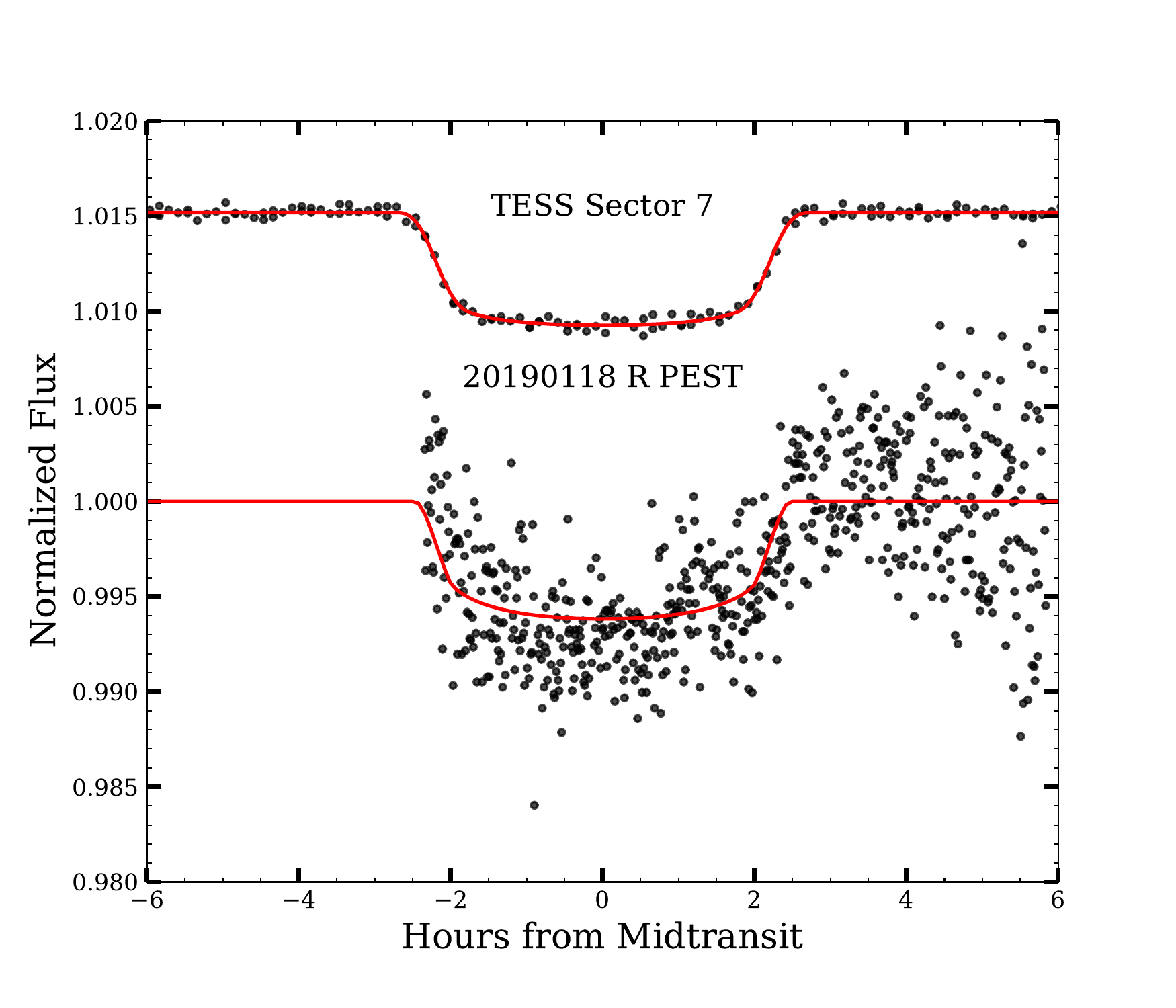}
\caption{The follow-up and \textit{TESS} light curves of \thisstarone. The KELT-FUN light curves phase-folded to the ephemeris determined in the global fit (Table \ref{tab:exofast_planetary}). Table \ref{tab:followup} contains information of all the KELT-FUN observations. The black points are the relative fluxes, while the red line shows the EXOFASTv2 model.}
\label{fig:all_lcs25} 
\end{figure}

\begin{figure}[!ht]
\vspace{.1in}
\centering
\includegraphics[width=1\linewidth]{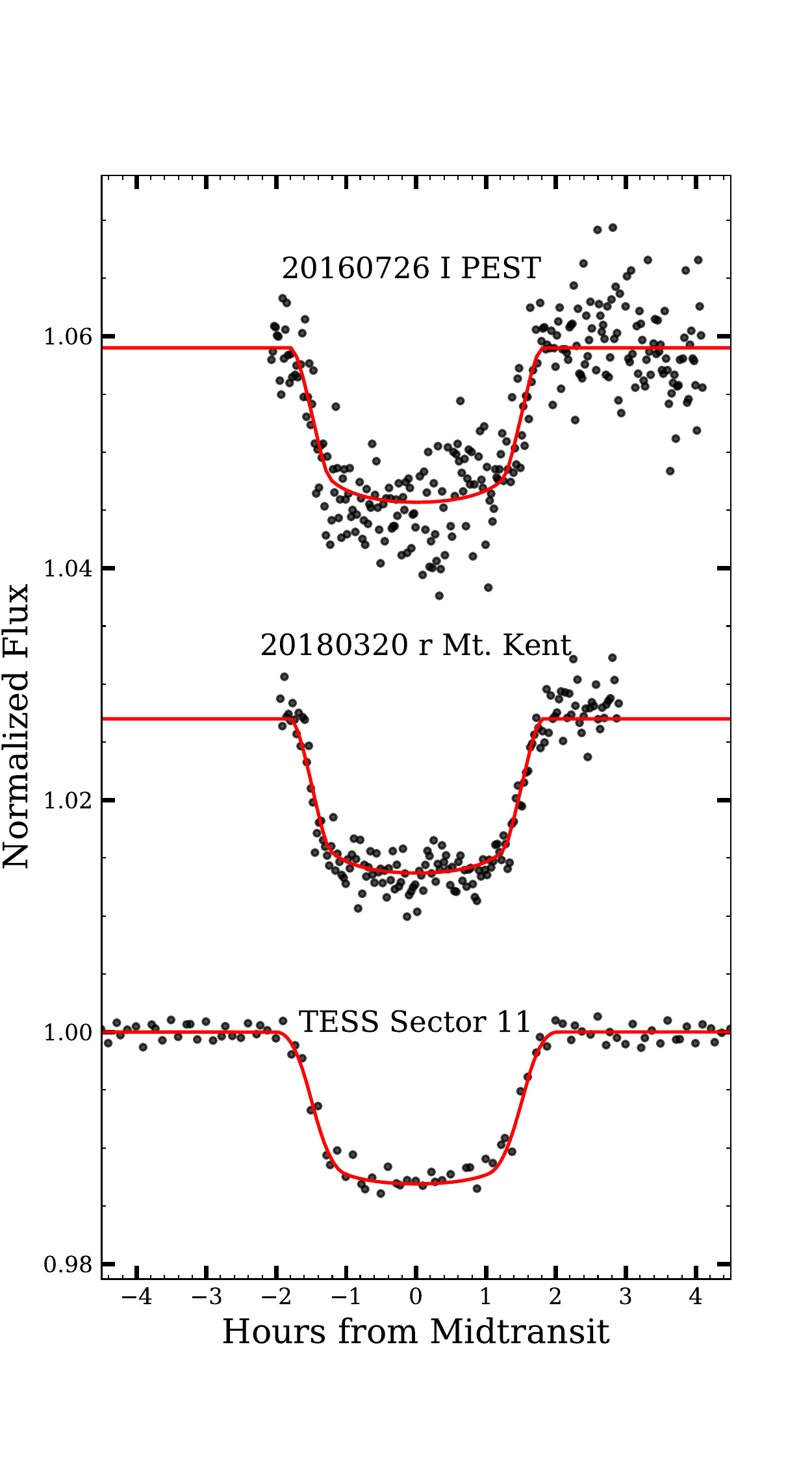}
\caption{The follow-up and \textit{TESS} light curves of \thisstartwo. The KELT-FUN light curves phase-folded to the ephemeris determined in the global fit (Table \ref{tab:exofast_planetary}). Table \ref{tab:followup} contains information of all the KELT-FUN observations. The black points are the relative fluxes, while the red line shows the EXOFASTv2 model.}
\label{fig:all_lcs26} 
\end{figure}


\subsection{Ground-based Photometry from the KELT Follow-up Network}
\label{sec:KFUN}

In order to confirm that the transit signals are due to bona fide planetary companions, rule out false positives such as eclipsing binaries, and refine the transit depth, duration and ephemeris of our candidates, we obtained photometric observations of \thisplanetone and \thisplanettwo from the KELT Follow-Up Network (KELT-FUN, \citealp{Collins:2018}). Some of the follow-up photometry was reduced using the \texttt{AstroImageJ} analysis software \citep{Collins:2017}. See Table~\ref{tab:followup} for technical information about the observatories that followed-up these systems. The KELT-FUN transits for both systems are shown in Figures ~\ref{fig:all_lcs25} and~\ref{fig:all_lcs26}. 

\begin{figure*}[!ht]
	\centering\vspace{.0in}
	\includegraphics[width=0.49\linewidth, trim={2.5cm 13cm 8.5cm 8cm}, clip]{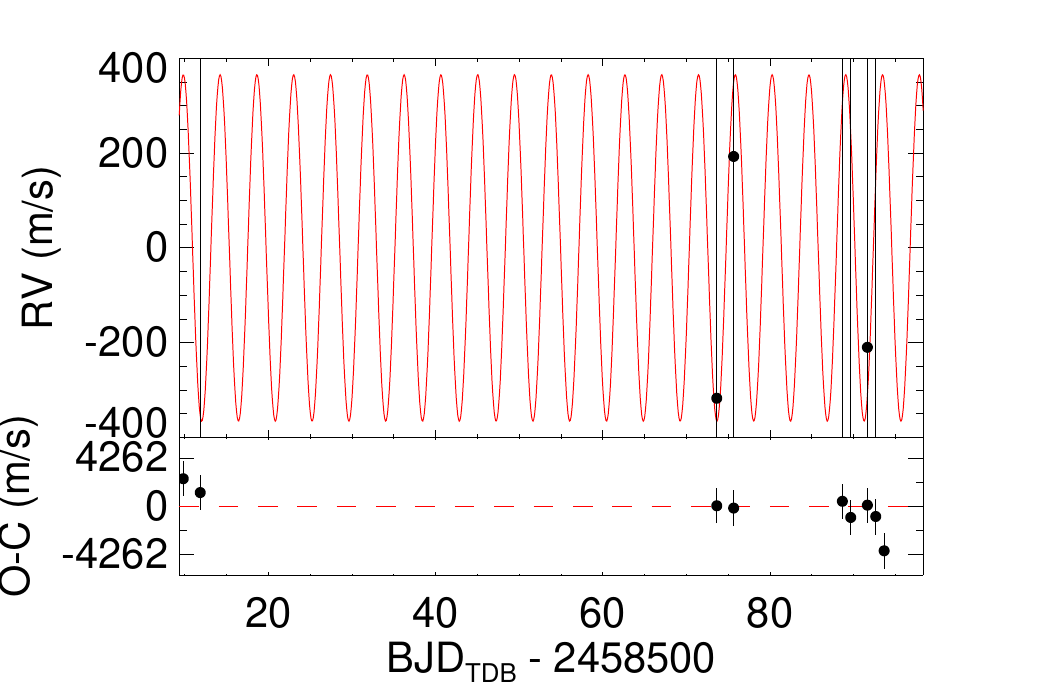}\includegraphics[width=0.49\linewidth, trim={2.5cm 13cm 8.5cm 8cm}, clip]{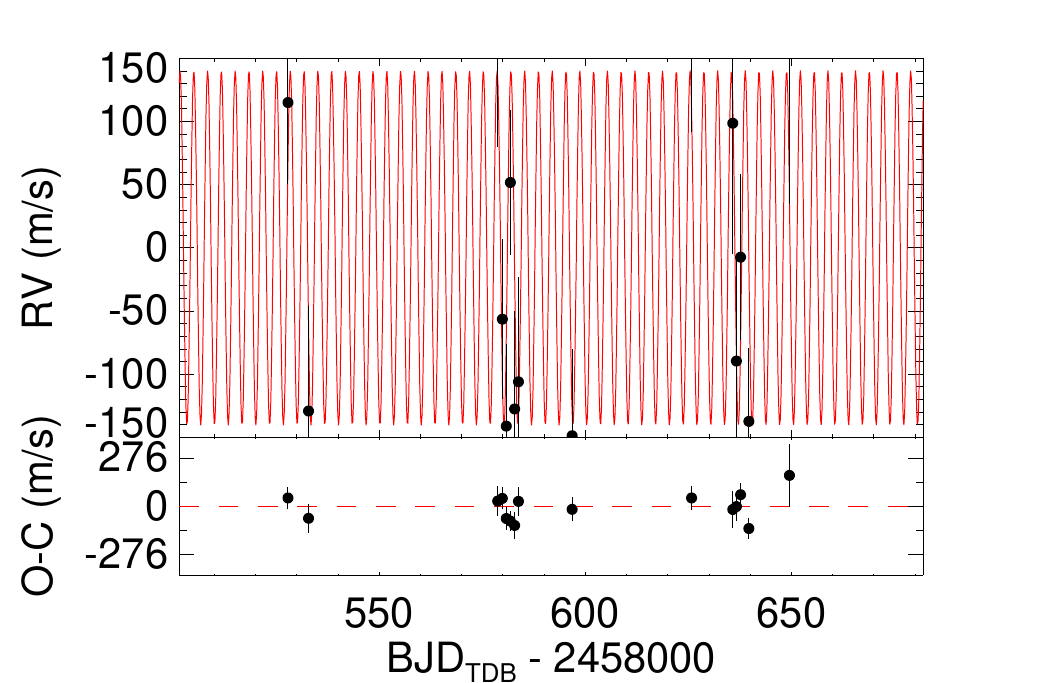}\\
	\includegraphics[width=.49\linewidth, trim={2.5cm 13cm 8.5cm 8cm}, clip]{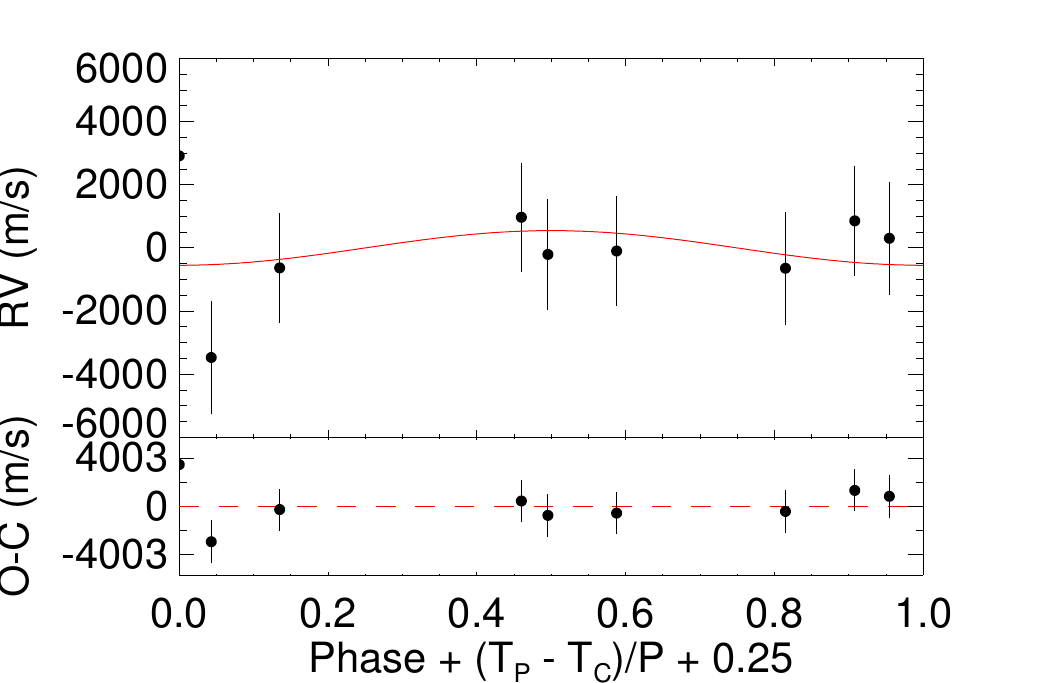}\includegraphics[width=.49\linewidth, trim={2.5cm 13cm 8.5cm 8cm}, clip]{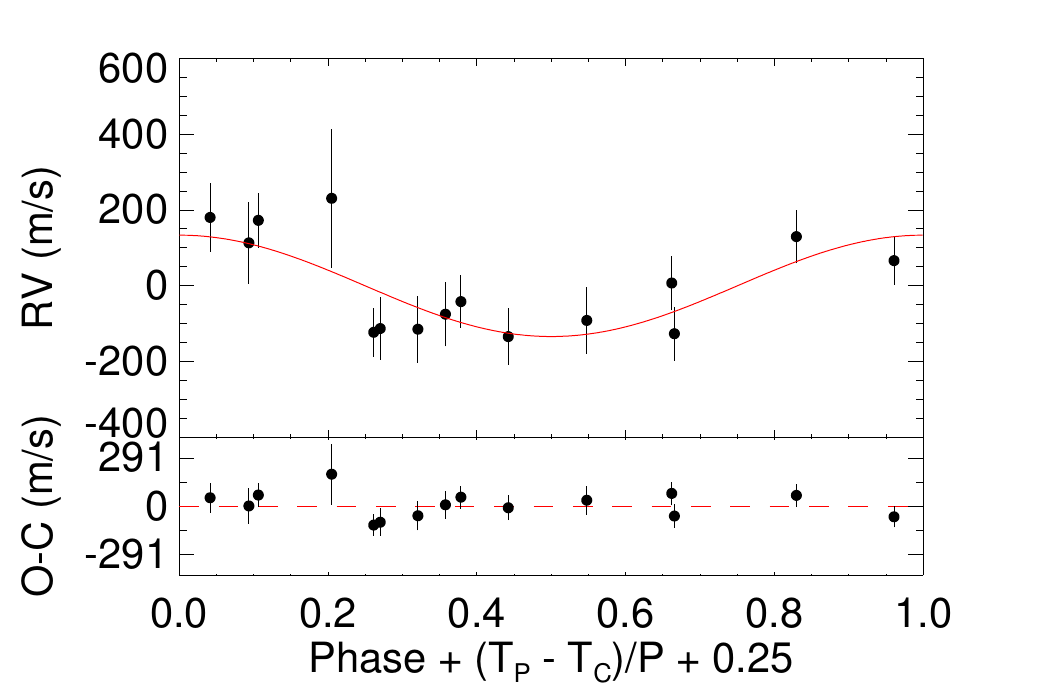}
    \caption{\textbf{(Top)} Radial velocity measurements for  \thisplanetone\ from TRES \textbf{(Left)} and  \thisplanettwo\ from CHIRON \textbf{(Right)}. \textbf{(Bottom)} The radial velocity measurements are phase-folded to the best determined period by EXOFASTv2. The EXOFASTv2 model is shown in red and the residuals to the best-fit are shown below each plot. While we measure the reflex Doppler signal from \thisstartwo to $\sim \sigma_K /K \simeq 30\%$, and thus are able to constrain the mass of \thisplanettwo, we do not obtain a definitive detection of the reflex Doppler signal from \thisstarone, and thus are only able to provide an upper limit on the planet mass.}
	\label{fig:RVs} 
\end{figure*}
 
\begin{deluxetable}{l l l l l}[bt]
\tabletypesize{\small}
\tablecaption{Relative Out-of-Transit Radial Velocities for KELT-25 from TRES \label{tab:K25_RV}}
\tablewidth{0pt}
\tablehead{
\colhead{\bjdtdb} & \colhead{RV (m s$^{-1}$)} & \colhead{$\sigma_{RV}$ (m s$^{-1}$)}}
\startdata
2458509.81953 & 3601.8 & 384.5\\
2458511.84309 & 1657.2 & 380.6\\
2458573.61546 & 480.7 & 457.1\\
2458575.63514 & 991.1 & 588.0\\
2458588.63382 & 1544.3 & 421.7\\
2458589.63349 & 54.5 & 452.6\\
2458591.62656 & 588.2 & 422.2\\
2458592.62509 & 41.9 & 593.6\\
2458593.63010 & -2782.7 & 620.8\\
\hline
\enddata
We note that while the radial velocities have been put on an absolute scale for the CHIRON data, the uncertainties are only relative to the mean.  There is an additional systematic uncertainty that would affect all of the data points by the same amount of roughly 40~$\rm m~s^{-1}$, due to the uncertainties of the standard radial velocity stars used to put the radial velocities on an absolute scale.
\end{deluxetable}

\begin{deluxetable}{l l l l l}[bt]
\tabletypesize{\small}
\tablecaption{Relative Out-of-Transit Radial Velocities for KELT-26 from CHIRON \label{tab:K26_RV}}
\tablewidth{0pt}
\tablehead{
\colhead{\bjdtdb} & \colhead{RV (m s$^{-1}$)} & \colhead{$\sigma_{RV}$ (m s$^{-1}$)}}
\startdata
2458527.84082 & 132.1 & 54.9\\
2458532.82920 & -112.1 & 75.8\\
2458578.72286 & 182.9 & 78.3 \\
2458579.85037 & -39.4 & 53.0\\
2458580.80996 & -124 & 55.3\\
2458581.79719 & 68.9 & 45.5\\
2458582.83253 & -110.5 & 69.3\\
2458583.76093 & -89 & 75.3\\
2458596.78738 & -131.7 & 59.3\\
2458625.76755 & 175.2 & 56.7\\
2458635.75959 & 115.8 & 97.9\\
2458636.64304 & -72.5 & 72.2\\
2458637.66033 & 9.7 & 56.1\\
2458639.66539 & -120.3 & 46.8\\
2458649.51090 & 233.2 & 178.2\\
\hline
\enddata
\end{deluxetable}


\subsubsection{Perth Exoplanet Survey Telescope} 
\label{sec:PEST}

The PEST home observatory was built in 2010 and has since helped discover dozens of exoplanets, including KELT candidates, mostly via the transit method. It is located in Perth, Australia, and is owned and run by Thiam-Guan Tan. The instrument is a 0.3m
Meade LX200 SCT f/10 and focal reducer yielding f/5. The camera is a SBIG ST-8XME with multiple filters including $I$, and it has a FOV of 31$\arcmin$ $\times$ 21$\arcmin$ and an image scale of 1.2$\arcsec$ per pixel. PEST observed a full transit of \thisplanetone on UT 2019 January 18 and a full transit of \thisplanettwo in the \textit{I} filter on UT 2016 July 26. 

\subsubsection{Mt Kent CDK700 Telescope} 
\label{sec:CDK700}

Photometric follow-up of \thisstartwo was taken with the University of Louisville's Shared Skies MKO-CDK700 telescope at Mt. Kent Observatory of the University of Southern Queensland, Australia. The instrument is a 0.7m Planewave corrected Dall-Kirkham (CDK) telescope with a Nasmyth focus. The telescope was used with an Apogee U16 CCD camera (Kodak KAF-16801E sensor). The CDK700 telescope observed a full transit of \thisplanettwo, acquiring 165 images in the Sloan $r^{\prime}$ filter on UT 2018 March 20.

\section{SPECTROSCOPIC OBSERVATIONS}
\label{sec:spectra}

\subsection{TRES Spectroscopy of KELT-25} \label{sec:TRES}

To constrain the planet mass and characterize the host star properties of \thisstarone, we made a series of spectroscopic observations with the Tillinghast Reflector Echelle Spectrograph (TRES;  \citealp{furesz:2008}) on the 1.5\,m telescope at the Fred Lawrence Whipple Observatory (FLWO) in Arizona, USA. TRES is a fiber-fed echelle spectrograph with a resolving power of $\lambda / \Delta \lambda \equiv R = 44,000$ spanning $3850 - 9100$\,\AA. A series of spectra were obtained for \thisstarone over 10 epochs from UT 2019 January 26 to UT 2019 May 20. These observations are reduced via the procedure described in \citet{Buchhave:2012}, and relative velocities were obtained via a multi-order cross correlation against an averaged observed spectral template, as per \citet{Quinn:2012}. The relative velocities are listed in Table~\ref{tab:K25_RV} and plotted in Figure~\ref{fig:RVs}. 

To establish the absolute systemic velocity of the system, we cross correlated the Mg b line order of one of the observed spectra against a synthetic template, and shifted all other velocities relative to this template. 

\subsection{CHIRON Spectroscopy of KELT-25 and KELT-26} \label{sec:CHIRON}

We obtained a series of spectroscopic observations of KELT-25 and KELT-26 with the CHIRON spectrograph on the SMARTS 1.5\,m telescope, located at Cerro Tololo Inter-American Observatory (CTIO), Chile. CHIRON is a fiber-fed echelle spectrograph, sliced via an image slicer, yielding a resolving power of $R=80,000$ over the wavelength range $4100-8700$\,\AA{} \citep{tokovinin:2013}. Wavelength calibrations are provided by bracketing Thorium-Argon (Th-Ar) arc lamp exposures. 

We used CHIRON to measure the spectroscopic orbit and characterize the host star of \thisplanettwo. A total of 15 CHIRON epochs covering all orbital phases of \thisplanettwo were obtained. Relative velocities were measured from each spectrum by deriving their stellar line broadening kernels via a Least Squares Deconvolution (LSD) analysis. These velocities are listed in Table~\ref{tab:K26_RV}, and Figure~\ref{fig:RVs} shows the radial velocities as a function of time and phase-folded by the photometric ephemeris.  

The absolute radial velocity was estimated by comparing the systemic velocity of \thisstartwo in the native CHIRON system to that of four radial velocity standard stars observed by CHIRON in order to determine a mean offset of $-1.455 \pm 0.037{\rm km~s^{-1}}$ of \thisstartwo systemic velocity relative to the CHIRON system.  We added the uncertainty in the systemic velocity in the CHIRON system of $0.025~{\rm km~s^{-1}}$ in quadrature to arrive at a final absolute systemic velocity of $-1.455 \pm 0.045{\rm km~s^{-1}}$. We corrected the individual CHIRON relative velocities and uncertainties in the same way.  

We also used CHIRON to observe the spectroscopic transit of KELT-25b on UT 2019 May 21. A total of 18 spectra were obtained covering the transit, with an integration time of 600\,s per exposure. Spectral line profiles were derived from each spectrum as per the procedure described in \S \ref{sec:PFS}. The spectroscopic shadow (i.e., Doppler Tomography signal) of the transiting companion is shown in Figure~\ref{fig:DTplots}. 

\subsection{Spectroscopic transit of KELT-25 with the Planet Finding Spectrograph}
\label{sec:PFS}

We monitored a transit of \thisplanetone with the Planet Finding Spectrograph (PFS; \citealp{Crane:2010}) on the 6.5\,m Magellan-Clay telescope at Las Campanas Observatory, Chile. A total of 19 spectra were obtained on UTC 2019 April 21 spanning 3.3 hours, each observation with an integration time of 600\,s. For our observations, PFS was fed by a $0\farcs3$ slit, yielding a spectral resolving power of $R=130,000$ over the wavelength region of $3910-7340$\,\AA. To enable better derivation of the stellar line profiles, the iodine cell was omitted from our observations. Wavelength calibrations were provided by Th-Ar hollow cathode lamp observations obtained at the beginning and end of the night. 

Stellar line profiles were derived from each spectrum as per \citet{CollierCameron2010} and \citet{Donati1997}, via a LSD analysis of the spectra against a non-rotating synthetic template spectrum generated via the ATLAS9 model atmospheres \citep{Castelli2003}. During the transit, the planet sequentially blocks parts of the rotating stellar disk, casting a shadow in our observed rotationally broadened line profiles of the star. When we subtract an averaged line profile from each observation, the residuals reveal the spectroscopic transit of the planet as a shadow traversing across the stellar surface. The line profile residuals and best fitting models are shown in Figure~\ref{fig:DTplots}. 

\subsection{Spectroscopic transit of KELT-26 with MINERVA-Australis}
\label{sec:MINERVA}

\begin{figure*}[!ht]
    \centering
    \begin{tabular}{ccc}
    \textbf{KELT-25b:CHIRON} & \textbf{KELT-25b:PFS} & \textbf{KELT-26b:MINERVA}\\ 
        \includegraphics[width=0.3\textwidth]{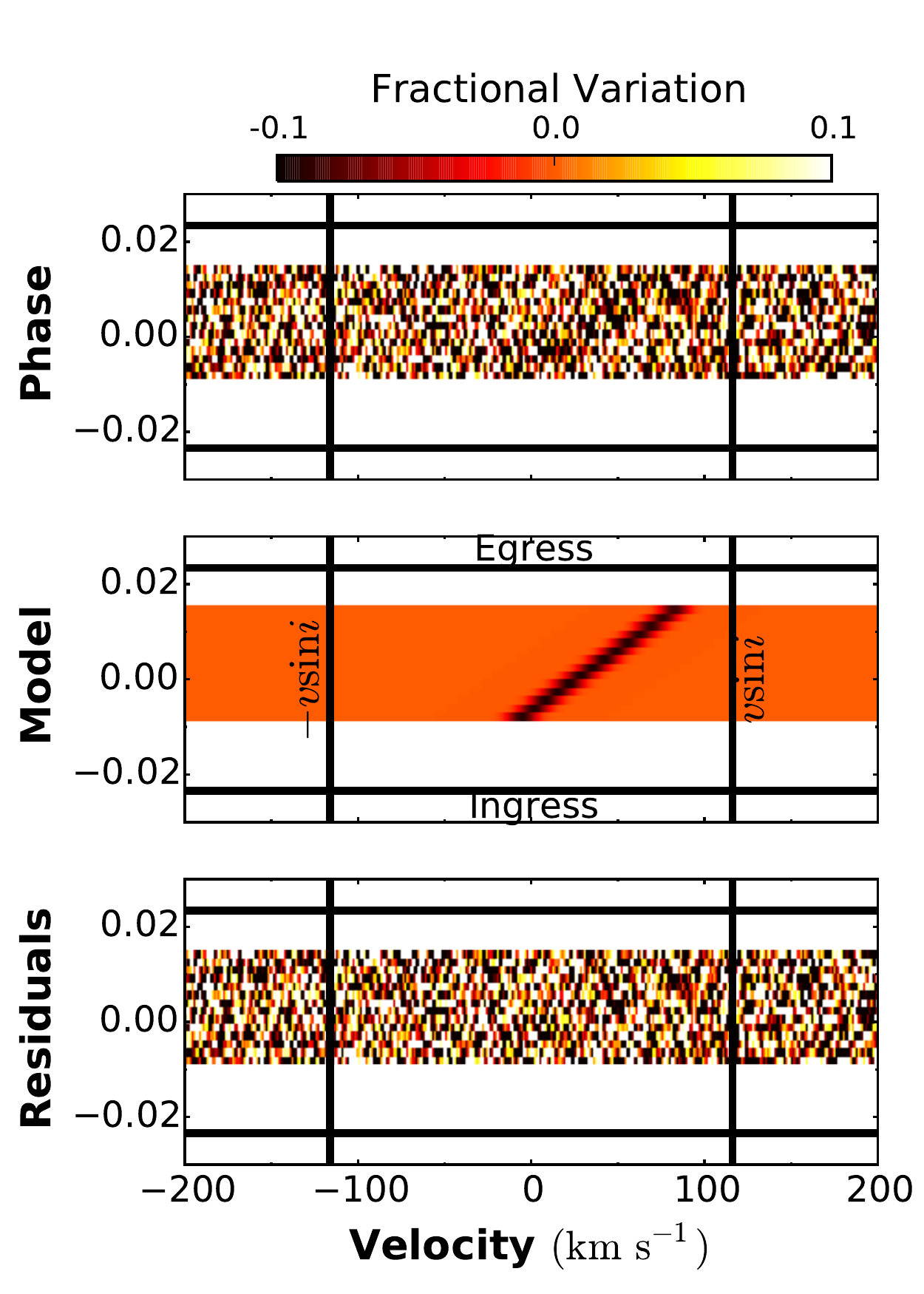} &
        \includegraphics[width=0.3\textwidth]{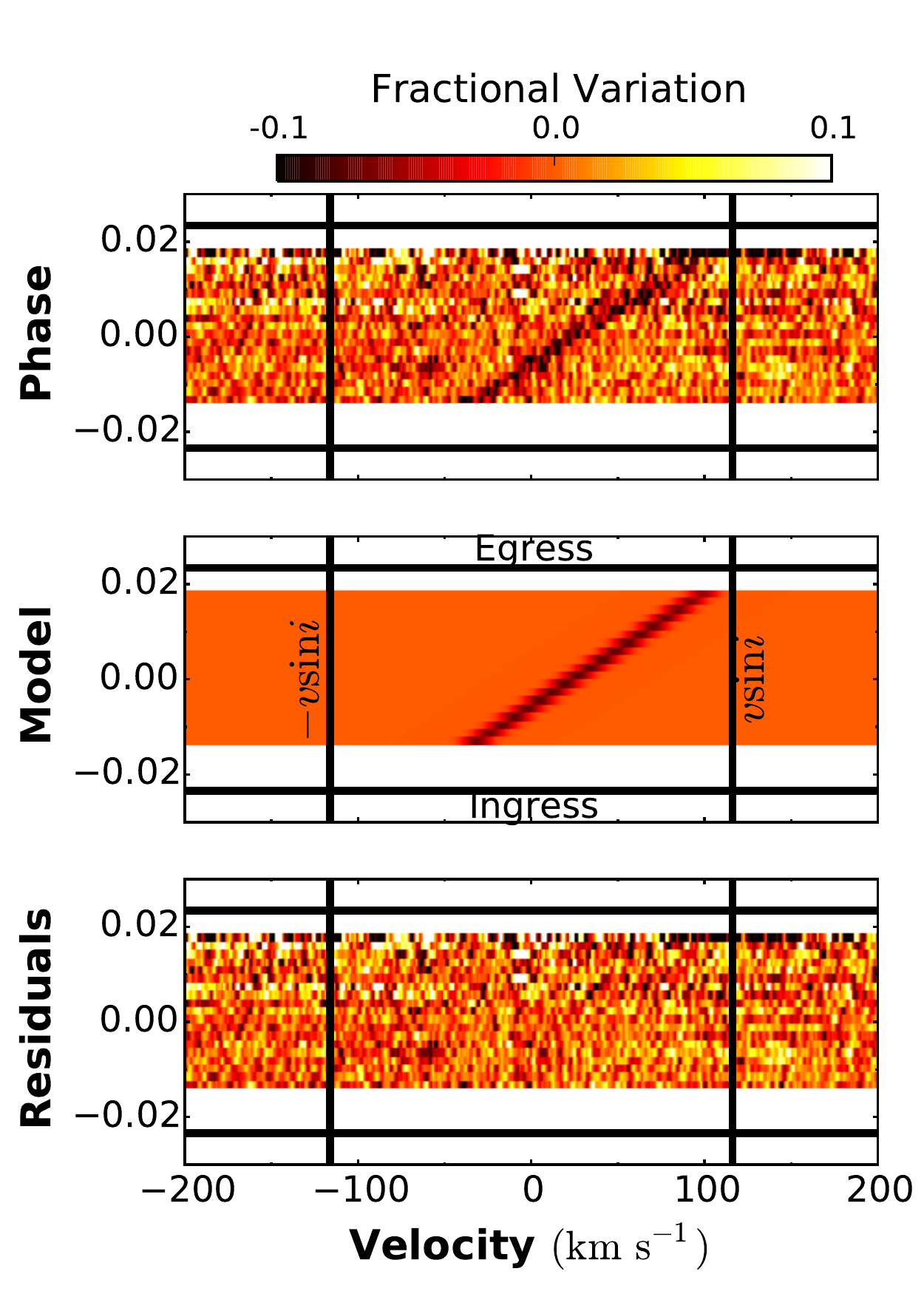} &         \includegraphics[width=0.3\textwidth]{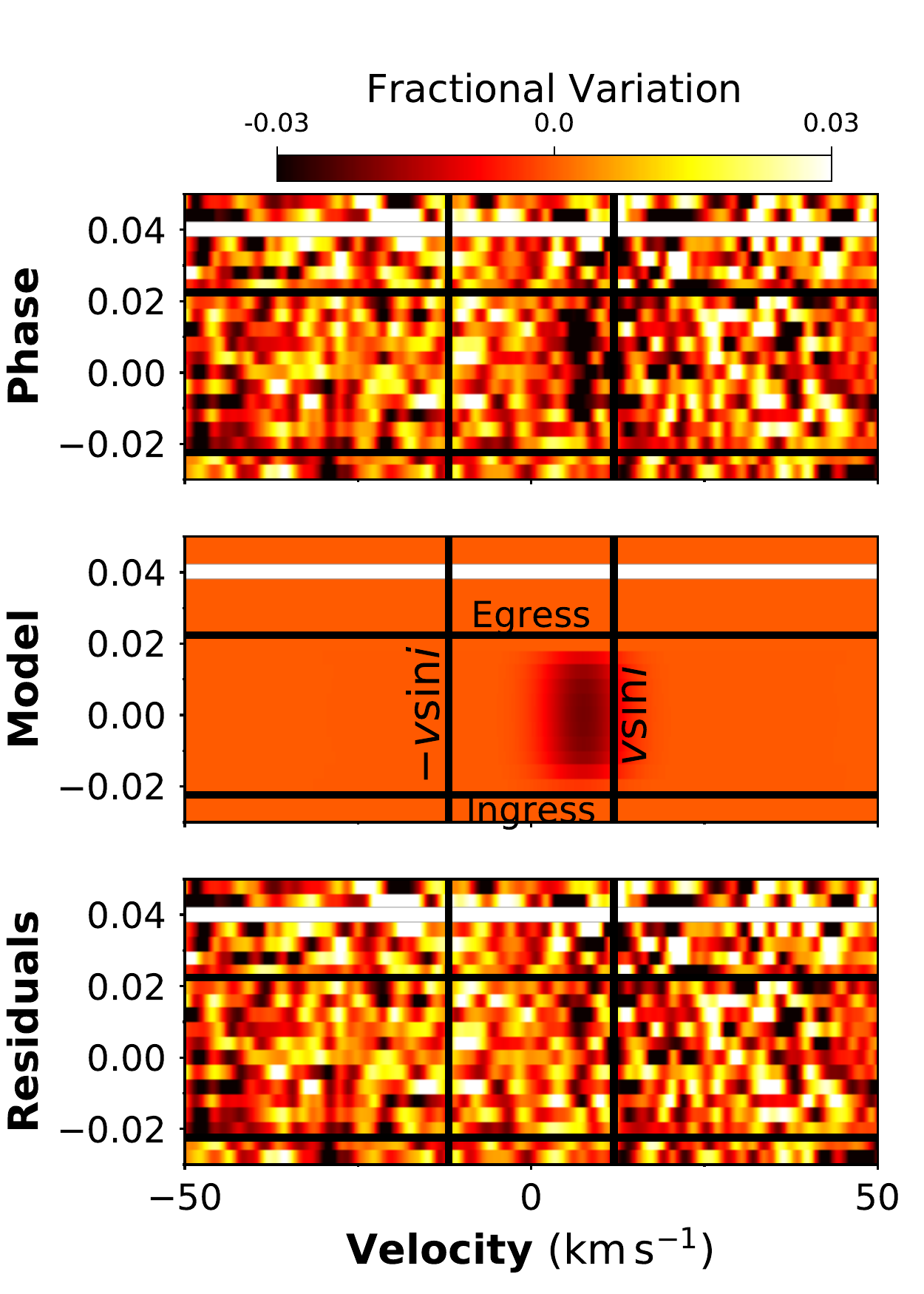}\\
    \end{tabular}
    \caption{The spectroscopic transit signals of \thisstarone from 
     CHIRON \textbf{(Left)} and PFS \textbf{(Middle)}, and \thisstartwo from \textsc{Minerva}-Australis \textbf{(Right)}. The spectroscopic transit signal shown for \thisstartwo from \textsc{Minerva}-Australis is the combined signal from two telescopes (See \S \ref{sec:MINERVA}). \textbf{Top}: The shadow cast by the planet appears as a dark trail on the line profile residuals. The horizontal axis plots the velocity space of the line profile, while the vertical axis plots the phase, with positive phase increasing upward. The limits of stellar rotation marked by the vertical lines, whereas the beginning and ending of each transit are show with the horizontal lines. \textbf{Middle} The best-fit model of the spectroscopic transit is shown. \textbf{Bottom}: The residuals after removal of the best fit model, showing a general lack of correlated noise in the line profile subtractions, leading to higher confidence in the detection and a lack of any observable stellar surface oscillations.}
    \label{fig:DTplots}
\end{figure*}

To measure the orbital obliquity and confirm the planetary status of \thisplanettwo, we obtained a series of spectroscopic observations during its transit on UT 2019 June 18 via the  MINiature Exoplanet Radial Velocity Array (\textsc{Minerva}-Australis) facility, located at the University of Southern Queensland's Mount Kent Observatory, Australia \citep{Addison:2019}. At the time of observations, \textsc{Minerva}-Australis was operating three active, 0.7m telescopes feeding into a single Kiwispec echelle spectrograph, yielding a resolving power of $R=80,000$ over the wavelength region of $5,000-6,300$\,\AA{}. We made use of data from the two of the three telescopes that yielded the highest signal-to-noise ratio spectra on the night of the transit. Spectral line profiles were derived from each spectrum via the LSD analysis described in Section~\ref{sec:PFS}. From these observations, we detect the spectroscopic shadow of the transit, finding that the path of the planet is offset from, but parallel to, the projected stellar equator. The spectroscopic transit of \thisplanettwo is shown in Figure~\ref{fig:DTplots}. 

\section{High-resolution imaging of \thisplanetone and \thisplanettwo}

\subsection{Gemini-South Zorro Speckle}
\label{sec:AO}

We obtained high resolution, speckle images of \thisstarone and \thisstartwo to search for nearby companions that could contaminate or dilute the light curves and thus affect the interpretation of the planetary radii, and to rule out sources of false positives like background eclipsing binaries.

Both stars were observed at the Gemini South Observatory using Zorro, a speckle interferometer residing at Gemini-South. Zorro observes in two band-passes simultaneously and is optimized for speckle observations. The observations occurred during instrument commissioning and were the first speckle interferometric science observations made by Zorro. Zorro is a dual-channel imager using two Electron Multiplying CCDs (EMCCDs) as the detectors and containing filter wheels providing bandpass limited observations\footnote{https://www.gemini.edu/sciops/instruments/alopeke/} (see \citealt{Scott:2018}). The Zorro data was reduced in the standard way as described in \citet{Howell:2011} and resulted in spatial reconstructions in each band-pass for \thisstarone and \thisstartwo providing high contrast, high resolution imaging results.

\begin{figure}[!ht]
\vspace{.1in}
\centering
\includegraphics[width=1\linewidth]{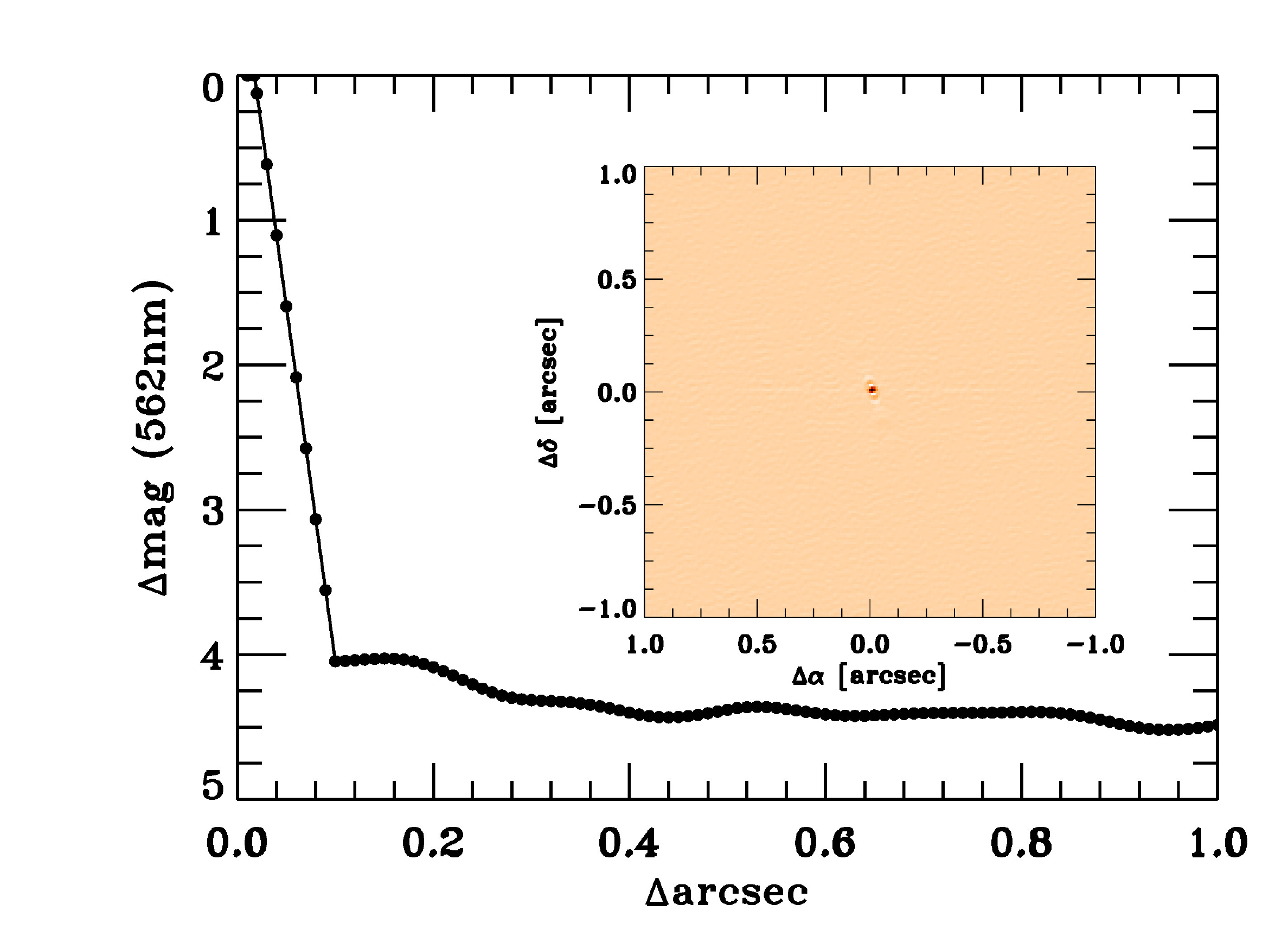}

\includegraphics[width=1\linewidth]{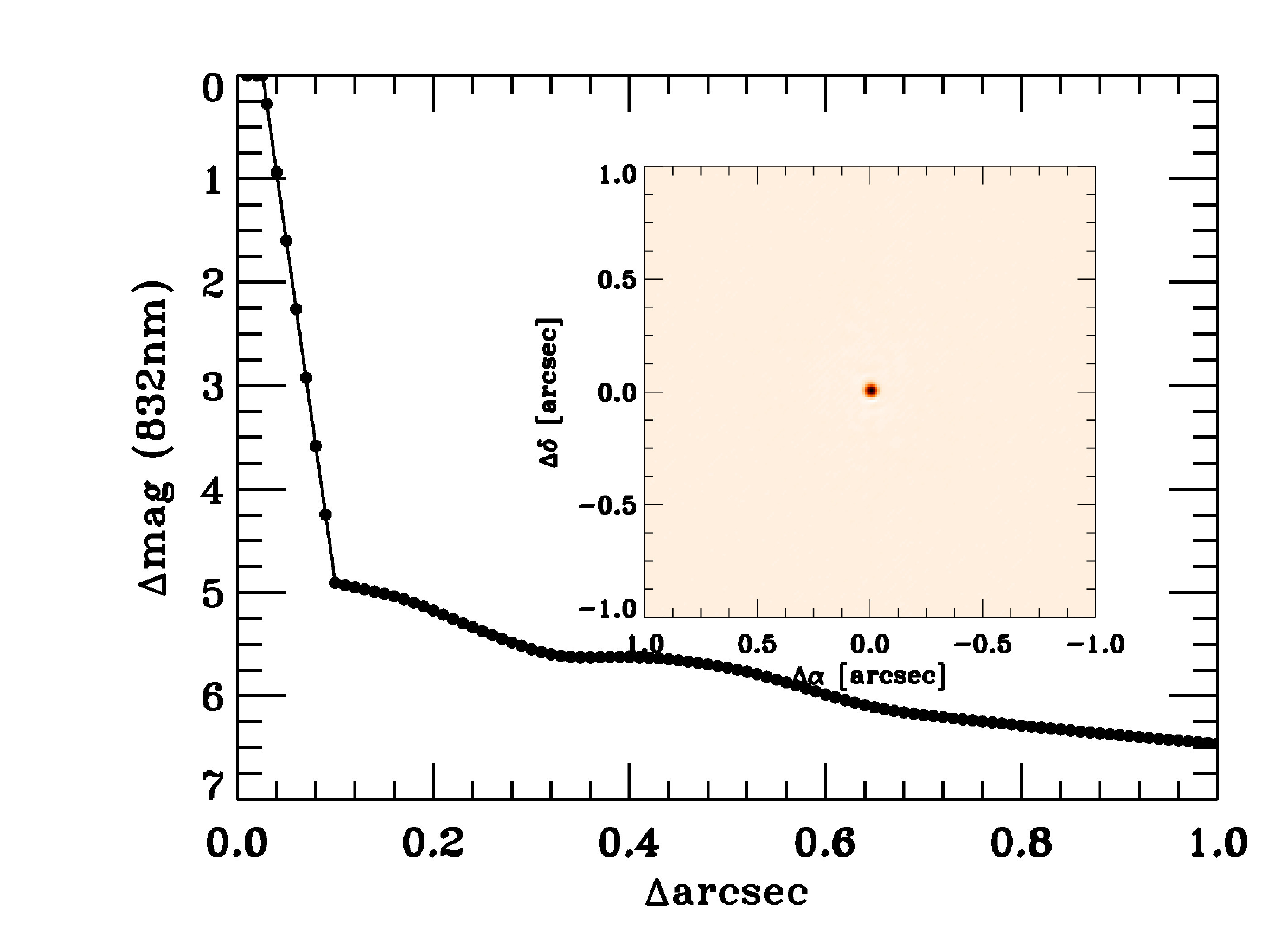}
\caption{Contrast curves and 562 nm image (inset, \textbf{Top}) and 832 nm image (inset, \textbf{Bottom}) for \thisstarone from Gemini-South with the Zorro instrument (Scott et al. 2019, in prep). }
\label{fig:speckle_K25} 
\end{figure}

\thisstarone was observed on UT 2019 May 18 in the blue (562/54 nm) and the red (832/40 nm) bandpasses simultaneously. \thisstarone and \thisstartwo observations consisted of 3 sets of 1000 frames with exposure times of 0.06 seconds each, coadded together during the data reduction process. Figure~\ref{fig:speckle_K25} shows the speckle reconstructed image for \thisstarone. The 562 nm observations (Figure~\ref{fig:speckle_K25}, Top) show no companion star from 17 mas out to $1{\arcsec}$ within 5 magnitudes of the source, and the 832 nm observations (Figure~\ref{fig:speckle_K25}, Bottom) confirm this as well in the red, starting from 28 mas and to a delta magnitude of 6-6.5.

\begin{figure}[!ht]
\vspace{.1in}
\centering
\includegraphics[width=1\linewidth]{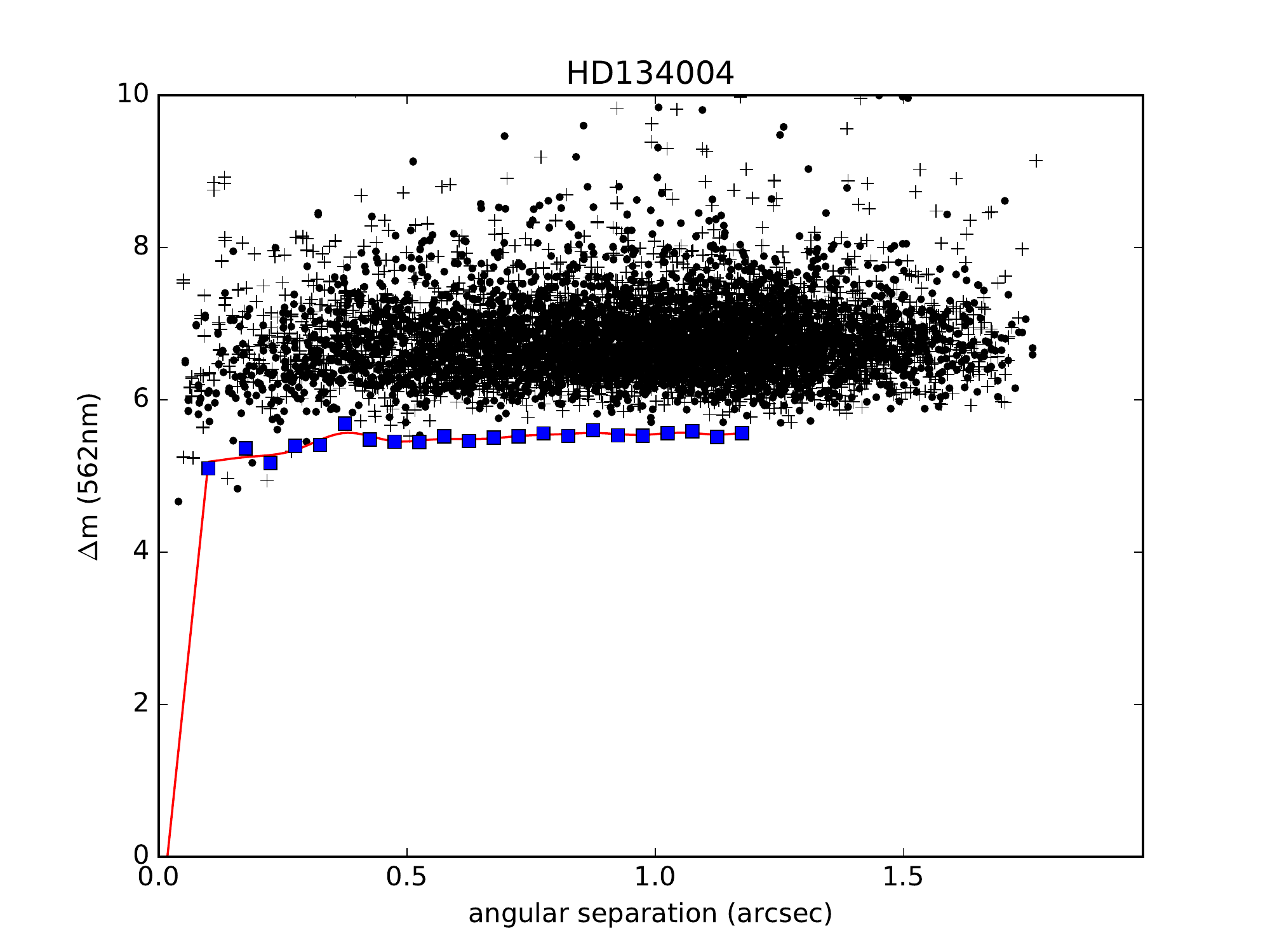}

\includegraphics[width=1\linewidth]{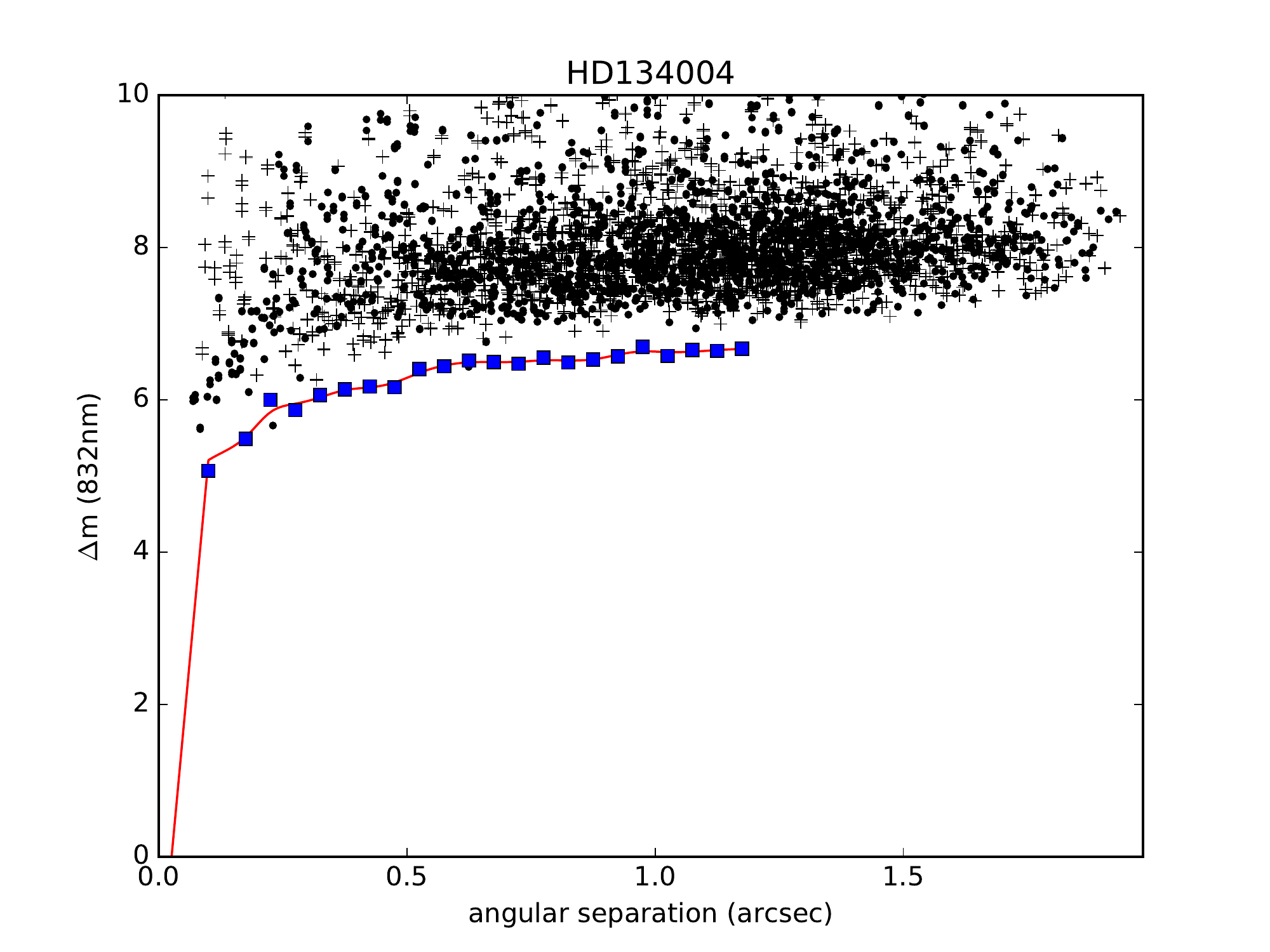}
\caption{Contrast curves and 562 nm image (inset, \textbf{Top}) and 832 nm image (inset, \textbf{Bottom}) for \thisstartwo from Gemini-South with the Zorro instrument (Scott et al. 2019, in prep).}
\label{fig:zorro_K26} 
\end{figure}

\thisstartwo was observed on UT 2019 May 21 in the blue and red bandpasses simultaneously. Figure~\ref{fig:zorro_K26} shows the constraints on possible stellar companions to \thisstartwo. No stellar companions are detected with angular separation from the primary from 17 mas (562 nm) and 28 mas (832 nm) out to $1.7{\arcsec}$  and for contrast limits of $\sim$4.2 mag (562 nm) and 5-7 mag (832 nm). The black solid line on the contrast curve marks the 5$\sigma$ detection limit.

\subsection{Southern Astrophysical Research Speckle (SOAR)}
\label{sec:soar}

\begin{figure}[!ht]
\vspace{.1in}
\centering
\includegraphics[width=1\linewidth]{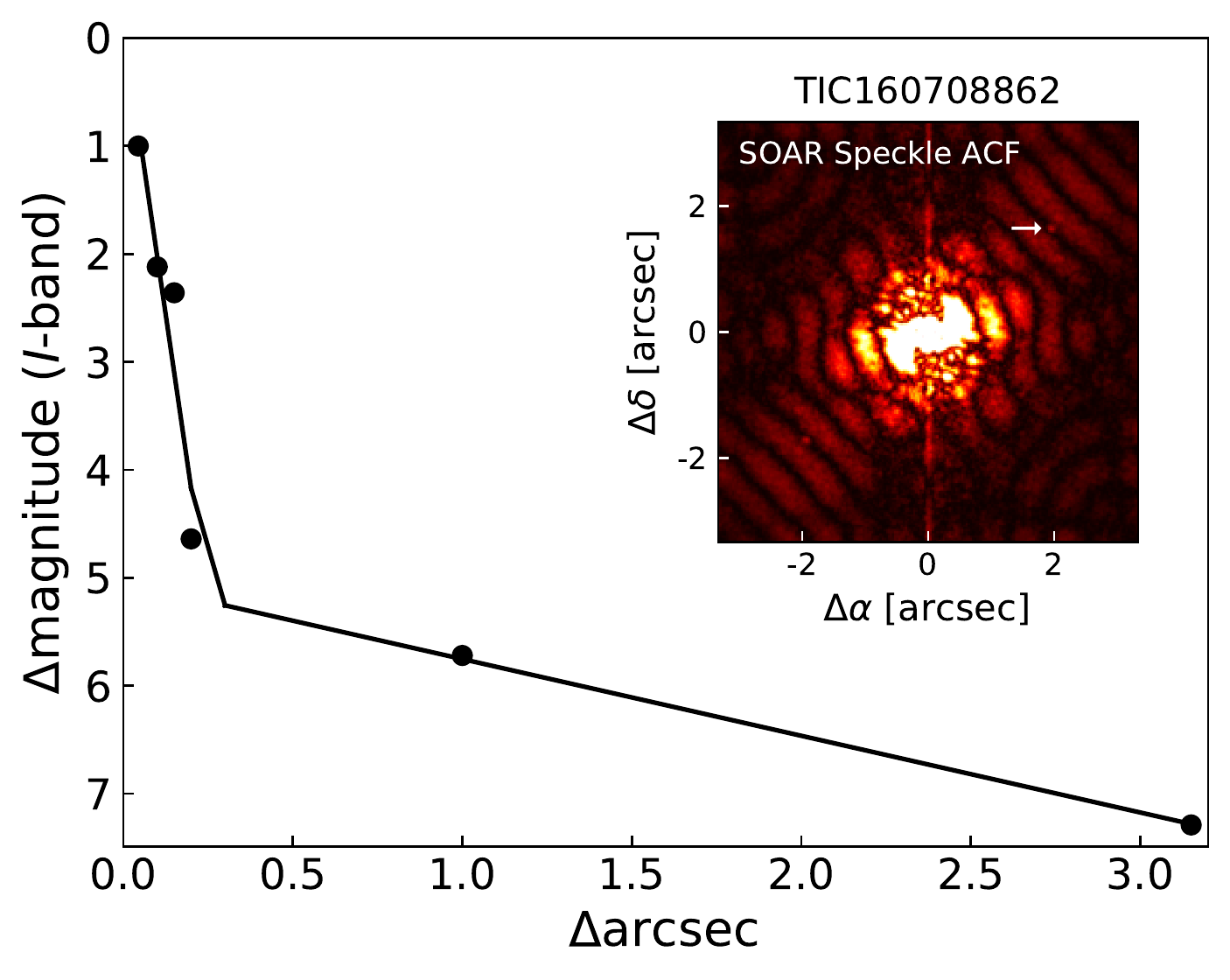}
\caption{$I_c$ band auto-correlation function of the speckle image for \thisstartwo from SOAR. The black points represent the 5$\sigma$ sensitivity limits for \thisstartwo. The inset shows the auto-correlation function.
A white arrow points to the location of the nearby companion. It is located 2.4096$\arcsec$ away from the target at PA = $311.2\degr$ and has a magnitude contrast of 7.1 mag in the $I_c$ band. The companion is mirrored in the ACF on the opposite side due to the speckle processing.}
\label{fig:k26_soar} 
\end{figure}

\thisstartwo was observed with the speckle camera at the 4.1 m Southern Astrophysical Research (SOAR) telescope on UT 2019 August 12. The instrument and data processing are covered in \citet{Tokovinin:2018}; a full description of the SOAR-\textit{TESS} survey can be found in \citet{ziegler:2019}. Figure~\ref{fig:k26_soar} shows the 5$\sigma$ limit of companion detection, and the inset shows the speckle auto-correlation function (ACF). The speckle ACF is symmetric, but the true quadrant was determined from the shift-and-add image. 

A nearby, faint companion was detected at an angular separation of 2.4096$\arcsec$ at position angle $311.15\degr$ and a magnitude difference of 7.1 mag in $I_c$ (central wavelength 824 nm, bandwidth 170 nm). This nearby source was not seen in the Gemini-South Zorro observation.

From the contrast magnitude of the neighbor in $I_c$, we can estimate its contribution to the flux in the diluted transit light curves. In this case, the contribution of the neighbor is proportional to the ratio of the flux of the secondary to the primary, or 0.14\% in $I_c$. This estimate agrees with \citet{matson:2019}, who found that stellar companions with magnitude contrasts of $\Delta m \la 7.8$ can only cause $< 0.1\%$ of a contribution to the flux. This dilution factor reflects the amount by which the true transit depth is diluted and therefore constrains the true radius of the planet. If the true depth is larger by 0.14\%, that would represent a $\sim$1 sigma difference between the true and reported depths (see Table~\ref{tab:exofast_planetary}).  Therefore, we should, in principle, remove the blended flux from the transit photometry and spectral energy distribution (SED). However, we do not know whether or not the companion is bound to the primary, and we do not have a flux in another band. Thus it is impossible to estimate the contribution to the flux due to the companion in any band other than $I_c$.  

The \textit{TESS} bandpass is centered on $I_c$, but it is $\sim$3 times wider in wavelength space \citep{Ricker:2015}, and essentially includes the $R$, $I_c$ and $z$ bandpasses. 
If the primary and the companion had the same SED in the \textit{TESS} bandpass, then the fractional flux contamination would be the same, i.e., $\sim 0.14\%$.  This is almost certainly not the case, as the companion is most likely to be either a foreground M star or a background giant. In either scenario, the companion would be fairly red, and thus would produce less fractional flux relative to the primary, which would have a much flatter SED in the \textit{TESS} bandpass. Thus, we take 0.14\% as a conservative upper limit to the blending in the \textit{TESS} bandpass.  

Given the blending from the companion, the true depth is larger and therefore the true radius of the planet is larger \citep{ciardi:2015}. The inferred radius of \thisplanettwo from the diluted \textit{TESS} and KELT-FUN light curves (all of which were blended by the neighbor in their apertures) is probably larger by at most $\sim 0.5\times 0.14\%$ or, equivalently, 0.001$R_{J}$. This is about 0.02$\sigma$ different from our reported value of the planetary radius, and thus it does not change our qualitative results. This is simply because the uncertainty on the radius of the planet is dominated by the uncertainty on the radius of the star, not the depth of the transit. 

Finally, we note that it is possible that this faint companion is an artifact, based on the fact that there are no sources at that position and magnitude found in either \textit{Gaia} DR2 or the TESS Input Catalog Version 8 (TIC-8; \citealp{stassun:2019}). Moreover, if it is a real source, at an angular separation of $\sim$2.5$\arcsec$, it is unlikely to be a bound companion (see, e.g., \citealt{matson:2019}).

Nevertheless, we encourage more AO/speckle observations in other photometric filters in order to  determine if the neighbor is real and bound, and to determine the spectral type of the neighbor and therefore its total flux contribution.  This will enable a more accurate measurements of the transit depth and planetary radius.

\section{Host Star Properties}
\label{sec:starprops}

\begin{figure}[!ht]
\centering 
\includegraphics[width=\columnwidth]{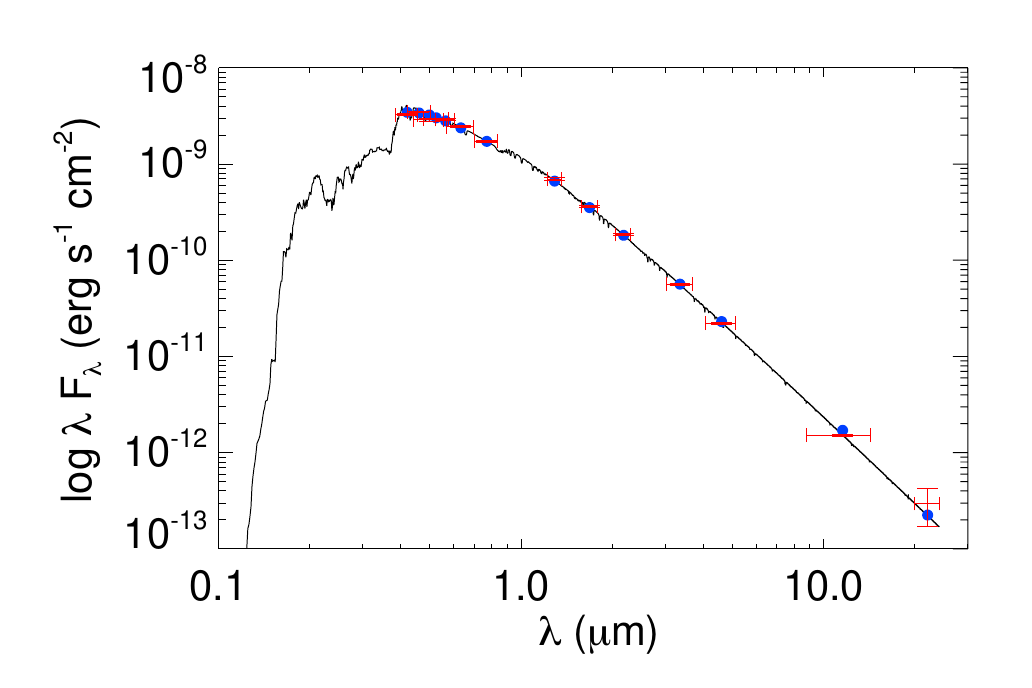}\\
\vspace{-0.3in}
\includegraphics[width=\columnwidth]{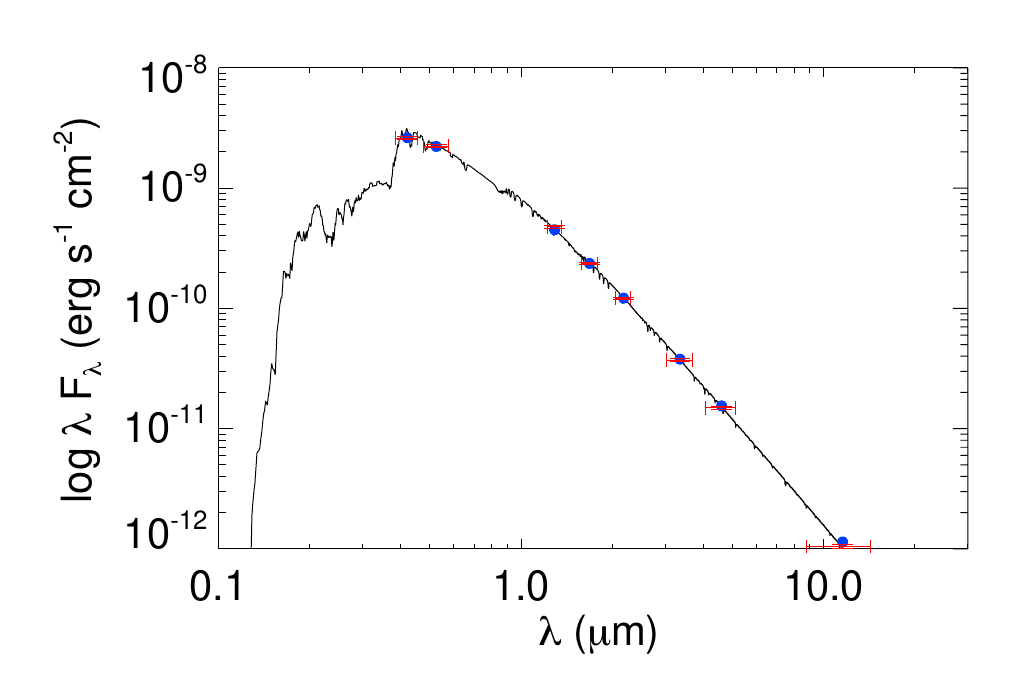}
\caption{The SED fit for \thisstarone \textbf{(Top)} and \thisstartwo \textbf{(Bottom)} from our EXOFASTv2 fit. The observed values are shown in red with 1$\sigma$ uncertainties while the predicted integrated fluxes are in blue. The final model is shown by the black line.}
\label{fig:sed_fit}
\end{figure}

\subsection{Location and Three-Dimensional Motion in the Galaxy, and Galactic Population}
\label{sec:uvw}
We determine the location in the Galaxy, 3-dimensional (UVW) space motion relative to the local standard of rest, and the inferred population of \thisstarone and \thisstartwo, using their proper motions, parallaxes, and absolute systemic velocities. Because we have poor metallicity constraints (albeit for different reasons; see the discussion below), we are unable to use this stellar property to help determine the stellar populations of the hosts.  However, given that they are both early A stars, we would be surprised if they have significantly subsolar metallicities.  

Using the proper motion, \textit{Gaia} parallax (corrected for the \citealt{stassun:2018a} systematic offset of $-82~\mu{\rm as}$), and the absolute systemic radial velocity determined as described in \S \ref{sec:TRES}, we compute $(U,V,W)=(-13.40\pm 0.58,-13.83\pm 0.84,-1.24\pm 0.19)~{\rm km~s^{-1}}$, correcting for the velocity of the Sun with respect to the local standard of rest as determined by \citet{Coskunoglu:2011}. These velocities imply that the probability of \thisstarone being in the thin disk relative to thick disk is 99.4\% using the classification scheme of \citet{Bensby:2003}.  The distance to \thisstarone is $427.0 \pm 7.8~{\rm pc}$ and it has Galactic coordinates of $(\ell,b)=(237.5\degr,-6.8\degr)$.  This implies that its vertical ($Z$) distance from the  sun is $Z-Z_\odot=-50.6~{\rm pc}$.  Given that the sun is located at $Z_\odot\simeq 30~{\rm pc}$ above the plane as determined by the local evolved stellar population according to \citet{Bovy:2017}, this means that \thisstarone is located only about $Z\simeq 20~{\rm pc}$ below the plane.  This is consistent with the scale height of early A stars as determined by \citet{Bovy:2017}.  

Using the same methodology, we computed $(U,V,W)=(-24.63\pm 0.37,6.75\pm 0.58,0.78\pm 0.20)~{\rm km~s^{-1}}$ for \thisstartwo.  These velocities imply that the probability of \thisstartwo being in the thin disk relative to thick disk is 99.3\% using the classification scheme of \citet{Bensby:2003}.  The distance to \thisstartwo is $417.78 \pm 10.5~{\rm pc}$ and it has Galactic coordinates of $(\ell,b)=(328.19\degr,+13.32\degr)$.  This implies that its vertical ($Z$) distance from the sun is $Z-Z_\odot=96.1~{\rm pc}$ and $Z\simeq 126~{\rm pc}$ above the plane.  This is roughly twice the typical scale height of an early A star as determined by \citet{Bovy:2017}, and thus is notable but not completely unexpected.  

Both \thisstarone and \thisstartwo have roughly the same Galactocentric distance at the sun.  Assuming that the distance from the sun to the Galactic center is roughly $R_0=8.2\simeq{\rm kpc}$ \citep{gravity:2019}, we estimate Galactocentric distances of $8.4~{\rm kpc}$ and $7.8~{\rm kpc}$ for \thisstarone and \thisstartwo, respectively. 

\begin{table}
\scriptsize
\setlength{\tabcolsep}{2pt}
\centering
\caption{Median values and 68\% confidence interval for the stellar parameters of \thisstarone and \thisstartwo derived from the global fit.}
\begin{tabular}{llcccc}
  \hline
  \hline
Parameter & Units & Values &Values & & \\
&& \thisstarone & \thisstartwo \\
\hline
~~~~$M_*$\dotfill &Mass (\msun)\dotfill & $2.18^{+0.12}_{-0.11}$ & $1.93^{+0.14}_{-0.16}$\\
~~~~$R_*$\dotfill &Radius (\rsun)\dotfill &$2.264^{+0.048}_{-0.052}$& $1.801^{+0.049}_{-0.048}$\\
~~~~$L_*$\dotfill &Luminosity (\lsun)\dotfill &$21.8^{+4.6}_{-1.8}$ & $16.4^{+3.8}_{-1.8}$\\
~~~~$\rho_*$\dotfill &Density (cgs)\dotfill &$0.263^{+0.025}_{-0.018}$ & $0.463^{+0.040}_{-0.038}$ \\
~~~~$\log{g}$\dotfill &Surface gravity (cgs)\dotfill &$4.064^{+0.032}_{-0.026}$ & $4.211^{+0.030}_{-0.033}$\\
~~~~$T_{\rm eff}$\dotfill &Effective Temperature (K)\dotfill &$8280^{+440}_{-180}$& $8640^{+500}_{-240}$\\
~~~~$[{\rm Fe/H}]$\dotfill &Metallicity (dex)\dotfill &$0.30^{+0.13}_{-0.21}$& $-0.06^{+0.30}_{-0.34}$\\
~~~~$[{\rm Fe/H}]_{0}$\dotfill &Initial Metallicity \dotfill &$0.34^{+0.11}_{-0.19}$&$0.03^{+0.25}_{-0.32}$\\
~~~~$Age$\dotfill &Age (Gyr)\dotfill &$0.46^{+0.14}_{-0.12}$& $0.43^{+0.31}_{-0.25}$\\
~~~~$EEP$\dotfill &Equal Evolutionary Phase \dotfill &$342.0^{+6.3}_{-7.1}$&$331^{+17}_{-28}$\\
~~~~$vsinI_*$\dotfill &Projected rotational velocity (km/s)\dotfill & $114.200\pm1.200$ &$12.280^{+7.80}_{-8.20}$\\
~~~~$V_{\rm line}$\dotfill &Unbroadened line width (m/s)\dotfill &$3700^{+2300}_{-2400}$& $6400^{+1800}_{-1900}$\\
~~~~$A_V$\dotfill &V-band extinction (mag)\dotfill &$0.104^{+0.16}_{-0.073}$& $0.103^{+0.14}_{-0.078}$\\
~~~~$\sigma_{SED}$\dotfill &SED photometry error scaling \dotfill &$2.50^{+0.78}_{-0.53}$& $1.76^{+0.81}_{-0.48}$\\
~~~~$\pi$\dotfill &Parallax (mas)\dotfill &$2.366^{+0.041}_{-0.042}$&$2.396\pm0.063$\\
~~~~$d$\dotfill &Distance (pc)\dotfill &$422.7^{+7.6}_{-7.3}$& $417\pm11$\\
\hline
\end{tabular}
\begin{flushleft} 
  \footnotesize{ 
    \textbf{\textsc{NOTES:}}
$^\dagger$The initial metallicity is the metallicity of the star when it was formed.\\
$^\ddagger$The Equal Evolutionary Phase corresponds to static points in a star's evolutionary history when using the MIST isochrones and can be a proxy for age. See \S2 in \citet{Dotter:2016} for a more detailed description of EEP.
               }
 \end{flushleft}
\label{tab:exofast_stellar}
\end{table}
\begin{table*}
\scriptsize
\centering
\caption{Median values and 68\% confidence interval for the physical parameters of \thisplanetone and \thisplanettwo from the global fit.}
\begin{tabular}{llcccc}
  \hline
  \hline
Parameter & Description (Units) & Values & Values & & \\
&& \thisstarone & \thisstartwo \\
\hline
~~~~$P$\dotfill &Period (days)\dotfill &$4.401131\pm0.000059$&$3.3448412\pm0.0000033$\\
~~~~$R_P$\dotfill &Radius (\rj)\dotfill &$1.642^{+0.039}_{-0.043}$&$1.940^{+0.060}_{-0.058}$\\
~~~~$M_P$\dotfill &Mass (\mj)\dotfill &$\color{red}(<64)$&$1.41^{+0.43}_{-0.51}$\\
~~~~$T_C$\dotfill &Time of conjunction (\bjdtdb)\dotfill & $2458493.31501\pm0.00037$ & $2457482.31209^{+0.00090}_{-0.00091}$\\
~~~~$T_0^\dagger$\dotfill &Optimal conjunction Time (\bjdtdb)\dotfill &$2458506.51840\pm0.00034$ &$2458321.86724^{+0.00038}_{-0.00039}$\\
~~~~$a$\dotfill &Semi-major axis (AU)\dotfill &$0.0681^{+0.0012}_{-0.0011}$&$0.0545^{+0.0013}_{-0.0015}$\\
~~~~$i$\dotfill &Inclination (Degrees)\dotfill &$85.37^{+0.55}_{-0.42}$ & $84.45^{+0.39}_{-0.41}$\\
~~~~$T_{eq}$\dotfill &Equilibrium temperature (K)\dotfill &$2306^{+100}_{-47}$&$2402^{+130}_{-71}$\\
~~~~$\tau_{\rm circ}$\dotfill &Tidal circularization timescale (Gyr)\dotfill &$-0.1^{+4.1}_{-3.9}$&$0.053^{+0.019}_{-0.020}$\\
~~~~$K$\dotfill &RV semi-amplitude (m/s)\dotfill &$\color{red}(<4687.8)$&$123^{+37}_{-45}$\\
~~~~$\log{K}$\dotfill &Log of RV semi-amplitude \dotfill &$\color{red}(<3.64)$&$2.09^{+0.11}_{-0.20}$\\
~~~~$R_P/R_*$\dotfill &Radius of planet in stellar radii \dotfill &$0.07450^{+0.00039}_{-0.00042}$&$0.11066^{+0.00090}_{-0.00087}$\\
~~~~$a/R_*$\dotfill &Semi-major axis in stellar radii \dotfill &$6.46^{+0.20}_{-0.15}$&$6.49\pm0.18$\\
~~~~$\delta$\dotfill &Transit depth (fraction)\dotfill &$0.005550^{+0.000059}_{-0.000062}$&$0.01225^{+0.00020}_{-0.00019}$\\
~~~~$\tau$\dotfill &Ingress/egress transit duration (days)\dotfill &$0.0192^{+0.0011}_{-0.0012}$&$0.0238^{+0.0016}_{-0.0014}$\\
~~~~$T_{14}$\dotfill &Total transit duration (days)\dotfill &$0.2051^{+0.0013}_{-0.0015}$&$0.1514\pm0.0016$\\
~~~~$T_{FWHM}$\dotfill &FWHM transit duration (days)\dotfill &$0.18586^{+0.00097}_{-0.00096}$&$0.12760^{+0.00098}_{-0.00099}$\\
~~~~$b$\dotfill &Transit Impact parameter \dotfill &$0.522^{+0.034}_{-0.048}$&$0.628^{+0.027}_{-0.029}$\\
~~~~$\delta_{S,3.6\mu m}$\dotfill &Blackbody eclipse depth at 3.6$\mu$m (ppm)\dotfill &$732^{+27}_{-23}$&$1664^{+82}_{-72}$\\
~~~~$\delta_{S,4.5\mu m}$\dotfill &Blackbody eclipse depth at 4.5$\mu$m (ppm)\dotfill &$872^{+26}_{-24}$&$1965^{+83}_{-74}$\\
~~~~$\rho_P$\dotfill &Density (cgs)\dotfill &$\color{red}(<18.3)$&$0.238^{+0.077}_{-0.088}$\\
~~~~$logg_P$\dotfill &Surface gravity \dotfill &$\color{red}(<4.77)$&$2.97^{+0.12}_{-0.20}$\\
~~~~$\lambda$\dotfill &Projected Spin-orbit alignment (Degrees)\dotfill &$23.4^{+3.2}_{-2.3}$&$91.3^{+6.5}_{-6.3}$\\
~~~~$\Theta$\dotfill &Safronov Number \dotfill &$-0.01^{+0.50}_{-0.48}$&$0.041^{+0.012}_{-0.015}$\\
~~~~$\fave$\dotfill &Incident Flux (\fluxcgs)\dotfill &$6.42^{+1.2}_{-0.51}$&$7.56^{+1.7}_{-0.86}$\\
~~~~$T_P$\dotfill &Time of Periastron (\bjdtdb)\dotfill &$2458493.31501\pm0.00037$&$2457482.31209^{+0.00090}_{-0.00091}$\\
~~~~$T_S$\dotfill &Time of eclipse (\bjdtdb)\dotfill &$2458495.51558\pm0.00036$ &$2457483.98451\pm0.00090$\\
~~~~$T_A$\dotfill &Time of Ascending Node (\bjdtdb)\dotfill &$2458496.61586\pm0.00035$&$2457484.82072\pm0.00090$\\
~~~~$T_D$\dotfill &Time of Descending Node (\bjdtdb)\dotfill &$2458494.41530\pm0.00036$&$2457483.14830\pm0.00090$\\
~~~~$M_P\sin i$\dotfill &Minimum mass (\mj)\dotfill &$\color{red}(<64)$&$1.40^{+0.43}_{-0.51}$\\
~~~~$M_P/M_*$\dotfill &Mass ratio \dotfill &$\color{red}(<0.028)$&$0.00070^{+0.00021}_{-0.00026}$\\
~~~~$d/R_*$\dotfill &Separation at mid transit \dotfill &$6.46^{+0.20}_{-0.15}$&$6.49\pm0.18$\\
~~~~$P_T$\dotfill &A priori non-grazing transit prob \dotfill &$0.1433^{+0.0035}_{-0.0043}$&$0.1370^{+0.0039}_{-0.0036}$\\
~~~~$P_{T,G}$\dotfill &A priori transit prob \dotfill &$0.1664^{+0.0041}_{-0.0051}$&$0.1710^{+0.0050}_{-0.0047}$\\
\smallskip\\\multicolumn{2}{l}{Telescope Parameters:}&TRES & CHIRON\\
~~~~$\gamma_{\rm rel}$\dotfill &Relative RV Offset (m/s)\dotfill &$720^{+810}_{-820}$&$-25595^{+25}_{-24}$\\
~~~~$\sigma_J$\dotfill &RV Jitter (m/s)\dotfill &$2240^{+1100}_{-620}$&$67^{+32}_{-23}$\\
\hline
\end{tabular}
 \begin{flushleft} 
  \footnotesize{ 
    \textbf{\textsc{\hspace{0.75in}NOTES:}}
$^\dagger$Minimum covariance with period.
All values in this table for the secondary occultation of \thisstarone b are predicted values from our global analysis. All values in red are 3$\sigma$ upper limits on mass dependent parameters for \thisplanetone.              
               }
 \end{flushleft}
\label{tab:exofast_planetary}
\end{table*}
\begin{table*}
\scriptsize
\centering
\caption{Median values and 68\% confidence interval for global model of \thisstarone and \thisstartwo from the global fit. \label{tab:stellarparams_other}}
\begin{tabular}{llcccccc}
  \hline
  \hline
\thisplanetone &&&\\
\multicolumn{2}{l}{Wavelength Parameters:}&R&TESS\smallskip\\
~~~~$u_{1}$\dotfill &linear limb-darkening coeff \dotfill &$0.292\pm0.049$&$0.159\pm0.029$\\
~~~~$u_{2}$\dotfill &quadratic limb-darkening coeff \dotfill &$0.348\pm0.049$&$0.260\pm0.035$\\
~~~~$A_T$\dotfill &Secondary eclipse depth (ppm)\dotfill &--&$187\pm46$\\
\multicolumn{2}{l}{Transit Parameters:}&PEST UT 2019-01-18 (R)&TESS Sector 7&TESS Sector 7 (secondary)\smallskip\\
~~~~$\sigma^{2}$\dotfill &Added Variance \dotfill &$0.00000779^{+0.00000062}_{-0.00000058}$&$-0.0000000863^{+0.0000000057}_{-0.0000000050}$&$-0.0000000885^{+0.0000000066}_{-0.0000000057}$\\
~~~~$F_0$\dotfill &Baseline flux \dotfill &$1.00362\pm0.00013$&$0.999826\pm0.000049$&$0.999811\pm0.000040$\\
\smallskip\\\multicolumn{2}{l}{Doppler Tomography Parameters:}&&\\
~~~~$\sigma_{DT}$\dotfill &Doppler Tomography Error scaling \dotfill &$0.9975^{+0.0100}_{-0.0098}$&$0.9839^{+0.0057}_{-0.0056}$\\
\hline
\smallskip\thisplanettwo &&&\\
\multicolumn{2}{l}{Wavelength Parameters:}&I&r'&TESS\smallskip\\
~~~~$u_{1}$\dotfill &linear limb-darkening coeff \dotfill &$0.185\pm0.048$&$0.207\pm0.045$&$0.212\pm0.044$\\
~~~~$u_{2}$\dotfill &quadratic limb-darkening coeff \dotfill &$0.264\pm0.051$&$0.279\pm0.048$&$0.280\pm0.047$\\
\smallskip\\\multicolumn{2}{l}{Transit Parameters:}&PEST UT 2016-07-26 (I)&CDK700 UT 2018-03-20 (r')&TESS UT 2019-04-07 (TESS)\smallskip\\
~~~~$\sigma^{2}$\dotfill &Added Variance \dotfill &$0.00000780^{+0.0000010}_{-0.00000089}$&$0.00000170^{+0.00000025}_{-0.00000021}$&$0.000000420^{+0.000000029}_{-0.000000027}$\\
~~~~$F_0$\dotfill &Baseline flux \dotfill &$0.99573\pm0.00021$&$0.99891\pm0.00015$&$1.000000\pm0.000026$\\
\smallskip\\\multicolumn{2}{l}{Doppler Tomography Parameters:}&MINERVA 3&MINERVA 4\\
~~~~$\sigma_{DT}$\dotfill &Doppler Tomography Error scaling \dotfill &$0.991\pm0.013$&$0.997\pm0.012$\\
\hline
 \end{tabular}
\end{table*}

\section{EXOFAST\lowercase{v}2 Global Fits for \thisstarone and \thisstartwo} 
\label{sec:exofast}
To constrain the system parameters, we modeled the available radial velocities, transit photometry, and multiband absolute photometry for \thisstarone and \thisstartwo using the exoplanet fitting suite, EXOFASTv2 \citep{Eastman:2013,Eastman:2019,eastman:2017}. 

\begin{figure}[!ht]
    \centering
    \includegraphics[width=\linewidth, trim={0.5cm 0cm 0cm 0.5cm }, clip]{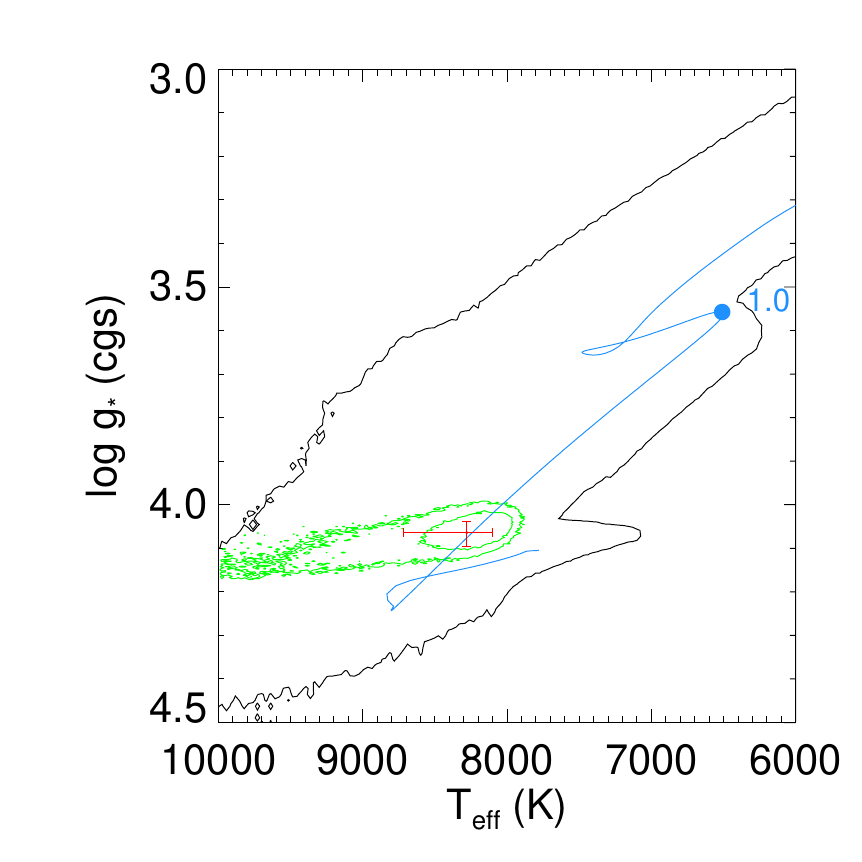}\\
\vspace{-0.2in}
    \includegraphics[width=\linewidth, trim={0.5cm 0cm 0cm 0.5cm }, clip]{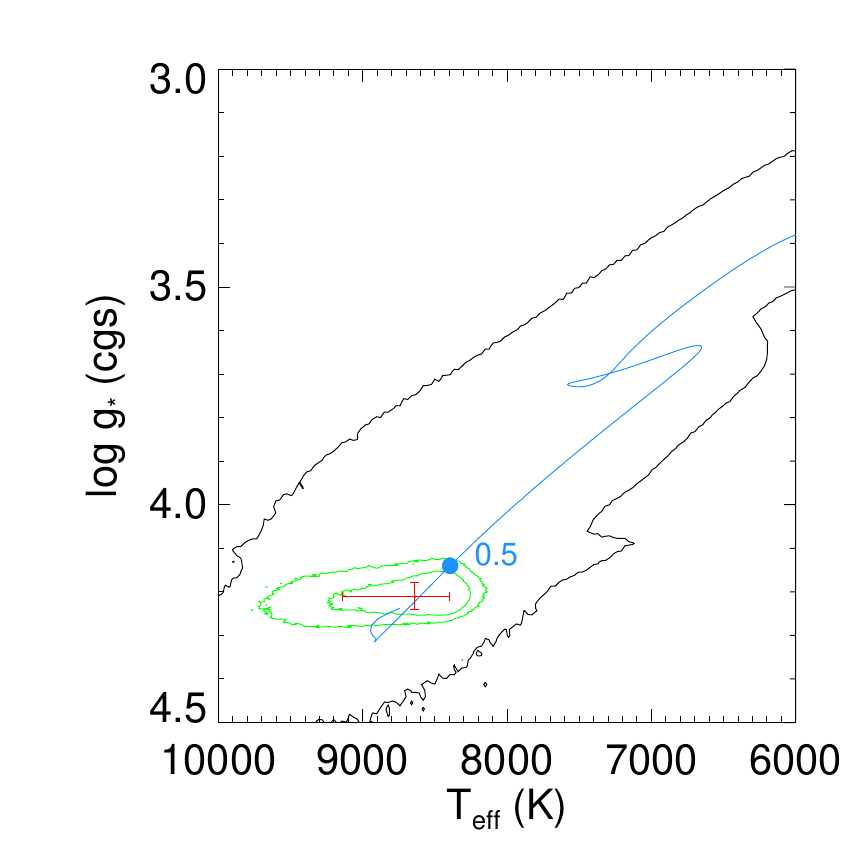}
    \caption{The best-fitting MIST track for \thisstarone\ \textbf{(Top)}  and  \thisstartwo\ \textbf{(Bottom)} shown by the blue line. The black line shows the 3$\sigma$ contours for the MIST evolutionary tracks. The red data point shows the median values and 1$\sigma$ uncertainties from our global fit, while the green contours correspond to the 3$\sigma$ errors. The blue points mark the location of 1.0 (Top) and 0.5 Gyr (Bottom) along the MIST track.}
    \label{fig:mistplots}
\end{figure}

For each system, we globally fit the radial velocities (see \S\ref{sec:spectra}), the \textit{TESS} and follow-up photometry (see \S\ref{sec:tess} and \S\ref{sec:KFUN}) and the Doppler tomographic shadows simultaneously (see \S\ref{sec:CHIRON}, \S\ref{sec:PFS}, and \S\ref{sec:MINERVA}). Within these fits, we determined the host star properties using a combination of spectroscopic priors, the spectral energy distribution, and the MESA Isochrones and Stellar Tracks (MIST) stellar evolution models \citep{Dotter:2016,Choi:2016,Paxton:2011,Paxton:2013,Paxton:2015}. For each system, we placed Gaussian priors of [Fe/H] $=0.0 \pm 0.5$ dex for the metallicity of the host stars, as we did not have reliable constraints from our available spectra. From an independent EXOFASTv2 analysis of the KELT photometry, we adopted a Gaussian prior on the period of \thisplanetone of 4.401093 $\pm$ 0.000073 days and on the epoch of $T_{c} = 2458493.3144$ \bjdtdb. Similarly, we placed a Gaussian prior on the orbital period ($P = 3.344886 \pm 0.000064$ days) and the epoch ($T_{c} = 2457482.315 \pm 0.008600$ \bjdtdb) of \thisplanettwo. We also placed priors on the $\vsini$ from spectroscopy for \thisstarone ($\vsini =111.277 \pm 1.422$ \kms) and \thisstartwo ($\vsini =9.9349 \pm 1.1333$ \kms) and used parallaxes from \textit{Gaia} DR2 (see Table~\ref{tab:LitProps}). 
In addition, we assumed a circular orbit for both systems and we fit for the depth of the secondary eclipse of \thisstarone observed in the lower right panel of Figure~\ref{fig:tess25}. We constrained $A_{V}$ to the maximum permitted line-of-sight extinction from \citet{Schlegel:1998}, and the temperature and stellar mass and radius from the SED best-fit values (see Figure~\ref{fig:sed_fit} for the empirical SEDs of both systems). The best-fitting evolutionary models are shown in Figure~\ref{fig:mistplots}, and the final fit parameters of the EXOFASTv2 analyses for both stars and their companions are shown in Tables~\ref{tab:exofast_stellar} and~\ref{tab:exofast_planetary}.

Although a global fit to the photometry, radial velocities (including the Doppler Tomography measurements), and the SED can completely constrain the properties of the system, we include constraints from the MIST stellar evolutionary models, which include reasonably well-understood stellar physics, in our global fit.  In this manner, we derive best-fit distances to both stars that are slightly different than those inferred from the \textit{Gaia} parallaxes alone. As a result, we find a posterior distribution of the distance of \thisstarone after the global fit of $422.5^{+7.6}_{-7.3}$ pc, which is $\sim 0.6$ sigma from that inferred from the {\it Gaia} parallax (see Table \ref{tab:LitProps}).  Similarly, for \thisstartwo, we derive a posterior distance after the global fit of $417\pm11$ pc, which is $\sim$0.06 sigma from the distance inferred from the {\it Gaia} DR2 parallax alone. We consider these differences to be completely consistent within the uncertainties. To calculate the location of \thisstarone and \thisstartwo within the Galaxy and their UVW space velocities in \S\ref{sec:uvw}, we opted to use the model-independent parallaxes and distances from {\it Gaia} DR2. Although in this case these distance measurements are completely consistent, the discovery of larger discrepancy in fits of other systems may have implications for the current stellar evolutionary models or {\it Gaia} data. Thus the empirical distances from {\it Gaia} DR2 can serve as a way to test and calibrate the models we use to derive fundamental stellar parameters, or uncover evidence of systematic errors in the {\it Gaia} data itself. 

\section{Discussion}
\label{sec:discussion}

\begin{figure}[!ht]
\vspace{.1in}
\centering
\includegraphics[width=1\linewidth]{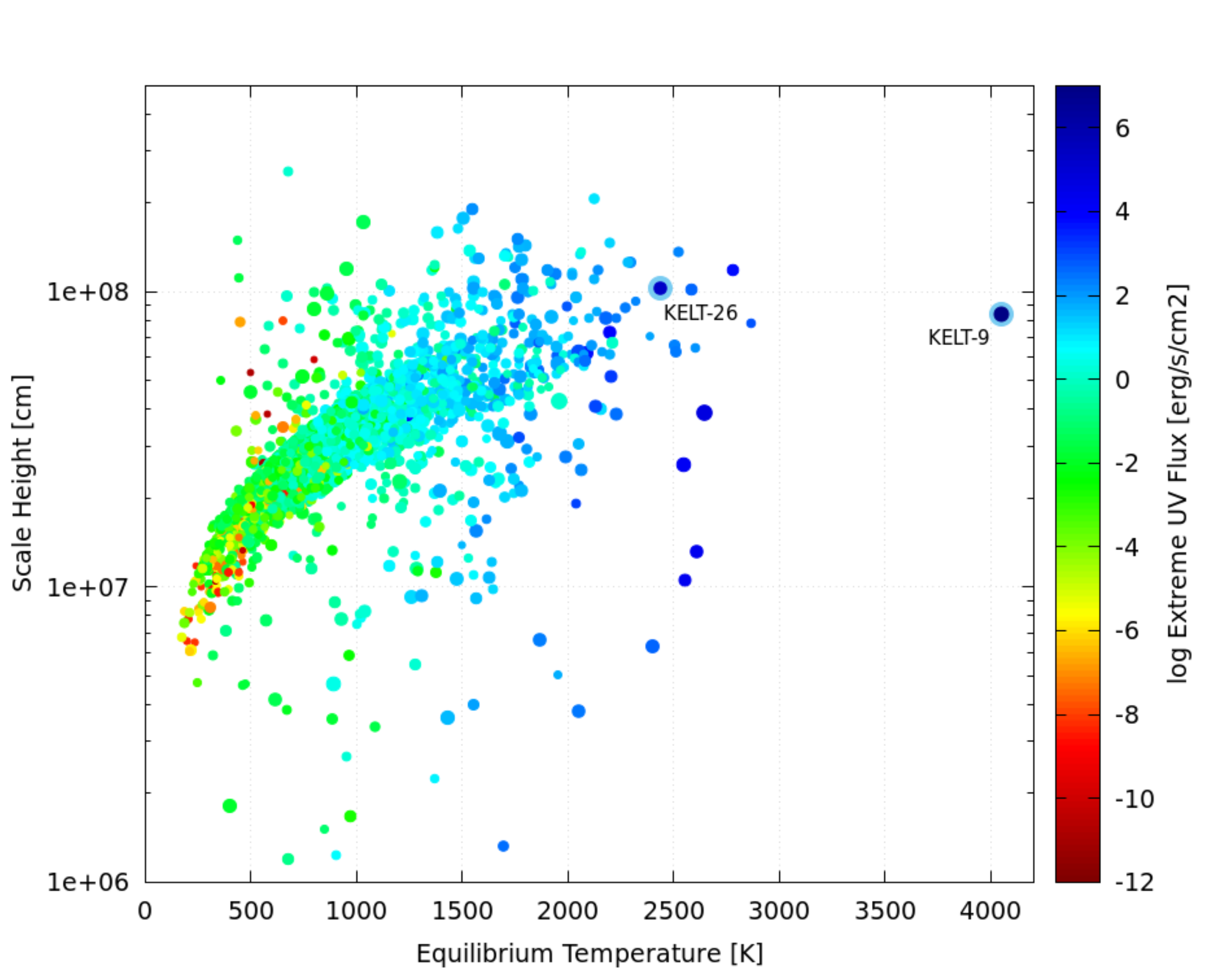}
\caption{This scatter-plot shows the atmospheric scale height versus equilibrium temperature for all the known transiting exoplanets with mass measurements. The color scale corresponds to the extreme ultraviolet radiation that planets receive from their host stars. The symbol sizes are inversely proportional to the magnitude in V-band of the host stars. At $T_{\rm eq} \sim$2402 K, \thisplanettwo stands out as one of the hottest known exoplanets, receiving extreme amounts of UV radiation, likely resulting in the exceptionally large radius of $R_p \sim 1.9\rj$, and subsequent large scale height.}
\label{fig:scale_height} 
\end{figure}

Both \thisplanetone and \thisplanettwo represent extreme transiting systems in a few key aspects. First, \thisplanetone and \thisplanettwo both orbit relatively bright (V$\sim$10 mag) and extremely hot hosts ($T_{\rm eff} \simeq 8300$ K and $\simeq$ 8700 K, respectively), and they also have short orbital periods ($P \sim 4.4$ and $P\sim 3.3$ days, respectively). Their proximity to their host stars and their stars' intrinsic luminosity mean that they receive extreme amounts of stellar radiation, paticularly high-energy radiation, resulting in high equilibrium temperatures -- assuming zero albedo and complete redistribution -- of $T_{eq}=2306$ K (\thisplanetone) and $T_{eq}=2402$ K (\thisplanettwo). Both \thisplanetone and \thisplanettwo join the recently-defined class of planets of ``Ultra Hot Jupiters'', which, similar to the prototype WASP-33b \citep{CollierCameron:2010}, are primarily planets on short-period orbits around early A stars.  They are thus among the hottest transiting exoplanets known. Indeed, the equilibrium temperatures (zero albedo, complete heat distribution) of these planets are commonly in excess of 2000 K, and it seems likely that their day-side temperatures would be markedly higher still (see Figure~\ref{fig:scale_height}), and likely have day-side temperatures that are much higher. 

Our EXOFASTv2 models (Table~\ref{tab:exofast_planetary}) indicate that the radii of both planets are also significantly inflated ($R_{p} = 1.64 R_{J}$ and $R_{p} = 1.94 R_{J}$). From an irradiation evolution analysis of these systems (see \S\ref{sec:orbevolution}), we conclude that these objects currently receive an insolation flux of around $5\times10^{9}$ $\rm erg$ $\rm s^{-1} \rm cm^{-2}$ (\thisplanetone) and $1\times10^{10}$ $\rm erg$ $\rm s^{-1} \rm cm^{-2}$ (\thisplanettwo).  Moreover, their orbital histories suggest that they probably have always been above the \citet{Demory:2011} insolation threshold of $2\times10^{8}$ $\rm erg$ $\rm s^{-1} \rm cm^{-2}$, which is an empirical threshold above which giant planets exhibit significant radius inflation. For this reason, it is not surprising that they are both highly inflated. The extreme temperatures of these companions and the optical and infrared brightness of their hosts (\thisstarone: $V =9.65$, $J=9.36$; \thisstartwo: $V=9.96$, $J=9.77$) mean that the prospects for detailed atmospheric characterization via transmission spectroscopy with James Webb Space Telescope (\textit{JWST}) or the Hubble Space Telescope ({\it HST}) are promising. Indeed, the \textit{TESS} light curve for \thisstarone (Figure \ref{fig:tess25}) demonstrates the weak but significant detection of the secondary eclipse of the system, as the planet moves behind the star.  Given the estimated equilibrium temperature of the star, this flux decrement is also certainly caused by the tail of the thermal emission from the planet. We estimate a secondary eclipse depth of $\sim 187\pm46$ ppm (see Table~\ref{tab:stellarparams_other}), implying a brightness temperature of the planet in the \textit{TESS} band of $\sim$$3396^{+140}_{-170}$K, which is substantially higher than the equilibrium temperature (assuming zero albedo and complete heat redistribution) of $2303^{+100}_{-47}$K.  

To better place these planets in context of all systems with measured rotation rates and projected spin-orbit misalignments, we highlight \thisstarone and \thisstartwo in Figures~\ref{fig:vsinIvsTeff} and~\ref{fig:lambdavsTeff} showing the distributions of $\vsini$ and spin-orbit misalignments $\lambda$ versus stellar temperature for A stars ($\vsini$ distribution) and all known planet hosts with measured spin-orbit angles. Figure~\ref{fig:scale_height} shows a plot of atmospheric scale height as a function of equilibrium temperature for all the known transiting exoplanets with measured masses. \thisplanettwo has a large scale height and receives extreme amounts of UV radiation from its host.
With \thisplanetone and \thisplanettwo, we have a large enough sample of A-stars with transiting exoplanets that we can begin to see emerging patterns in the population (see Table~\ref{tab:Astarplanets}). Perhaps one such interesting trend is the gap in planet equilibrium temperatures between roughly 2600 K and 4000 K, visible in Figure~\ref{fig:scale_height}. We still do not understand whether this gap is real or the result of selection effects. All the confirmed planets around A stars have short periods (1.22 < $P$ < 4.79 days) and transit bright host stars ($V \la 10$), which, as previously remarked, make them attractive targets for atmospheric characterization with the upcoming \textit{JWST}. Before \textit{JWST} launches, however, \textit{TESS} may be able to observe these planets in transit, in some cases even resolving their secondary eclipses, as with \thisplanetone. These observations can constrain the brightness temperature of these planets and therefore provide insights into the heat distribution mechanisms of their atmospheres. 

\begin{table*}
\footnotesize

\caption{A-star transiting planet hosts from the literature ordered by decreasing stellar temperature \label{tab:Astarplanets}}
\centering
\begin{tabular}{lccccccccc}
\hline\hline
Planet & $M_{p}$ & $R_{p}$ & $T_{\rm{eff}}$ & $P$ & $L_{\star}$   & $T_{eq}$ & $\lambda$ & SpT & Ref. \\
& ($M_{J}$) & ($R_{J}$) & (K) & (days) & ($L_{\odot}$) & (K) & (deg) & \\ 
\hline
KELT-9b & 2.88$\pm$0.84 & $1.891^{+0.061}_{-0.053}$ & 10170$\pm$450 & 1.48 & $53^{+13}_{-10}$ & 4050$\pm$180 & -84.8$\pm$1.4 & A0 & 1\\
KELT-20b/MASCARA-2b & < 3.5 (3$\sigma$) & $1.735^{+0.07}_{-0.075}$ & $8730^{+250}_{-260}$ & 3.47  &$12.7^{+2.2}_{-1.9}$&  $\sim$2250 & $3.4\pm 2.1$& A2 & 2 \\
\thisplanettwo & $1.41^{+0.43}_{-0.51}$ & $1.940^{+0.060}_{-0.058}$ & $8640^{+500}_{-240}$ & 3.34 & $16.4^{+3.8}_{-1.8}$ & $2402^{+130}_{-71}$ & $91.3^{+6.5}_{-6.3}$ & A3m &3 \\
HAT-P-70b & $<6.14$ & $2.011_{-0.114}^{+0.051}$ & $8450_{-690}^{+540}$ & $2.74$ & $16.7_{-4.6}^{+5.3}$ & $2562_{-52}^{+43}$ & $116.5_{-3.8}^{+3.5}$ & A3V& 4 \\
\thisplanetone & $<$64 & $1.642^{+0.039}_{-0.043}$ & $8280^{+440}_{-180}$ & 4.40 & $21.8^{+4.6}_{-1.8}$ & $2306^{+100}_{-47}$ & $23.4^{+3.2}_{-2.3}$ & A4 & 5 \\
WASP-189b & $2.13 \pm 0.28$& $1.374 \pm 0.082$& $8000 \pm 100$& 2.72 & $1.293 \pm 0.045$&  $2641 \pm 34$& $89.3 \pm 1.4$& A6IV-V & 6\\
HATS-70b & $12.9_{-1.6}^{+1.8}$ & $1.384_{-0.074}^{+0.079}$ & $7930_{-820}^{+630}$ & $1.89$ & $12.0_{-3.4}^{+5.5}$ & $2730_{-160}^{+140}$ & $8.9_{-4.5}^{+5.6}$ & A6V & 7 \\
MASCARA-4b & $3.1 \pm 0.9$ & $1.53_{-0.04}^{+0.07}$ & $7800 \pm 200$ & $2.82$ & $12.23 \pm 0.655$ & $2100 \pm 100$ & $247.5_{-1.7}^{+1.5}$ & A7V& 8\\
Kepler-13Ab & $\sim$ 9.2 $\pm$ 1.1 & $1.512\pm 0.035$ & $7650\pm250$ & 1.76  & -&  $2550\pm80$ & $58.6 \pm 2.0$ & A8V &9 \\
KELT-21b & < 3.91 (3$\sigma$) & $1.586^{+0.039}_{-0.040}$ & $7598^{+81}_{-84}$ & 3.61  &  $8.03^{+0.54}_{-0.53}$& $2051^{+29}_{-30}$ & $-5.6^{+1.7}_{-1.9}$ & A8V & 10\\
MASCARA-1b & $ 3.7 \pm 0.9$ & $1.5 \pm 0.3$ & $7554 \pm 150$ & 2.14  & $13.1 \pm 3$ &  $2570^{+50}_{-30}$ & $69.5 \pm 3$ & A8V & 11 \\ 
HAT-P-57b & $1.41\pm1.52$& $1.74 \pm 0.36$ & $7500\pm250$ & 2.46  & $6.4\pm 1.1$ &  2200& $-16.7$-$3.3$ or $2.76$-$57.4$ & A8V &12 \\
KELT-19Ab & $1.62^{+0.25}_{-0.20}$ & $1.83\pm0.10$ & 7500$\pm$110 & 4.61  &$9.5^{+1.2}_{-1.1}$& $\sim$1935 & $-179.7^{+3.7}_{-3.8}$ & Am &13 \\
KELT-17b & $1.31^{+0.28}_{-0.29}$ & $1.525^{+0.065}_{-0.060}$ & 7454 $\pm$ 49 & 3.08 &$7.51^{+0.62}_{-0.55}$ & $2087^{+32}_{-33}$  &-115.9$\pm$4.1 &A9V& 14 \\
WASP-33b & 4.1 & $1.497\pm0.045$ & 7430$\pm$100  & 1.22 & -& $2710\pm50$ & $251.6 \pm 0.7$ & A9V & 15 \\
HAT-P-69b & $3.54_{-0.60}^{+0.61}$ & $1.714 \pm 0.028$ & $7394_{-600}^{+360}$ & $4.79$ & $10.0_{-0.9}^{+1.8}$ &$1930_{-230}^{+80}$ & $16.5_{-1.9}^{+2.1}$ & A9V &16 \\

\hline
\end{tabular}
\begin{flushleft}
 \footnotesize{ \textbf{\textsc{NOTES:}}

References: 1. \citet{Gaudi:2017} 2. \citet{Lund:2017,talens:2018}  3. This work 4. \citet{Zhou:2019} 5. This work 6. \citet{anderson:2018} 7. \citet{Zhou:2019} 8. \citet{Dorval:2019} 9. \citet{shporer:2011,Esteves:2015,Johnson:2014} 10. \citet{Johnson:2018} 11. \citet{Talens:2017} 12.  \citet{Hartman:2015} 13. \citet{Siverd:2018} 14. \citet{Zhou:2016} 15. \citet{CollierCameron2010} 16. \citet{Zhou:2019b}}

\end{flushleft}
\end{table*}

\subsection{\thisplanettwo: A giant planet orbiting a likely Am star with a likely significant transit asymmetry}
\label{sec:discussK26}

\thisplanettwo orbits a relatively young ($\sim$ 430 Myr), slowly rotating A star (\vsini = 12.2 km s$^{-1}$). This rotational velocity is rather atypical for an early A star, as such stars tend to be much faster rotators on average. From the Doppler tomographic observations (see \S\ref{sec:spectra}), we also measured the projected spin-orbit angle of this system, and conclude that it is on an orbit that is consistent with being exactly parallel to the projected stellar equator, with $\lambda$ = $91.3^{+6.5}_{-6.3}$. However, the orbit is likely not polar, because the impact parameter of planet's transit is $b\simeq 0.6$ and therefore the planet is not transiting exactly across the pole of the star. The projected spin-orbit alignment $\lambda$ need not be the true true obliquity $\psi$, which is a more fundamental quantity of the system. However, the latter is harder to constrain, as explained in \citet{Johnson:2018}, because it requires knowledge of both the planetary orbital inclination \textit{i} as well as the stellar spin axis $I_{*}$. The observation of an orbit that has $\lambda \sim 90\degr$ with the unusually slow rotation period of \thisstartwo ($\vsini =12.2$ km s$^{-1}$) may imply that the star is perhaps spinning faster but we are observing it nearly pole-on.  The phase-folded \textit{TESS} light curve of the primary (Figure \ref{fig:tess26}) appears to show a slight asymmetry, such that the planet is first passing over a region of the star with higher-than average surface brightness, whereas the planet later passes over a region of the star with lower surface brightness.  This would be expected from gravity darkening, assuming the planet first passes over or near the pole, and then over the lower surface brightness equator \citep{Barnes:2009}. Gravity darkening has already been observed with \textit{TESS} for two hot Jupiters: HAT-P-69b (TOI 625.01) and HAT-P-70b (TOI 624.01) \citep{Zhou:2019}, and is also clear in the unpublished \textit{TESS} light curve of KELT-9 (\citet{wong:2019}, Wachiraphan et al.; Ahlers et al., in preparation). However, as we discuss below, the host star \thisstartwo also appears to be variable at the $\sim$few millimagnitude level at a period that is nearly commensurate (1:18) with the period of the planet.  This variability may also be causing the slight asymmetry.  

A Lomb-Scargle periodogram (\citealt{Lomb:1976,Scargle:1982}) of the light curve of \thisstartwo shows a significant peak at a period of 0.185 days, with an amplitude of 0.115\%, or 1.25 mmag.  This period is nearly 1/18 (1/18.06, to be precise) of the period of the planet.  We do not know if this is simply a coincidence, as the mass of the companion is likely too small to induce periodic oscillations on its host star.  Both \thisstarone and \thisstartwo are inside the instability strip, where one might expect to find $\delta$ Scuti pulsations.  Indeed, the period and amplitude of the variability of \thisstartwo are consistent with other $\delta$ Scuti variables. A more detailed study of the nature of the intrinsic variability of \thisstartwo is beyond the scope of this paper.

\begin{figure}[h]
\vspace{.1in}
\centering
\includegraphics[width=1\linewidth]{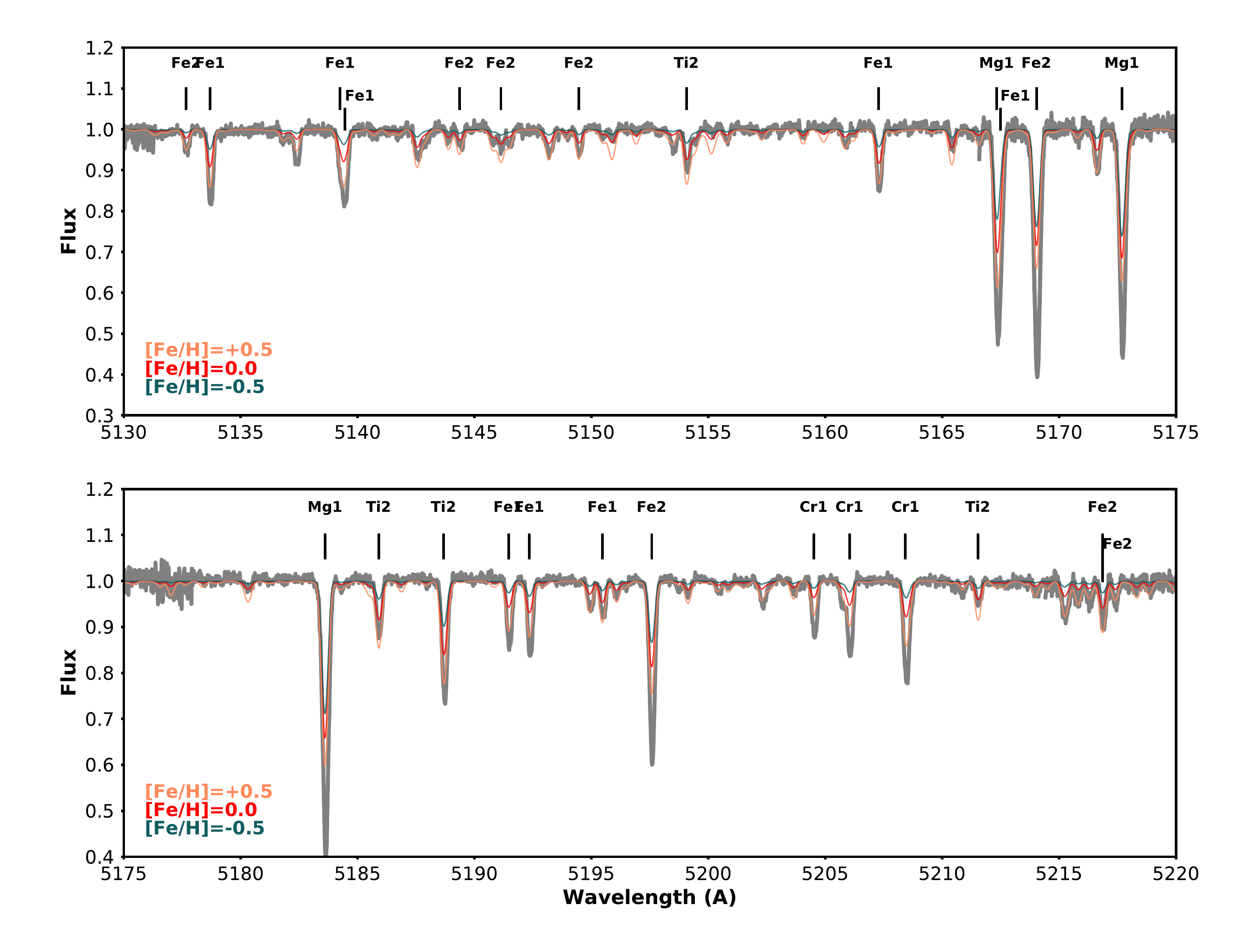}
\caption{A portion of the CHIRON spectrum near the Mgb region for \thisstartwo is shown in grey.  The other lines show a set of $\teff=9000$ K, $\loggstar=4.25$ synthetic spectra with [Fe/H] of $-0.5$ (green), 0.0 (red), and $+0.5$ (orange) dex, generated with the ATLAS9 model atmospheres \citep{Castelli2003}, demonstrating that a single [Fe/H] cannot simultaneously fit all of observed spectral features.}
\label{fig:K26spec}
\end{figure}

On the other hand, the spectrum of \thisstartwo shows evidence of being an Am star (or "metallic-line A star").  Am stars typically rotate much more slowly than A stars of the same effective temperature.  This is usually attributed to a stellar companion that has spun down the star or otherwise "stolen" its angular momentum at birth.  However, we find no evidence of a stellar companion which would affect the spin rate of \thisstartwo, and the planetary companion is not sufficiently massive to play this role. Am stars are typically identified by the fact that the star does not appear to have a consistent metallicity when measured using absorption lines formed at different depths in the photosphere. This can be seen in Figure \ref{fig:K26spec}, where it is clear that for models with a fixed $T_{\rm eff}$ and varying [Fe/H], no single model can simultaneously fit all of the spectral lines.

\subsection{\thisplanetone: A substellar object transiting a rapidly rotating, young A star and likely cluster member}\label{sec:discussK25}

With a $\vsini$ of 114.2 km $\rm s^{-1}$, \thisstarone is rotating much faster than \thisstartwo, and this has implications for the dynamical history of its potential planet. In this case, the classical scenario of hot Jupiters spiraling towards their host stars is reversed. The stellar tidal dissipation causes the semi-major axis of \thisplanetone's orbit to gradually increase (see \S\ref{sec:orbevolution}). As a result, \thisplanettwo will avoid getting engulfed by its host star, at least until the star leaves the main sequence. The Doppler tomography shadows (Figure~\ref{fig:DTplots}) suggest that it is on a prograde, aligned orbit ($\lambda$ = $23.4^{+3.2}_{-2.3}$).  To estimate the age of the systems, we show a modified Hertzsprung-Russell diagram (log$g_{*}$ vs $T_{\rm eff}$) in Figure~\ref{fig:mistplots}. From the MIST stellar evolutionary models, and taking the 1-sigma upper limit on the mass for \thisstarone, we infer an age of $0.46^{+0.14}_{-0.12}$ Gyr. 
Using the same models, we obtain an age for \thisstartwo of $0.43^{+0.31}_{-0.25}$ Gyr.

\subsubsection{Is \thisstarone a member of a stellar cluster, association, or moving group?}

We cross-matched existing \textit{TESS} Objects of Interest (TOIs) to the catalog of clusters presented in \citet{kounkel:2019}, and we found a match between TOI-626.01 (\thisstarone) and one of the putative clusters identified in that paper as Theia 449. In that paper, they identified 1900 clusters from an analysis of the distribution of sources in 5-dimensional space (three-dimensional position and two-dimensional (e.g., transverse) velocity) in {\it Gaia} DR2. They then performed a clustering analysis on {\it Gaia} sources within $|b| < 30\degr$ of the Galactic plane and parallaxes with $\pi > 1$ mas using a Python implementation of HDBSCAN (Hierarchical Density-Based Spatial Clustering of Applications with Noise, McInnes et al 2017). They estimated ages of their clusters and associations applying a combination of machine learning and isochronal fitting techniques to determine the ages of their sources to a precision of $\sim$0.15 dex.

For the cluster Theia 449, \citet{kounkel:2019} report a mean Galactic latitude of $b =-7.13\degr$, a mean parallax of $2.37 \pm 0.82$ mas, an age of $0.162$ Gyr and mean radial velocity of $23.54 \pm 18.21$ km $\rm s^{-1}$. From our MIST models, we determine KELT-25 is a young A star, with an age of $0.46^{+0.14}_{-0.12}$ Gyr, which is just $\sim$2-sigma discrepant with the mean reported age of the cluster. Our derived properties for \thisstarone are thus in general agreement with the average properties of Theia 449. However, the broader issue of whether Theia 449 truly represents a collection of coeval stars is outside the scope of this paper. Clarification of the status of Theia 449 could help resolve the age, metallicity, and formation environment of \thisstarone.

\subsubsection{Is \thisplanetone a planet or brown dwarf?}
\label{sec:BD}

Given its extremely fast rotation, we were only able to constrain the mass of \thisplanetone to a $3\sigma$ upper limit of $\sim 64~\mj$, or a $1\sigma$ upper limit of $5.46~\mj$.  We argue that \thisplanetone is likely to be a planet or low-mass brown dwarf based on several lines of reasoning.  First, substellar companions at the upper end of the allowed mass range and with this period are known to be relatively rare (the so-called "brown dwarf desert" (e.g., \citealp{Grether:2006}).  Second, no brown dwarfs (BDs) are known that are as highly inflated as \thisplanetone \citep{Zhou:2019b}.  Inspection of Figure 9 of \citet{Zhou:2019b} reveals that not only do no BD have radii as large as \thisplanetone, only about a dozen lower-mass transiting planets have radii this large.  Given the large surface gravity expected if \thisplanetone had a mass significantly above the deuterium burning limit, this is strong circumstantial evidence that it is less massive.  Finally, there is no evidence of Doppler beaming \citep{loeb:2013} or ellipsoidal variability \citep{Drake:2003} in the \thisplanetone \textit{TESS} light curve, which would likely be expected if \thisplanetone had a mass substantially above the deuterium burning limit. The weight of evidence indicates that \thisplanetone is most likely a giant planet or very low-mass brown dwarf.  

\subsubsection{Intrinsic Variability of KELT-25}
\label{sec:variability}

As mentioned previously, both \thisstarone and \thisstartwo are inside the instability strip.  Only about 40\% of stars within this range of $\teff$ show $\delta$ Scuti pulsations (\citealt{Murphy:2019}; see their Fig. 11).  While \thisstartwo does show variability consistent with $\delta$ Scuti pulsations (see \S\ref{sec:discussK26}), we find no evidence of intrinsic variability in the \thisstarone \textit{TESS} light curve.

\subsection{Tidal Evolution and Irradiation History}
\label{sec:orbevolution}

\begin{figure}[ht!]
\vspace{.1in}
\centering
\includegraphics[width=1\linewidth]{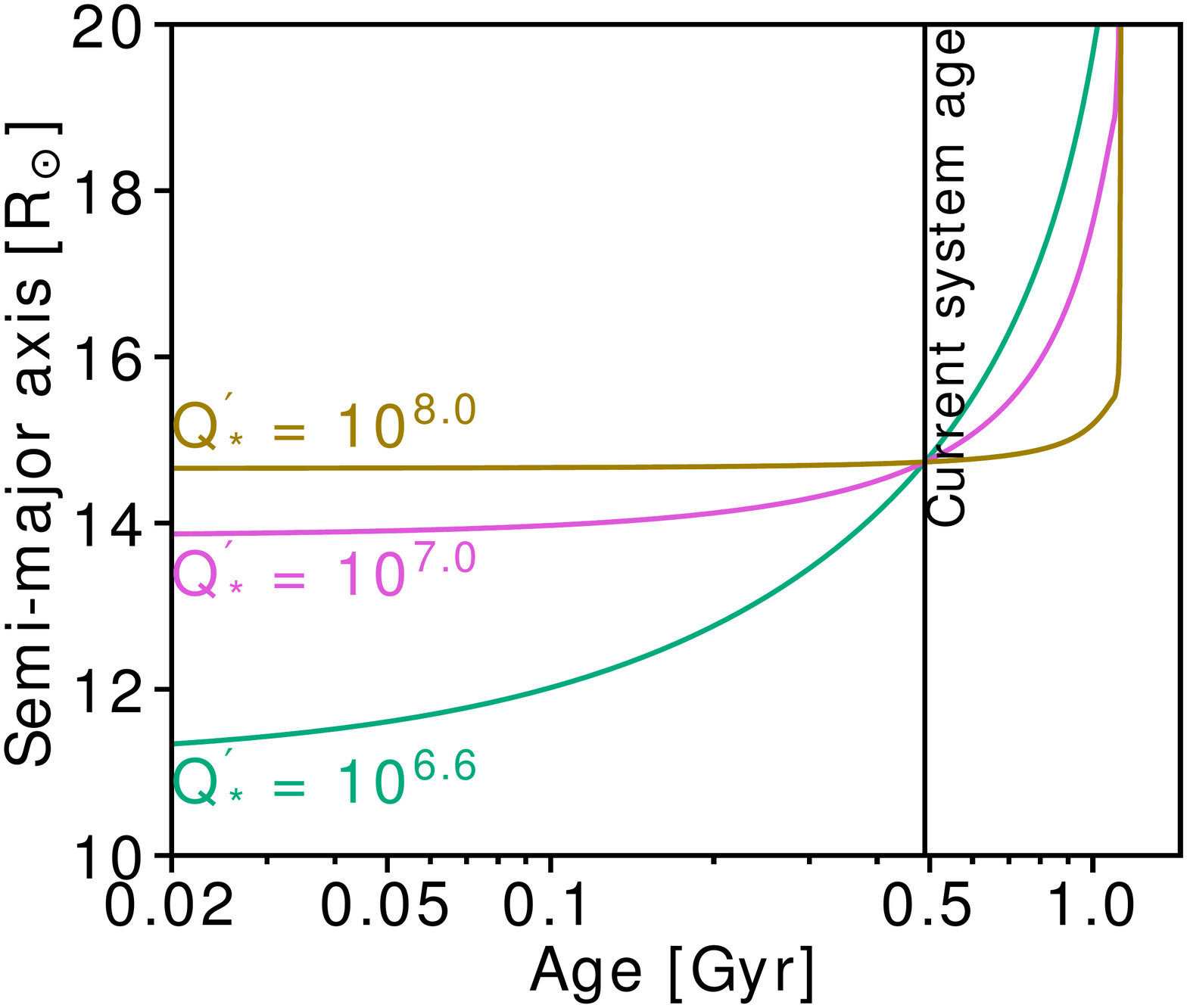}

\includegraphics[width=1\linewidth]{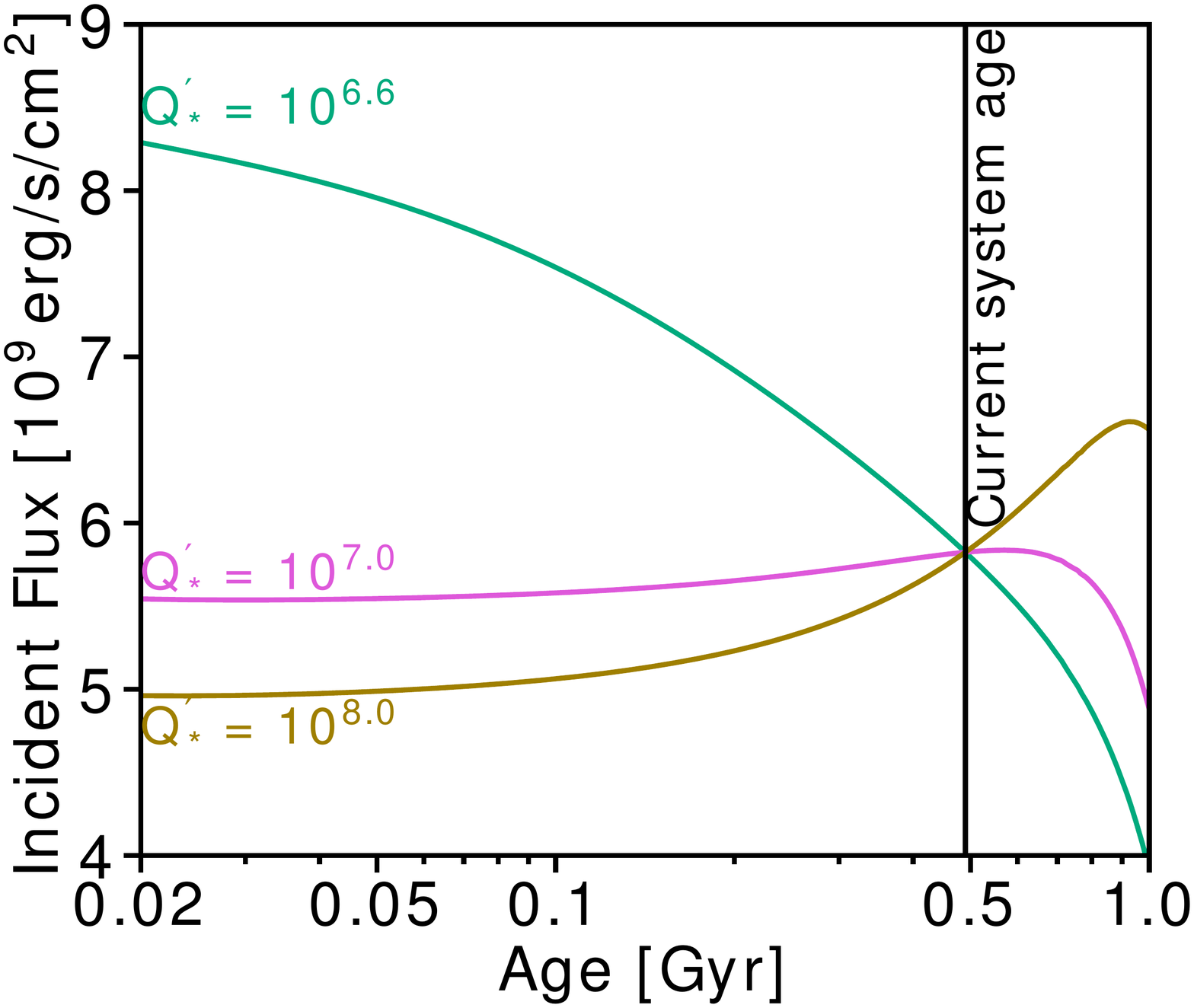}
\caption{Evolution of the semi-major axis in units of stellar radius (\textbf{Top}) and incident flux (stellar radiation) as a function of stellar age (\textbf{Bottom}) of \thisplanetone for constant values of $Q'_{\star}$ between $Q'_{\star}=10^{6.6}$ (turquoise line), $Q'_{\star}=10^7$ (pink) and $Q'_{\star}=10^8$ (gold). For any given dissipation parameter, the semi-major axis increases with time, while the incident flux decreases.
}
\label{fig:evolution25} 
\end{figure}

\begin{figure}[!ht]
\vspace{.1in}
\centering
\includegraphics[width=1\linewidth]{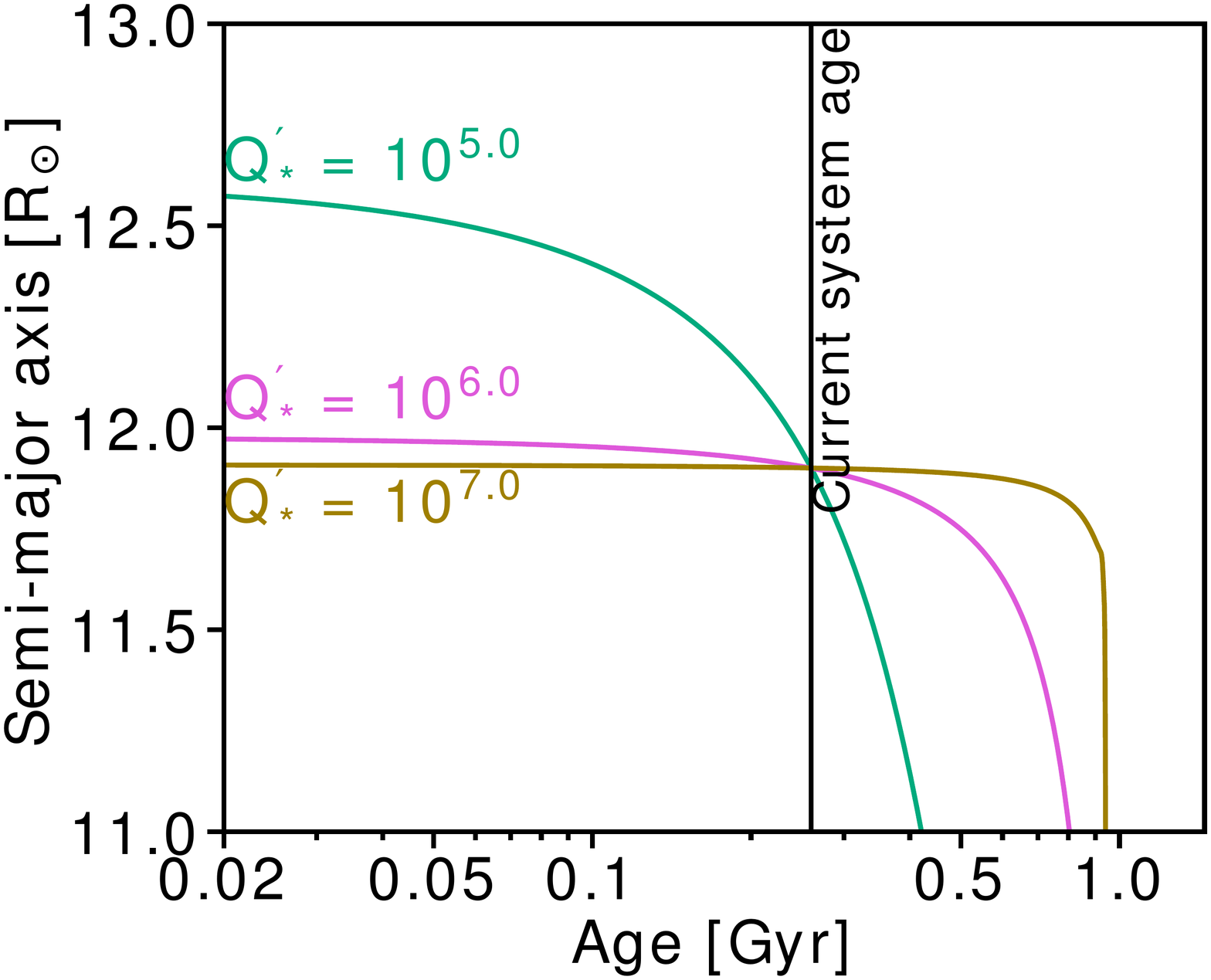}

\includegraphics[width=1\linewidth]{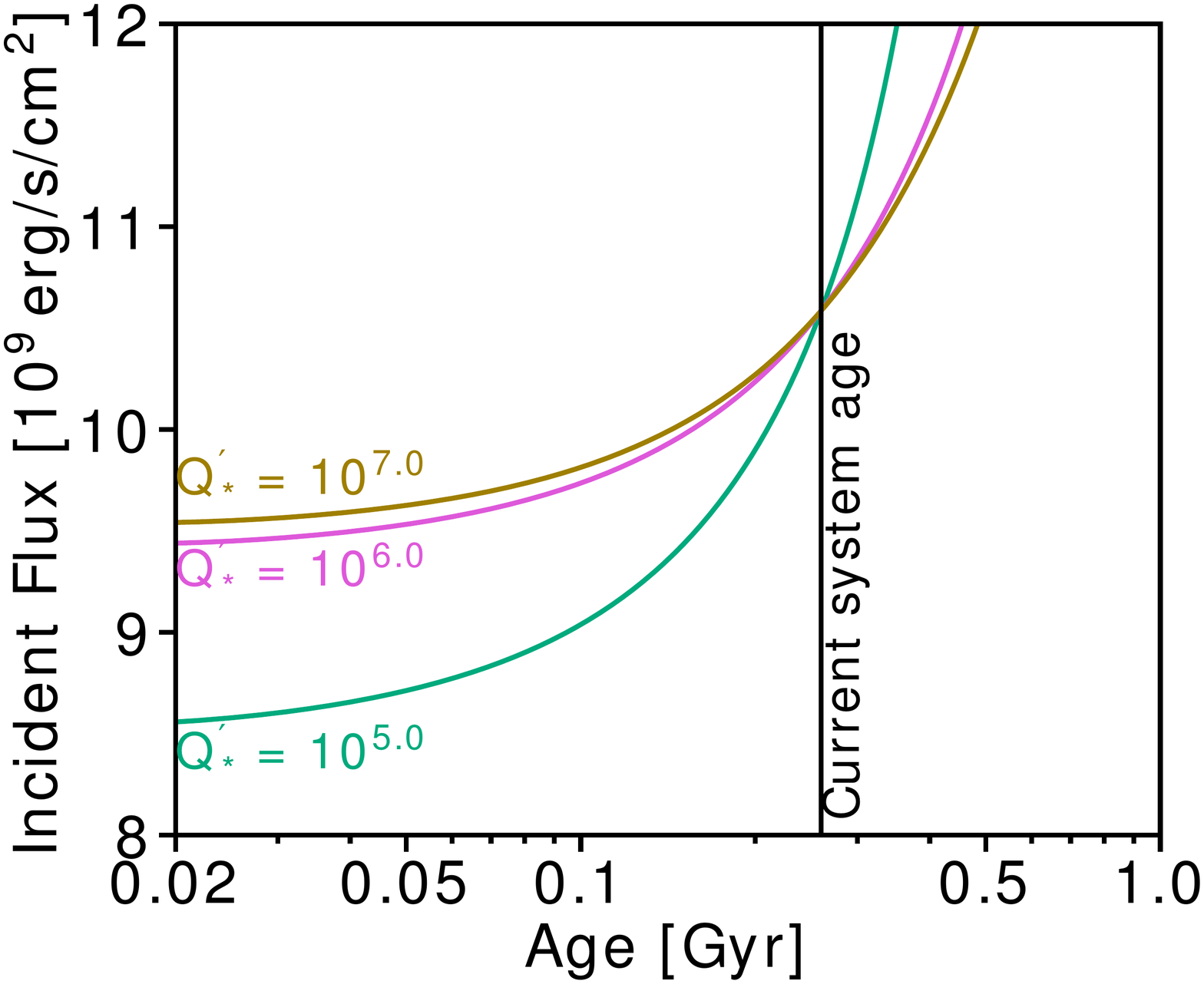}
\caption{Evolution of the semi-major axis in units of stellar radius (\textbf{Top}) and incident flux (stellar radiation) as a function of stellar age (\textbf{Bottom}) of \thisplanettwo for constant values of $Q'_{\star}$ between $Q'_{\star}=10^{5}$ (turquoise line), $Q'_{\star}=10^6$ (pink) and $Q'_{\star}=10^7$ (gold).
}
\label{fig:evolution26} 
\end{figure}

\begin{figure}[!ht]
    \vspace{.0in}
    \centering\includegraphics[width=0.99\columnwidth, trim = 0.0in 0.0in 0.0in 0.0in]{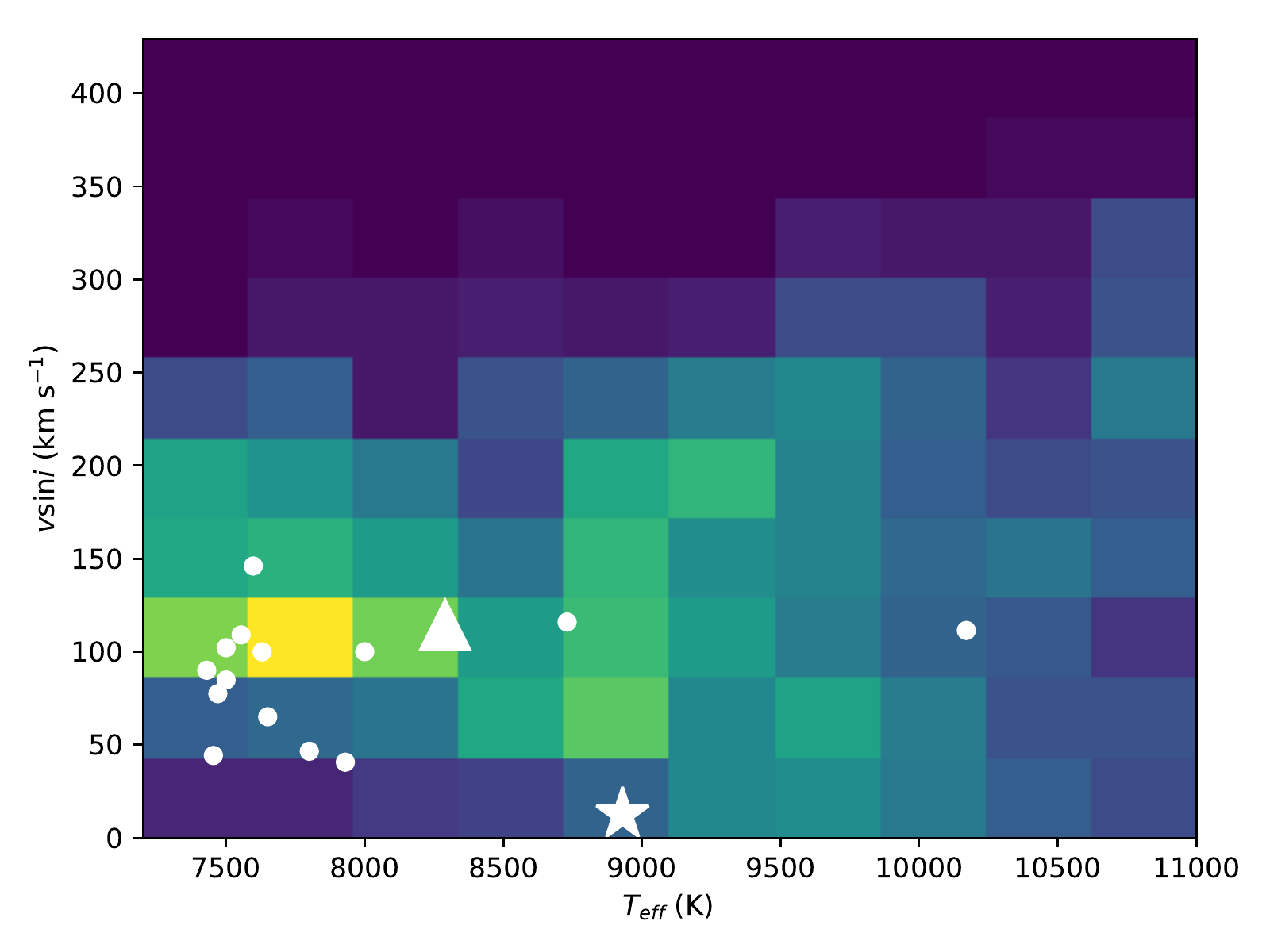}
    \caption{Distribution of rotational velocities in units of km s$^{-1}$ as a function of stellar effective temperature for all the measured A-type planet hosts in the literature. The color scale is proportional to the fraction of A-stars at that effective temperature that lie within each bin - warmer colors indicate a higher fraction of A-stars. The sample is from \citet{zorec:2012}. The big triangle and star represent \thisstarone and \thisstartwo, respectively. \thisstartwo displays an unusually slow rotation for its temperature, which could be the result of the orientation of its spin-axis along our line of sight rather than an intrinsic slow rotation.}
    \label{fig:vsinIvsTeff}
\end{figure}

\begin{figure}[!ht]
    \vspace{.0in}
    \centering\includegraphics[width=0.99\columnwidth, trim = 0.0in 0.0in 0.0in 0.0in]{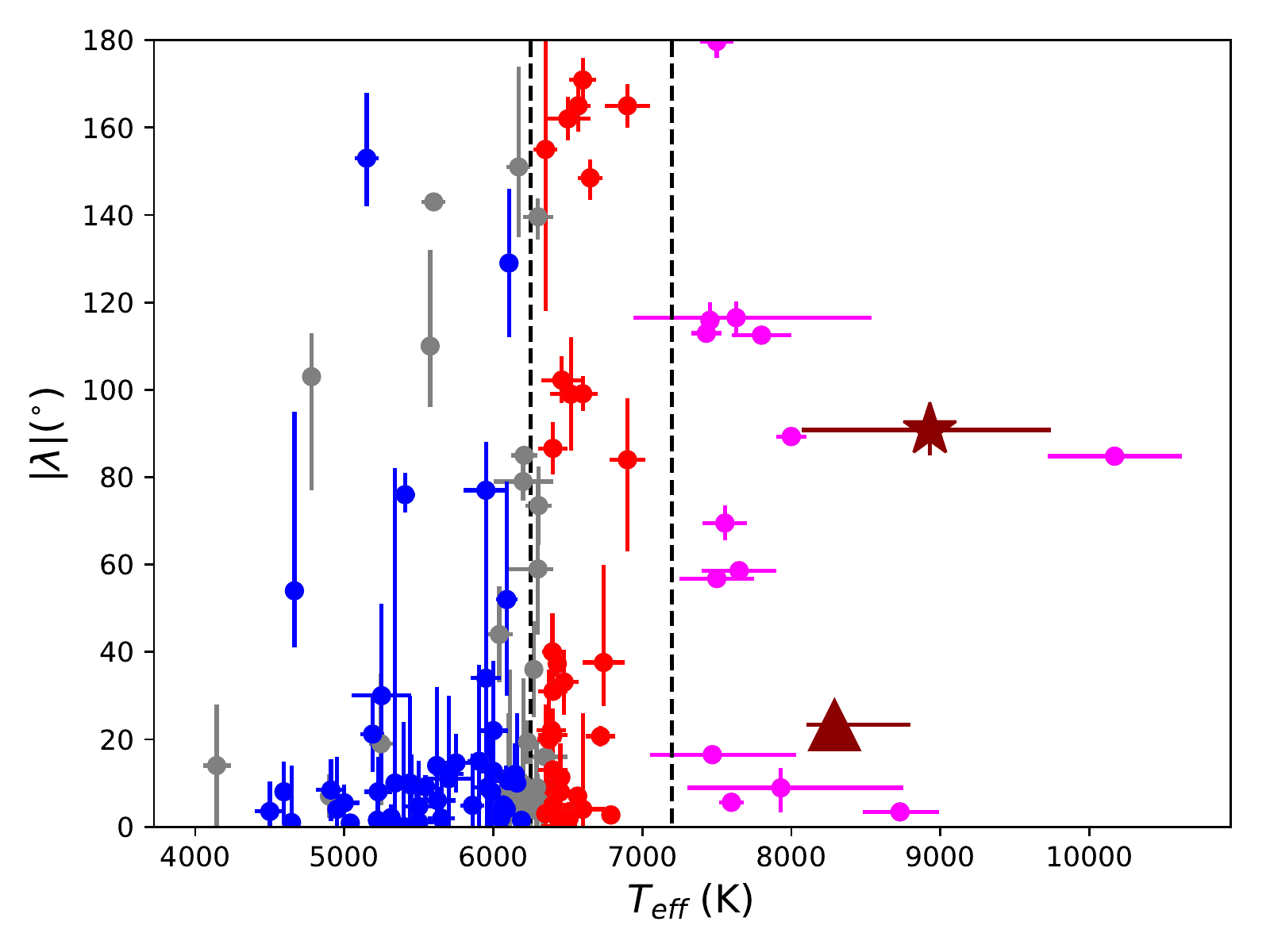}
    \caption{Distribution of projected spin-orbit misalignments $\lambda$ as a function of stellar effective temperature for all the measured hot Jupiters in the literature (the format of this plot is derived from \citealp{Winn:2010}). Planets around cool stars ($T_{\rm eff} <6250$ K) are represented by blue dots; the red dots represent hot stars ($T_{\rm eff} >6250$ K), while those with uncertainties in $\lambda$ $> 20\degr$ are colored in gray. The dashed vertical line marks the location of the Kraft break (Left) and the approximate dividing line between F and A spectral types. A-stars are shown in magenta. The crimson star and triangle depict \thisplanettwo and \thisplanetone, respectively. The literature sample was taken from John Southworth's TEPCat Rossiter-McLaughlin catalog\footnote{https://www.astro.keele.ac.uk/jkt/tepcat/}}
    \label{fig:lambdavsTeff}
\end{figure}

We estimated the orbital and irradiation evolution of \thisstarone and \thisstartwo; in particular, we calculated the history of the companions' semi-major axis and irradiation as a function of stellar age using the Planetary Orbital Evolution due to Tides (POET; \citealp{penev:2014}). POET assumes a constant tidal phase lag or quality factor $Q_{\star}$, a circular orbit, and no perturbations in the orbits due to unseen stellar or planetary companions. We further assumed that the tides raised by the planet or substellar companion are negligible and that the evolution of the planet’s orbit is therefore dominated by the dissipation of tidal perturbations in the star (as explained in \citealp{Rodriguez:2019B}). We accounted for the changes in stellar radius and luminosity in time by using a MIST stellar evolutionary track corresponding to the best-fit stellar properties (see \S\ref{sec:exofast}). Because of the large uncertainties in the knowledge of the tidal dissipation in stars and the tidal quality factor, $Q_{\star}$, we consider different constant values of the dissipation parameter $Q’_{\star}$, namely,  $Q’_{\star}= 10^{6.6}$, $10^{7}$ and $10^{8}$ for \thisstarone and $Q’_{\star}= 10^{5}$, $10^{6}$ and $10^{7}$ for \thisstartwo, where $Q’_{\star}$ is just proportional to $Q_{\star}$.
With these assumptions, we proceeded to calculate the past and future evolution of the semi-major axis (in units of the stellar radius) as a function of the age of the system. For KELT-26b, we can see in Figure~\ref{fig:evolution26} (Top) that for every assumed dissipation parameter $Q’_{\star}$, the planet’s orbit moderately decays until the present age of the system. Beyond that point, the future evolution of the planet’s orbit strongly depends on $Q’_{\star}$: for $Q’_{\star}= 10^{5}$, the planet gradually falls into the star within 500 Myr. For higher values of $Q’_{\star}$, the planet would take longer to be engulfed by its host star, perhaps surviving the entire stellar lifetime. As a consequence of its decaying orbit, \thisplanettwo's stellar radiation increases for all assumed dissipation parameters, as expected. We further note that \thisplanettwo has remained subjected throughout its lifetime to radiation above the $2\times10^{8}$ erg$\rm s^{-1} \rm cm^{-2}$ insolation threshold established in \citet{Demory:2011}. This likely explains why \thisplanettwo is presently significantly inflated, with $R_{p}=1.9R_{J}$. 
In contrast to \thisstartwo, because \thisstarone is so rapidly rotating ($v\sin{I_*} = 114.2$ km $\rm s^{-1}$) and the stellar rotation period is probably shorter than the companion's orbital period, the object's semi-major axis increases with time, rather than decreases, for all physical values of $Q'_{\star}$. The lowest value of $Q’_{\star}$ of $10^{6.6}$ predicts the fastest orbital evolution. If $Q’_{\star}$ is close to $10^{8}$, the decreasing insolation due to the planet's expanding orbit will be offset by an increase in stellar radiation as the evolving host expands, resulting in a net increase in incident flux. For all these calculations, the $3\sigma$ upper limit on the mass of \thisplanetone was assumed. Because we did not have the moment of inertia for \thisstarone, we could not compute its rotational history, so all the orbital paths were calculated assuming that the spin period of the star has remained constant throughout its lifetime. Assuming that $Q_{\star}$ is constant, and as long as the orbital period has always remained longer than the stellar spin period, the assumption of a constant rotation does not affect the results. However, if that is not the case, the direction of the evolution would be reversed. Figures~\ref{fig:evolution25} and~\ref{fig:evolution26} show the semi-major and irradiation evolution of both systems. 


\section{Conclusion}
\label{sec:conclusion}
 
In this paper, we presented the discovery of \thisplanetone, an ultra-hot, sub-stellar companion in a 4.4-day orbit around a young, rapidly-rotating A star; and \thisplanettwo, a puffy Ultra Hot Jupiter on a highly inclined, 3.3-day orbit around a young, slowly rotating Am star. Both were also observed by the \textit{TESS} mission. These companions both have exceptionally high equilibrium temperatures and their host stars are bright, making them excellent candidates for follow-up observations. With a rotational velocity of \vsini = 114.2 km s$^{-1}$, \thisstarone is among the most rapidly rotating A stars with transiting companions, while \thisstartwo is in contrast among the slowest. The highly inflated radius of \thisplanettwo can provide constraints on empirical mass-radius relations for giant planets. The orbital evolution of \thisplanetone suggests that the semi-major axis is increasing over time, a rather unusual trend for hot Jupiters, which could provide insights into migration mechanisms for these giant planets. With now roughly a dozen exoplanets detected around A-stars, we begin to have a more comprehensive sample that enables a better understanding of the physical properties, formation and evolution of these systems.

\software{EXOFASTv2 \citep{Eastman:2013, Eastman:2017}, AstroImageJ \citep{Collins:2017}, SPC \citep{Buchhave:2010}}
\facilities{FLWO 1.5m (Tillinghast Reflector Echelle Spectrograph, TRES); Kilodegree Extremely Little Telescope (KELT); MINiature Exoplanet Radial Velocity Array (MINERVA); Las Cumbres Observatory at Tenerife (LCO TFN); University of Louisville Manner Telescope (ULMT, Mt. Lemmon); KeplerCam (FLWO 1.2m); Stacja Obserwacji Tranzyt\'{o}w Egzoplanet w Suwa\l{}kach (SOTES); CROW Observatory; Koyama Astronomical Observatory (KAO); Gemini-South Zorro}

\acknowledgements

J.E.R. was supported by the Harvard Future Faculty Leaders Postdoctoral fellowship. Work by G.Z. is provided by NASA through Hubble Fellowship grant HST- HF2-51402.001-A awarded by the Space Telescope Science Institute, which is operated by the Association of Universities for Research in Astronomy, Inc., for NASA, under contract NAS 5-26555.  D.J.S. is supported by the Penn State University's Eberly Research Fellowship. The Center for Exoplanets and Habitable Worlds is supported by the Pennsylvania State University, the Eberly College of Science, and the Pennsylvania Space Grant Consortium.

Support for this work was provided by NASA through Hubble Fellowship grant HST-HF2-51399.001 awarded to J.K.T. by the Space Telescope Science Institute, which is operated by the Association of Universities for Research in Astronomy, Inc., for NASA, under contract NAS5-26555.

MNG acknowledges support from MIT's Kavli Institute as a Juan Carlos Torres Fellow.

C.Z. is supported by a Dunlap Fellowship at the Dunlap Institute for Astronomy \& Astrophysics, funded through an endowment established by the Dunlap family and the University of Toronto. K.G.S. acknowledge support from the Vanderbilt Office of the Provost through the Vanderbilt Initiative in Data-intensive Astrophysics. T.N and A.Y. are also grateful to Mizuki Isogai, Akira Arai, and Hideyo Kawakita for their technical support on observations at Koyama Astronomical Observatory. CDK was supporteed by the Swarthmore College Provost's Office. This work is partly supported by JSPS KAKENHI Grant Numbers JP18H01265 and JP18H05439, and JST PRESTO Grant Number JPMJPR1775. J.L.-B. acknowledges support from FAPESP (grant 2017/23731-1). K.P. acknowledges support from NASA grants 80NSSC18K1009 and NNX17AB94G.

Funding for the \textit{TESS} mission is provided by NASA's Science Mission directorate.

This research has made use of the Exoplanet Follow-up Observation Program website, which is operated by the California Institute of Technology, under contract with the National Aeronautics and Space Administration under the Exoplanet Exploration Program.

Resources supporting this work were provided by the NASA High-End Computing (HEC) Program through the NASA Advanced Supercomputing (NAS) Division at Ames Research Center for the production of the SPOC data products.

This paper includes data collected by the \textit{TESS} mission, which are publicly available from the Mikulski Archive for Space Telescopes (MAST)

This work has made use of data from the European Space Agency (ESA) mission {\it Gaia} (\url{https://www.cosmos.esa.int/gaia}), processed by the {\it Gaia} Data Processing and Analysis Consortium (DPAC, \url{https://www.cosmos.esa.int/web/gaia/dpac/consortium}). Funding for the DPAC has been provided by national institutions, in particular the institutions participating in the {\it Gaia} Multilateral Agreement. This work makes use of observations from the LCO network. This research has made use of the NASA Exoplanet Archive, which is operated by the California Institute of Technology, under contract with the National Aeronautics and Space Administration under the Exoplanet Exploration Program.

Based on observations obtained at the Gemini Observatory, which is operated by the Association of Universities for Research in Astronomy, Inc., under a cooperative agreement with the NSF on behalf of the Gemini partnership: the National Science Foundation (United States), National Research Council (Canada), CONICYT (Chile), Ministerio de Ciencia, Tecnolog\'{i}a e Innovaci\'{o}n Productiva (Argentina), Minist\'{e}rio da Ci\^{e}ncia, Tecnologia e Inova\c{c}\~{a}o (Brazil), and Korea Astronomy and Space Science Institute (Republic of Korea). Some of the Observations in the paper made use of the High-Resolution Imaging instrument ‘Zorro at Gemini-South). Zorro was funded by the NASA Exoplanet Exploration Program and built at the NASA Ames Research Center by Steve B. Howell, Nic Scott, Elliott P. Horch, and Emmett Quigley.

This research made use of Lightkurve, a Python package for Kepler and TESS data analysis (Lightkurve Collaboration, 2018).


\textsc{Minerva}-Australis is supported by Australian Research Council LIEF Grant LE160100001, Discovery Grant DP180100972, Mount Cuba Astronomical Foundation, and institutional partners University of Southern Queensland, UNSW Australia, MIT, Nanjing University, George Mason University, University of Louisville,
University of California Riverside, University of Florida, and The University of Texas at Austin.

We respectfully acknowledge the traditional
custodians of all lands throughout Australia, and recognise their continued cultural and spiritual connection to the land, waterways, cosmos, and community. We pay our deepest respects to all Elders, ancestors and descendants of the Giabal, Jarowair, and Kambuwal
nations, upon whose lands the \textsc{Minerva}-Australis facility at Mt Kent is situated.
Resources supporting this work were provided by the NASA High-End Computing (HEC) Program through the NASA Advanced Supercomputing (NAS) Division at Ames Research Center for the production of the SPOC data products.


\bibliographystyle{apj}

\bibliography{KELT-25}



\end{document}